\documentclass[conference,compsoc]{IEEEtran}
\usepackage{amsmath,amssymb,amsfonts}
\usepackage{amsthm}
\newtheorem{definition}{Definition}
\usepackage{algorithmic}
\usepackage{graphicx}
\usepackage{microtype}
\usepackage{textcomp}
\usepackage{xcolor}
\usepackage[normalem]{ulem}
\usepackage{multirow}
\usepackage{colortbl}
\usepackage{hyperref}
\usepackage{makecell}
\usepackage{xparse}
\usepackage{xspace}
\usepackage{pifont}
\def\BibTeX{{\rm B\kern-.05em{\sc i\kern-.025em b}\kern-.08em
    T\kern-.1667em\lower.7ex\hbox{E}\kern-.125emX}}
\usepackage{soul}
\usepackage{tabularx}
\usepackage{booktabs}
\usepackage{algorithm}
\usepackage{collcell}
\hypersetup{
  hidelinks,
  pdfborder={0 0 0},
  hypertexnames=false
}
\definecolor{heatlow}{HTML}{4A7FB5}
\definecolor{heathigh}{HTML}{C85B5B}
\definecolor{tablepanel}{HTML}{E7EDF6}
\definecolor{tablehead}{HTML}{F3F5F8}
\definecolor{bestcell}{HTML}{E8F3E8}
\definecolor{secondcell}{HTML}{FFF2D6}
\newcommand{\best}[1]{\cellcolor{bestcell}\textbf{#1}}
\newcommand{\second}[1]{\cellcolor{secondcell}\underline{#1}}

\newcommand{\mypara}[1]{\noindent{\bf {#1}}\xspace}
\newcommand{\mysubpara}[1]{\noindent{\itshape #1}\xspace}
\IEEEoverridecommandlockouts
\newcommand{\method}{RAPID}

\newcommand{\mFID}{FID}
\newcommand{\mQA}{QA}
\newcommand{\mPK}{PK}
\newcommand{\mAES}{AES}
\newcommand{\mLP}{LP}
\newcommand{\mSS}{SS}
\newcommand{\mDE}{DE}
\newcommand{\mPN}{PN}
\newcommand{\mLG}{LG}
\newcommand{\mLtwo}{L2}

\begin{document}

\title{RAPID: A Real-Time Defense Against Unauthorized Model Distillation for Text-to-Image Services}

\author{
\IEEEauthorblockN{
Zihan Wang\textsuperscript{1},
Boheng Li\textsuperscript{2},
Rui Zhang\textsuperscript{1},
Wenshu Fan\textsuperscript{1},
Qingchuan Zhao\textsuperscript{3},\\
Tianwei Zhang\textsuperscript{2},
Hongwei Li\textsuperscript{1},
and Guowen Xu\textsuperscript{1,*}
}

\IEEEauthorblockA{
\textsuperscript{1}
University of Electronic Science and Technology of China,
Chengdu, China
}

\IEEEauthorblockA{
\textsuperscript{2}
Nanyang Technological University,
Singapore
}

\IEEEauthorblockA{
\textsuperscript{3}
City University of Hong Kong,
Hong Kong SAR, China\\
zihanwang@std.uestc.edu.cn, guowen.xu@uestc.edu.cn
}

}
\maketitle

\begin{abstract}
Diffusion-based text-to-image (T2I) models are increasingly used for visual content creation, making their proprietary generation capability a valuable intellectual property asset. 
However, this capability is vulnerable to black-box output-based distillation, where an adversary queries the service, collects prompt-image pairs, and trains an unauthorized substitute model that mimics its generation behavior.
Existing perturbation-based defenses mainly protect generated images by applying sample-wise optimization, making the perturbed images disruptive to unauthorized training. 
Although effective, such sample-wise optimization introduces substantial computation and latency, significantly reducing the usability of online T2I services.
A natural solution is to integrate the defensive perturbation into the generation process itself, such as the VAE decoder, allowing the protected model to generate defended images directly without sample-wise online optimization.
However, existing sample-wise optimization designs struggle to transfer to the shared decoder setting.
We empirically find that the defensive shared decoder induces a substantially smaller latent shift than sample-wise optimization, suggesting that objective reachability matters more than destructiveness in the shared decoder setting.
To overcome this limitation, we propose \method{}, a self-referenced latent maximization framework that removes external dependencies and directly encourages the same model update to induce consistently disruptive effects across all training samples, thereby improving reachability.
We further introduce a reconstruction-guided color regularization that blocks the latent shortcut and reinforces the visual disruption.
Extensive experiments on four T2I models and four datasets, with comparisons against five representative baselines, show that \method{} consistently degrades the generation quality of the substitute model while preserving service visual fidelity.
Our work establishes a paradigm for real-time protection against unauthorized distillation in deployed T2I systems.

\end{abstract}

\section{Introduction}

\begin{figure}[t]
    
    \centering
    \includegraphics[width=\columnwidth]{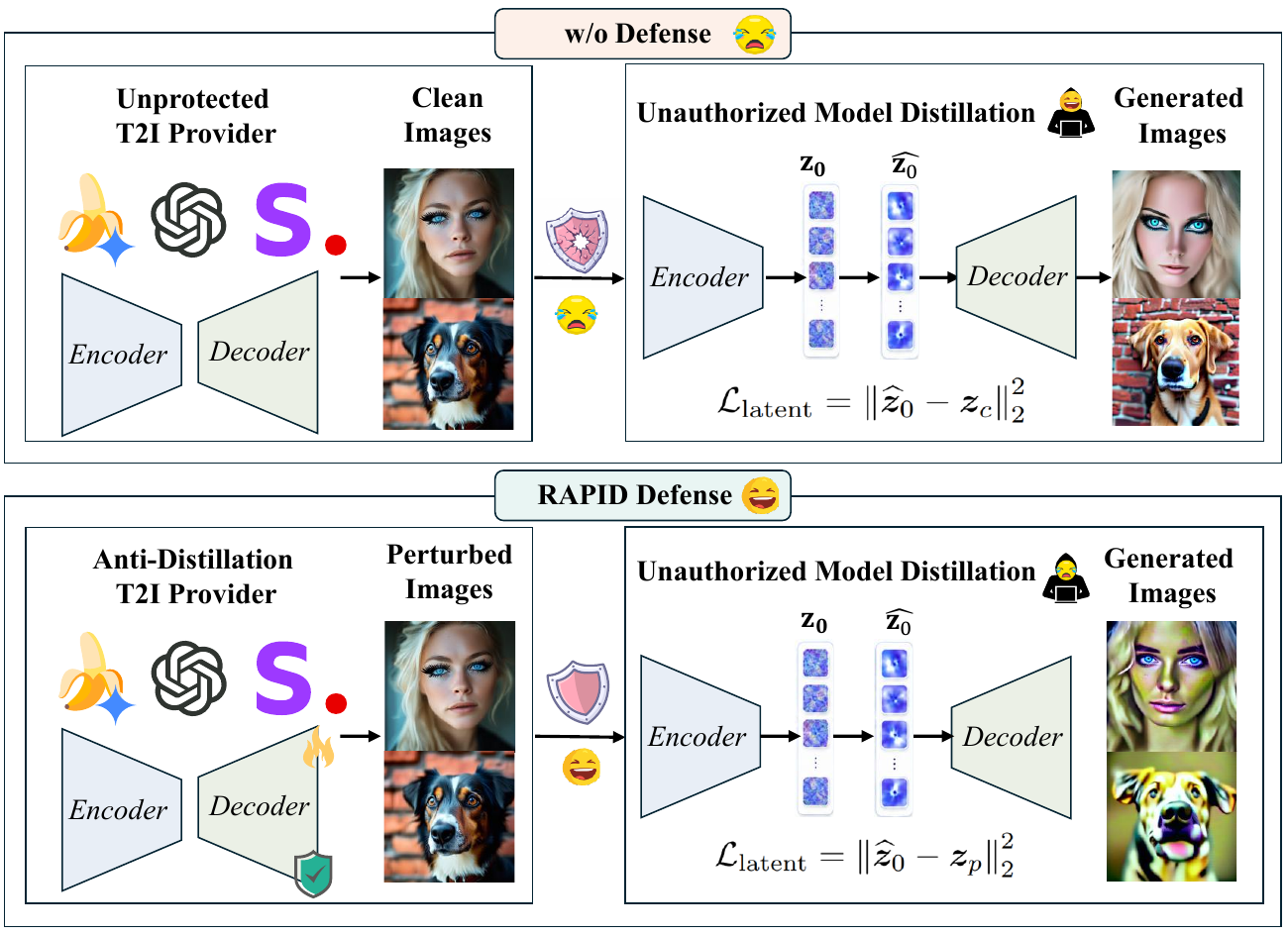}
    \vspace{-10pt}
    \caption{Application scenario and overview of \method{}. A provider adopting \method{} can substantially degrade the generation quality of the distilled model, whereas an unprotected provider cannot.}
    \label{fig:scenario}
\end{figure}
Diffusion-based text-to-image (T2I) generation has become a foundation of commercial image-generation services, with systems such as Stable Diffusion, DALL-E, Imagen, Midjourney, and Adobe Firefly~\cite{rombach2022high,ramesh2022dalle2,saharia2022imagen,midjourneyTerms2026,adobeGenAI2024} achieving high-quality, controllable synthesis through large-scale diffusion modeling~\cite{ho2020ddpm,ko2023visualArtists,epstein2023art}.
Building such systems demands substantial resources: curating and cleaning billion-scale image-text datasets~\cite{schuhmann2022laion5b}, designing complex training pipelines, and sustaining large-scale distributed computing over extended periods.
These investments make their outstanding generative capabilities valuable intellectual property.

However, there is a shortcut that avoids such substantial overhead.
Existing studies on black-box model stealing and distillation show that model behavior can often be approximated through query access alone~\cite{tramer2016stealing,jagielski2020high,orekondy2019knockoff,oliynyk2023modelStealingSurvey}. 
By repeatedly querying the target service and collecting prompt-image pairs, a malicious user can train an unauthorized substitute diffusion model to imitate the victim service's prompt alignment, visual style, output distribution, subject and style handling, and specialized generation behaviors~\cite{orekondy2019knockoff,salimans2022progressive,meng2023distillation,kim2023bksdm}.
Recent provider-facing reports also identify API-based distillation as an operational risk~\cite{anthropicDistillation2026,deepseekOpenAI2025}.
This threat is difficult to prevent because it can be mounted through black-box access to publicly returned outputs, and the behavior of a distiller is indistinguishable from that of a normal user. 
Unless large-scale industrial-level distillation is performed, it is difficult to distinguish such behavior~\cite{tramer2016stealing,orekondy2019knockoff,anthropicDistillation2026}. \looseness=-1

Recognizing this risk, major generative-AI service providers, including OpenAI, Google Gemini, Midjourney, and Stability AI, explicitly prohibit the use of generated outputs or API services for reverse engineering, developing competing services, or training competing models~\cite{openaiTerms2026,googleGeminiTerms2026,midjourneyTerms2026,stabilityTerms2025}.
However, policy restrictions alone do not technically prevent model distillation. A technical defense must reduce the training utility of the output stream collected by attackers while preserving the visual quality received by benign users.

Existing defenses against unauthorized T2I training can be broadly divided into post-verification, anomaly query detection, and preventive image perturbation. 
Post-verification methods, including watermarking and fingerprinting, embed detectable signals into model parameters or generated outputs so that a provider can later test whether a suspicious model or output derives from protected assets~\cite{uchida2017embedding,adi2018turning,zhang2020passport,fernandez2023stable,wen2023treering,diffusionshield2023,hmark2025,siren2025}. 
These methods support attribution and accountability, but the defender must first observe the suspicious model or its outputs before verification becomes possible. 
Anomaly query detection distinguishes benign user queries from suspicious distillation-oriented queries. 
For example, in Anthropic's distillation-detection setting for Fable 5, queries identified as distillation attempts may be routed to a weaker model Opus 4.8~\cite{anthropic_fable5_antidistillation_2026}. 
However, false positives may incorrectly flag benign queries and thereby significantly degrade the user experience~\cite{degradation1,degradation2}.
Although perturbation-based defenses offer direct protection for copyrighted content, they are primarily designed for offline settings involving fixed images, subjects, styles, or editing scenarios, in which each image is carefully optimized before release. 
Consequently, these methods typically require a separate perturbation to be learned for each image, incurring substantial computational overhead and resource consumption~\cite{liang2023advdm,shan2023glaze,shan2024nightshade,vanle2023antidreambooth,salman2023photoguard}. 
Such a sample-wise paradigm leads to a deployment mismatch for online services, which must protect all public outputs while preserving low latency, high throughput, and feasible GPU capacity~\cite{dean2013tail}, as verified in Section~\ref{subsec:CH1}.\looseness=-1

Motivated by previous work~\cite{fernandez2023stable}, a natural and effective solution is to internalize protection into the VAE decoder; this forward-pass design shifts protection from sample-wise post-processing to the deployed generation path, eliminating the online protection overhead.
Nevertheless, existing sample-wise optimization designs show limited effectiveness in disrupting query-based service distillation, even though they exhibit strong disruptive capability in sample-wise offline settings, as empirically verified in our experiments.
To investigate the underlying cause of this failure, we conduct a motivation study and obtain two key \textit{observations}. 
First, under shared decoder optimization, the optimization bottleneck shifts from destructiveness to ensuring the reachability of the objective. 
Second, a large latent shift does not necessarily induce stronger visual disruption. \looseness=-1

Motivated by these observations, we propose \method{}, a forward-pass protection mechanism against T2I distillation.
As illustrated in Figure~\ref{fig:scenario}, this forward-pass design relocates the protection mechanism from sample-wise post-processing to the generative pathway of the deployed decoder.
To understand why sample-wise optimization objectives exhibit poor effectiveness when adapted to a shared decoder, we examine the fundamental differences between sample-wise perturbation optimization and shared decoder finetuning. 
Specifically, we identify two key constraints governing the optimization process: the \textit{parameter-sharing restriction} and the \textit{visual-preservation restriction}. 
Based on these constraints, we formulate the \emph{Reachable Set} as a central requirement for reachability, which comprises visually feasible shared decoder updates capable of satisfying the objective up to a prescribed threshold $\tau$.
Guided by this, we further analyze two failure modes of existing methods: target dependence and stochasticity dependence.
To address these limitations, we design a self-referenced objective that eliminates external dependence and better matches the shared decoder structure.
Moreover, we introduce a reconstruction-guided color regularization term to avoid the latent shortcut and enhance disruption effectiveness against model distillation.
Together, these designs produce perturbations that are more reachable for the shared decoder and more effective at disrupting downstream distillation.\looseness=-1

Extensive experiments on four T2I models and four datasets, together with comparisons against five representative baselines, show that \method{} consistently degrades substitute model generation quality while preserving benign service visual fidelity.
Our work establishes a new service-level paradigm for real-time protection against unauthorized distillation in deployed T2I systems.
Our contributions are summarized as follows:\looseness=-1

\begin{itemize}
    \item We introduce forward-pass protection for T2I distillation defense, which embeds defensive perturbations into the deployed generation pipeline through decoder finetuning, thereby eliminating the online overhead of sample-wise protection.
    \item We provide a reachability analysis for shared decoder optimization by characterizing the reachable set of feasible decoder updates. 
    This analysis motivates our self-referenced latent maximization objective. We further introduce reconstruction-guided color regularization to mitigate latent shortcuts and composite visual constraints to preserve visual quality. \looseness=-1
    \item We conduct a comprehensive evaluation on four T2I models and four datasets, together with comparisons against five representative baselines. 
    The results show that \method{} achieves the strongest disruption across most evaluated configurations while preserving benign service visual fidelity.
\end{itemize}

\section{Preliminaries and Related Work}

\subsection{Latent Diffusion Models}

Latent diffusion models (LDMs) reduce the cost of high-resolution image synthesis by running the diffusion process in a learned compressed representation rather than directly in the pixel space~\cite{ho2020ddpm,rombach2022high,yang2023diffusionSurvey,croitoru2023diffusionVision}.
An autoencoding module, commonly implemented as a variational autoencoder, first learns an encoder $E$ that maps an image $x$ into a latent code $z=E(x)$ and a decoder $D$ that reconstructs the image from this latent representation~\cite{kingma2014vae,rombach2022high}.
The denoising network is then trained on noisy latents, often with text conditioning, to predict a clean latent representation whose semantics match the conditioning prompt~\cite{ho2020ddpm,rombach2022high,ramesh2022dalle2,saharia2022imagen}.
Most modern T2I systems, such as Stable Diffusion, follow this design as latent-space generation enables efficient high-resolution synthesis while largely preserving controllability and visual quality.~\cite{rombach2022high,podell2023sdxl}.
At inference time, an LDM starts from a sampled noisy latent and repeatedly applies the denoiser or sampler to obtain a prompt-aligned latent~\cite{song2021ddim,rombach2022high}.
The VAE decoder then converts this latent into the final image returned by the service.
This separation between latent generation and image decoding is central to our threat setting.
The denoiser and internal latents remain private to the provider, but the decoded image is released to the user and can therefore become training supervision for an unauthorized downstream model.
Consequently, the decoder is the last provider-controlled component before the output crosses the black-box service boundary, making it a natural location for service-internal protection.
\looseness=-1

\subsection{Query-Based Model Distillation}

Query-based model distillation converts public model access into a training signal for an unauthorized substitute model.
Prior work on model stealing, also widely referred to as model distillation, has shown that an adversary can approximate a victim model's functionality by querying a prediction API and training a local model on the observed input-output pairs~\cite{tramer2016stealing,orekondy2019knockoff,jagielski2020high,oliynyk2023modelStealingSurvey}.
In the T2I setting, the victim service $G_v$ receives a prompt $p_i$ and returns an image $y_i=G_v(p_i)$.
An attacker can similarly build a stolen dataset $\mathcal{D}_{steal}=\{(p_i,y_i)\}_{i=1}^{n}$ through repeated service queries~\cite{wang2023diffusiondb}.
The attacker then trains an unauthorized substitute model $G_a$ on these collected pairs, often using an open-source diffusion backbone or a distilled diffusion architecture as the starting point~\cite{hinton2015distilling,salimans2022progressive,meng2023distillation,kim2023bksdm}.
The goal of this attack is broader than copying individual returned images.
The attacker seeks to transfer the victim service's prompt alignment, visual style, output distribution, and specialized generation behavior into another model.\looseness=-1

\subsection{Defenses Against Unauthorized Distillation}

Existing defenses against unauthorized training are mainly designed for dataset-level offline protection or service-level abuse response, and can be broadly divided into post-verification methods~\cite{fernandez2023stable,wen2023treering,siren2025,adi2018turning,hmark2025}, service-side anomaly detection~\cite{juuti2019prada,anthropicDistillation2026}, and preventive perturbation~\cite{liang2023advdm,shan2023glaze,shan2024nightshade,salman2023photoguard,vanle2023antidreambooth} methods.\looseness=-1

\mypara{Post Verification.}
Post-verification methods, mainly watermarking and fingerprinting, embed detectable signals into model parameters, model behaviors, protected datasets, or generated outputs so that a provider can later test whether a suspicious model or output is derived from protected assets~\cite{uchida2017embedding,adi2018turning,zhang2020passport,diffusionshield2023,hmark2025,siren2025}.
Although these methods effectively support attribution and accountability, they are primarily retrospective: evidence can be verified only after a suspected model or its outputs are observed.
As a result, they do not prevent collection at the source and offer limited protection against misuse that remains unobserved.
Moreover, ordinary output watermarks may not reliably survive finetuning, retraining~\cite{radioactiveWatermarks2025}, and whether watermark traces remain detectable after distillation is uncertain.
Thus, post-verification alone is insufficient for preventing unauthorized T2I distillation. \looseness=-1

\mypara{Service-Side Anomaly Query Detection.}
Another line of defense monitors the public API interface and attempts to detect extraction behavior from usage traces.
Such systems can score accounts or sessions by query volume, timing, prompt diversity, repeated exploration of prompt or image-embedding space, near-duplicate requests, and deviations from benign query distributions; suspicious clients may then be throttled, challenged, logged for investigation, or denied high-throughput access~\cite{juuti2019prada,anthropic_fable5_antidistillation_2026}.
For example, Anthropic recently reported that Claude Fable 5 uses a detector to identify requests related to distillation and automatically falls back to Claude Opus 4.8 when such requests are flagged~\cite{anthropic_fable5_antidistillation_2026}.
However, these detectors depend on behavioral assumptions that a defense-aware attacker can manipulate.
An adaptive attacker can spread queries across accounts, mix extraction prompts with ordinary traffic, sample prompts from public distributions, slow down request rates, or use active-learning strategies that appear closer to benign usage~\cite{pal2020activethief}.
Moreover, this does not prevent model distillation at its source; rather, it serves more as an auxiliary defense mechanism.
Therefore, anomaly detection is valuable for operational monitoring and abuse triage, but it is not enough for source-level protection against model distillation. \looseness=-1

\mypara{Preventive Image Perturbation.}
Preventive image-perturbation methods utilize adversarial optimization to modify released images so that downstream generative training, personalization, or editing receives a harmful or unreliable signal~\cite{liang2023advdm,shan2023glaze,shan2024nightshade,vanle2023antidreambooth,liu2024metacloak,salman2023photoguard}.
One line of work perturbs latent representations.
For example, Glaze~\cite{shan2023glaze} moves artwork toward a misleading target style in latent space, and PhotoGuard~\cite{salman2023photoguard} includes an encoder-side protection that drives an editing model toward an incorrect representation.
Such latent-space objectives are comparatively stable because they optimize against a fixed representation path rather than unrolling the full downstream training process.
Another line of work directly targets diffusion learning or the personalization process.
AdvDM~\cite{liang2023advdm}, Anti-DreamBooth~\cite{vanle2023antidreambooth}, and MetaCloak~\cite{liu2024metacloak} optimize perturbations against diffusion training.
These diffusion-process objectives are more expensive and less stable because they must approximate downstream training dynamics; Anti-DreamBooth, for instance, uses bilevel optimization that alternates between adversarial-image updates and surrogate personalization training.\looseness=-1
Overall, although existing perturbation-based defenses are effective, they primarily operate as offline, sample-wise protections for fixed images, subjects, styles, or editing targets. 
These methods incur substantial computational and memory overhead, making them unsuitable for online service scenarios where additional per-output optimization is not permitted.
We therefore study an intrinsic alternative, where protection is learned offline and embedded into the deployed generation path, such that each public output is protected within a single forward pass.

\section{Threat Model}

\mypara{Scenario.}
We study black-box model distillation against a public T2I service.
The defender is the service owner or operator who provides generation access to users through a public interface while keeping the model weights, training data, internal latents, and defense parameters private. 
The attacker is a malicious user who accesses the same public interface as normal users.
By repeatedly querying the service, the attacker records the submitted prompts and the corresponding returned images, thereby constructing a prompt-image dataset for unauthorized training.

\mypara{Attacker Capabilities.}
The attacker has only black-box access to the deployed service. 
The attacker can choose arbitrary prompts, issue repeated queries, store returned images, and train a substitute model using the collected prompt-image pairs.
We consider a strict scenario in which the attack is not subject to a query budget, since stealing a state-of-the-art closed-source T2I model can bring substantial economic benefits. \looseness=-1

\mypara{Defender Capabilities.}
The defender has full control over the deployed T2I model.
Specifically, the defender can access and modify model parameters, train the full model or selected modules, and apply service-side output processing before returning images to users.
The defender does not know the training strategy adopted by the downstream attacker. \looseness=-1

\mypara{Defender Goals.}
The defender has two defense goals: \textit{quality-preserving} for benign users and \textit{disrupting effectiveness} for unauthorized T2I training.
\textit{Quality-preserving} means the protection should minimize visible quality loss in public service outputs.
Protected images should remain close to normal service outputs and avoid obvious color shifts, high-frequency noise, local artifacts, or semantic drift that would degrade user experience.
\textit{Disrupting effectiveness} means the same protected images should disrupt their usefulness as supervision for model distillation.
When an attacker trains a substitute model on collected protected outputs, the resulting model should exhibit degraded generation quality compared with a model trained on clean outputs.\looseness=-1

\section{Challenge and Motivating Studies}
\label{subsec:motivationstudy}
We analyze the challenges of applying existing perturbation-based methods to distillation scenarios and provide insights for method design through motivation studies. \looseness=-1

\subsection{Challenge I: Excessive Online Overhead}\label{subsec:CH1}
Sample-wise protection optimizes a separate perturbation for each image as:
\begin{equation}
    \delta_i^*=\arg\min_{\delta_i}\mathcal{L}_{attack}(x_i+\delta_i).
\end{equation}
Here, $\delta_i^*$ is the optimized perturbation for image $x_i$, $\delta_i$ is the perturbation variable, and $\mathcal{L}_{attack}$ is the downstream attack-oriented loss minimized by the sample-wise optimizer.
This sample-wise optimization allows each image to receive an optimized perturbation tailored to its downstream objective, such as style cloaking, personalization disruption, prompt-specific poisoning, or diffusion-editing disruption~\cite{shan2023glaze,vanle2023antidreambooth,shan2024nightshade,salman2023photoguard,liang2023advdm}, thereby exhibiting strong disruption effectiveness.
However, the same sample-wise design creates a deployment mismatch for online T2I services.
The perturbation must be optimized after each image is generated, typically through repeated forward and backward passes on surrogate models.
Adding this iterative loop to every public output increases GPU cost, average latency, and tail-latency unpredictability. \looseness=-1

To evaluate the online protection cost of prior methods, we conduct experiments on an RTX A6000 on Glaze~\cite{shan2023glaze}, Nightshade~\cite{shan2024nightshade}, PhotoGuard~\cite{salman2023photoguard}, and one-step adversarial optimizers known for their efficiency, FGSM~\cite{goodfellow2015explaining} on 1024$\times$1024 resolution images.
The results are shown in Table~\ref{tab:latency_overhead}.
Although FGSM uses only a single optimization step, it still incurs 4.53 seconds of latency and 33 GB of memory overhead.
Other optimization-based methods increase latency to approximately two minutes.
More importantly, in high-throughput service deployments, such overhead scales proportionally with the service volume, leading to substantial resource waste and ultimately becoming unacceptable to both users and service providers.
Such overhead makes sample-wise optimization impractical for online T2I services, where each user query must be answered with low latency and high throughput~\cite{dean2013tail}.
This limitation motivates a forward-pass protection mechanism that avoids query-time optimization altogether.
Specifically, we internalize perturbation generation into the T2I model VAE decoder, so that protected images are produced during the standard forward generation process without adding sample-wise online optimization, eliminating the excessive overhead. \looseness=-1

\begin{table}[t]
\centering
\caption{Online protection overhead. $s$ denotes seconds and MB denotes megabytes.}
\label{tab:latency_overhead}
\scriptsize
\setlength{\tabcolsep}{2.5pt}
\renewcommand{\arraystretch}{1.10}
\begin{tabular}{@{}lccc@{}}
\toprule
\rowcolor{tablepanel}\multicolumn{4}{c}{\textbf{Additional Online Protection Overhead}} \\
\midrule
\rowcolor{tablehead}\textbf{Method} & \textbf{Opt Steps} & \textbf{Latency} & \textbf{CUDA Memory} \\
\midrule
\textbf{FGSM} & 1 & $4.53$s & $33097.27$MB \\
\textbf{PhotoGuard} & 200 & $118.94$s & $33587.55$MB \\
\textbf{Nightshade} & 200 & $120.27$s & $33587.55$MB \\
\textbf{Glaze} & 200 & $120.15$s & $33587.55$MB \\ \midrule
\textbf{\method{} (Ours)} & 0 & $0$s & $0$MB \\

\bottomrule
\end{tabular}
\end{table}

\subsection{Challenge II: Limited Disruption Effectiveness}\label{subsec:CH2}
The second challenge is that existing optimization designs provide limited downstream disruption after being adapted to an intrinsic decoder-level defense.
As shown in Figure~\ref{fig:generation_quality_comparison}, direct adaptations of representative protection objectives to intrinsic defense, including targeted optimization methods such as PhotoGuard~\cite{salman2023photoguard}, Glaze~\cite{shan2023glaze}, and Nightshade~\cite{shan2024nightshade}, as well as proxy diffusion loss optimization methods such as AdvDM~\cite{liang2023advdm} and Anti-DreamBooth~\cite{vanle2023antidreambooth}, lead to limited protection effectiveness, which is insufficient to disrupt unauthorized model distillation.
To investigate the reason why conventional perturbation objectives fail in this setting, we design the following motivating study using the CelebA~\cite{liu2015faceattributes} dataset, with PixArt~\cite{chen2024pixartsigma} as the protected source model, and SSD~\cite{segmind2023ssd1b} as the downstream distillation surrogate.
For each objective, we optimize protected images under the same visual budget, and the settings are aligned with those in the main experiments.
After the protection training, we feed each protected image $x_p$ into the downstream surrogate VAE to obtain its latent representation $E_s(x_p)$ and reconstructed image $\hat{x}_p = D_s(E_s(x_p))$, where $E_s$ and $D_s$ denote the surrogate VAE encoder and decoder.
For comparison, we also include a self-referenced latent maximization attack that maximizes downstream latent discrepancy, as well as our \method{}.
To quantify downstream disruption from both latent and visual perspectives, we compare the protected images $x_p$ with the clean outputs $x_r$, which are generated from the same latents using a clean decoder, in terms of both latent deviation and VAE reconstruction deviation.
Specifically, we measure latent deviation using MSE, and quantify reconstruction-visible deviation using \mLtwo{}, LPIPS, CIEDE2000, SSIM, and PSNR between the reconstructed version of $x_p$ and the clean reference image $x_r$. \looseness=-1

Table~\ref{tab:motivation_gap} shows the motivating study results.
The key observations are as follows: \looseness=-1

\mysubpara{$\bigstar$Observation I: Reachability matters more than nominal disruptiveness.}
As shown in Table~\ref{tab:motivation_gap}, PhotoGuard, Glaze, Nightshade, AdvDM, and Anti-DreamBooth induce only limited downstream latent gaps under the same visual budget.
These objectives are disruptive in their original sample-wise settings, where each perturbation is optimized independently, but this disruptiveness does not directly transfer to our shared decoder setting.
For example, PhotoGuard yields a latent gap of only 37.63, whereas Latent Max reaches 115.24.
This observation suggests that the critical factor is not only whether an objective is destructive, but whether the optimization design is reachable by the shared decoder optimization.
If their rewarded directions cannot be jointly expressed by one shared decoder, their downstream effect remains weak even when the original sample-wise objective is strong.\looseness=-1

\mysubpara{$\bigstar$Observation II: A large latent gap does not necessarily imply large visible disruption.}
Direct latent maximization produces the largest latent discrepancy, reaching 115.24.
However, the larger latent distance does not necessarily translate into stronger visual disruption, as its reconstruction-visible metrics remain much weaker than those of \method{}. 
Moreover, although Nightshade yields a smaller latent gap of 63.01 compared with 115.24 for Latent max, its visual discrepancy remains comparable or even more pronounced. \looseness=-1

This result suggests that latent-only maximization may exploit a shortcut in the downstream latent space, where the latent representation changes substantially while the reconstructed image quality remains comparatively less affected or even unaffected. 
We attribute this phenomenon to the constrained nature of shared decoder optimization: the model tends to find solutions that enlarge latent discrepancies while inducing limited visual changes, as these solutions are easier to reach within the feasible, narrow decoder optimization space. 
We refer to this behavior as the \textit{latent shortcut}. 
This motivates an optimization strategy that explicitly avoids this shortcut, thereby converting targeted perturbations into more effective downstream reconstruction disruption. 
\looseness=-1

\begin{table}[t]
\centering
\caption{Motivating study on disruption effectiveness. \mLG{} denotes latent MSE gap; \mLtwo, \mLP, \mDE, \mSS, and \mPN{} denote L2, LPIPS, CIEDE2000, SSIM, and PSNR, respectively.}
\label{tab:motivation_gap}
\footnotesize
\setlength{\tabcolsep}{2.2pt}
\renewcommand{\arraystretch}{1.10}
\begin{tabular}{@{}lrrrrrr@{}}
\toprule
\rowcolor{tablepanel}\multicolumn{7}{c}{\textbf{Motivating Disruption Study}} \\
\midrule
\rowcolor{tablehead}\textbf{Method} & \textbf{\mLG}$\uparrow$ & \textbf{\mLtwo}$\uparrow$ & \textbf{\mLP}$\uparrow$ & \textbf{\mDE}$\uparrow$ & \textbf{\mSS}$\downarrow$ & \textbf{\mPN}$\downarrow$ \\
\midrule
\textbf{Clean} & 1.93 & 0.0010 & 0.0001 & 0.06 & 0.999 & 60.74 \\
\midrule
\textbf{AdvDM} & 30.50 & 0.0169 & 0.0431 & 1.48 & 0.959 & 35.56 \\
\textbf{Anti-DB} & 28.88 & 0.0114 & 0.0446 & 1.18 & 0.964 & 39.07 \\
\textbf{PhotoGuard} & 37.63 & 0.0158 & 0.0254 & 1.36 & 0.969 & 36.35 \\
\textbf{Glaze} & 41.50 & 0.0194 & 0.0953 & 1.91 & 0.914 & 34.41 \\
\textbf{Nightshade} & 63.01 & \second{0.0353} & 0.1335 & \second{3.47} & 0.872 & \second{29.19} \\
\textbf{Latent max} & \best{115.24} & 0.0256 & \second{0.1365} & 2.80 & \best{0.829} & 31.94 \\ \midrule
\textbf{\method{}} & \second{103.55} & \best{0.0410} & \best{0.2086} & \best{7.17} & \second{0.830} & \best{27.89} \\

\bottomrule
\end{tabular}%
\end{table}

\subsection{Challenge III: Poor Visual Quality}\label{subsec:CH3}
\begin{figure}[t]
\centering
\setlength{\tabcolsep}{1.2pt}
\resizebox{0.8\columnwidth}{!}{%
\begin{tabular}{@{}cccc@{}}
\includegraphics[width=0.10\textwidth]{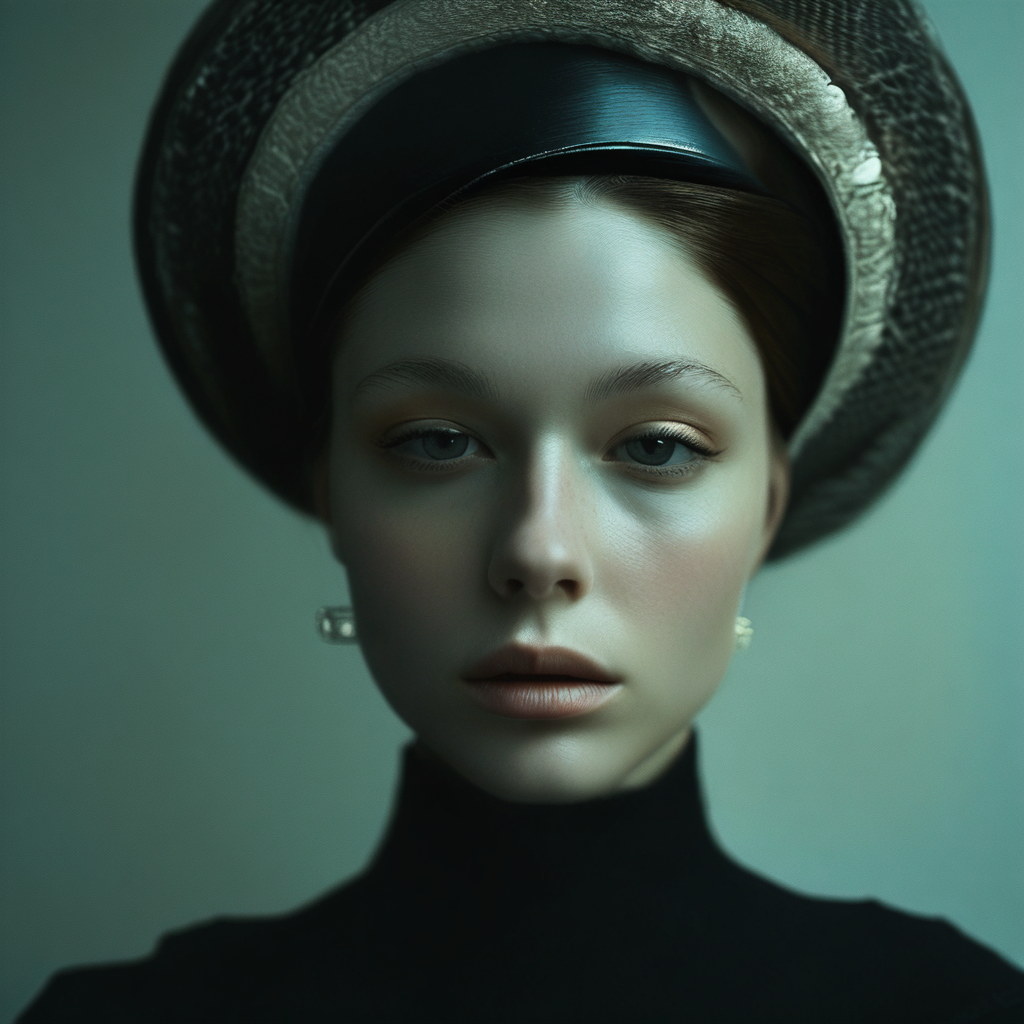} &
\includegraphics[width=0.10\textwidth]{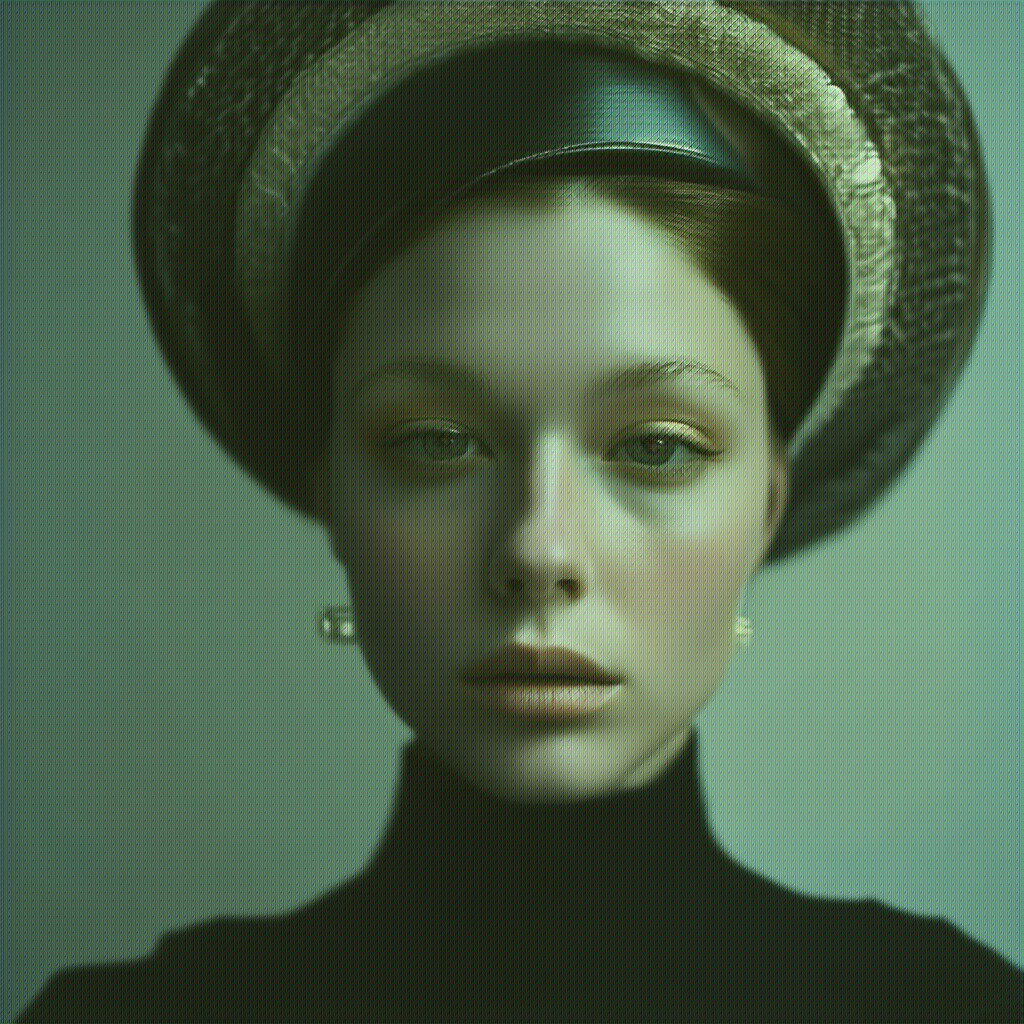} &
\includegraphics[width=0.10\textwidth]{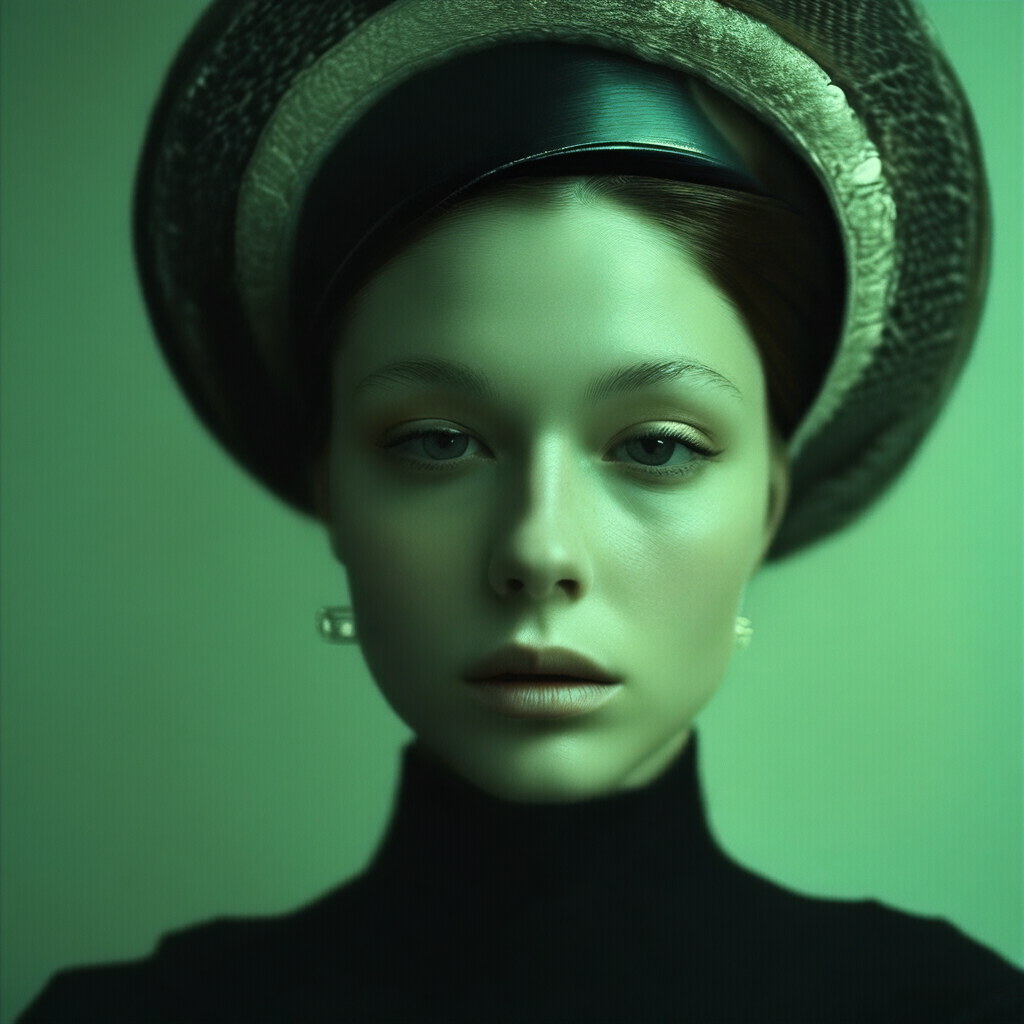} &
\includegraphics[width=0.10\textwidth]{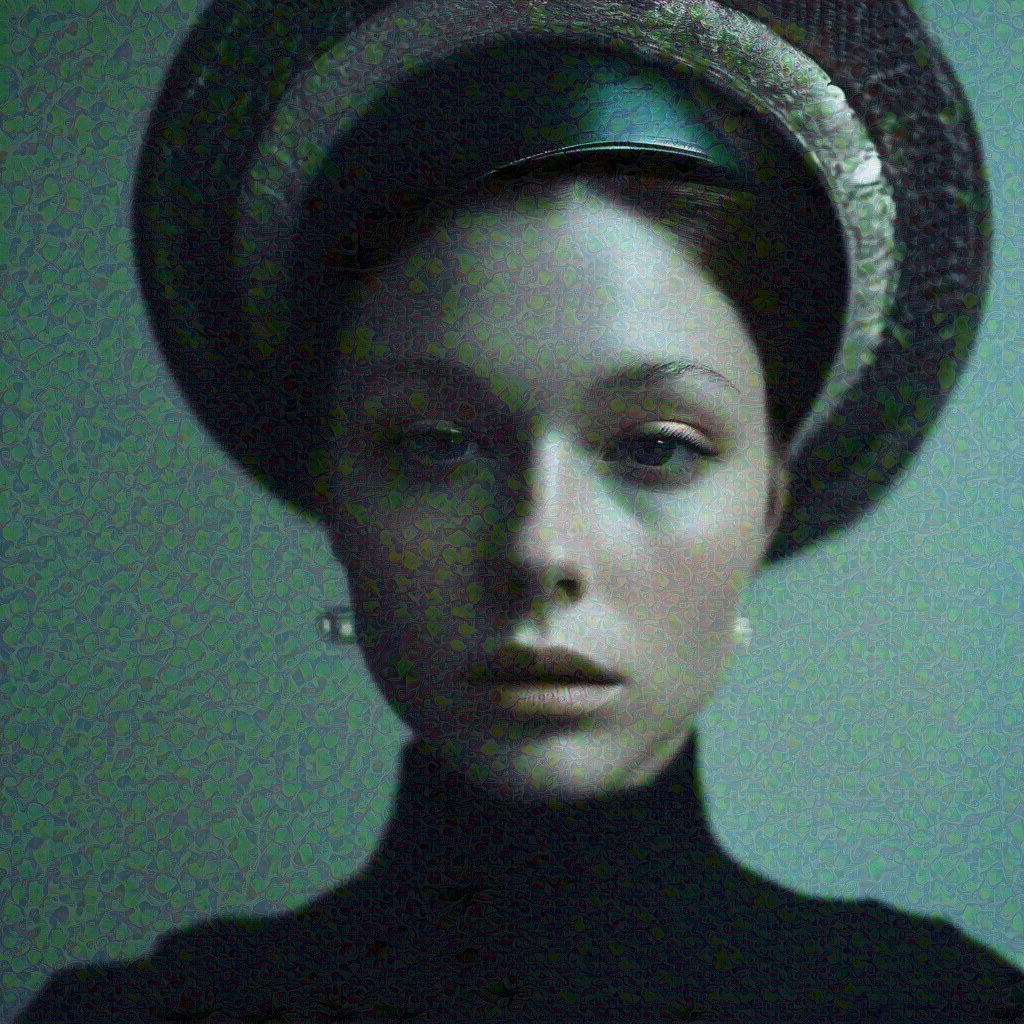} \\
\includegraphics[width=0.10\textwidth]{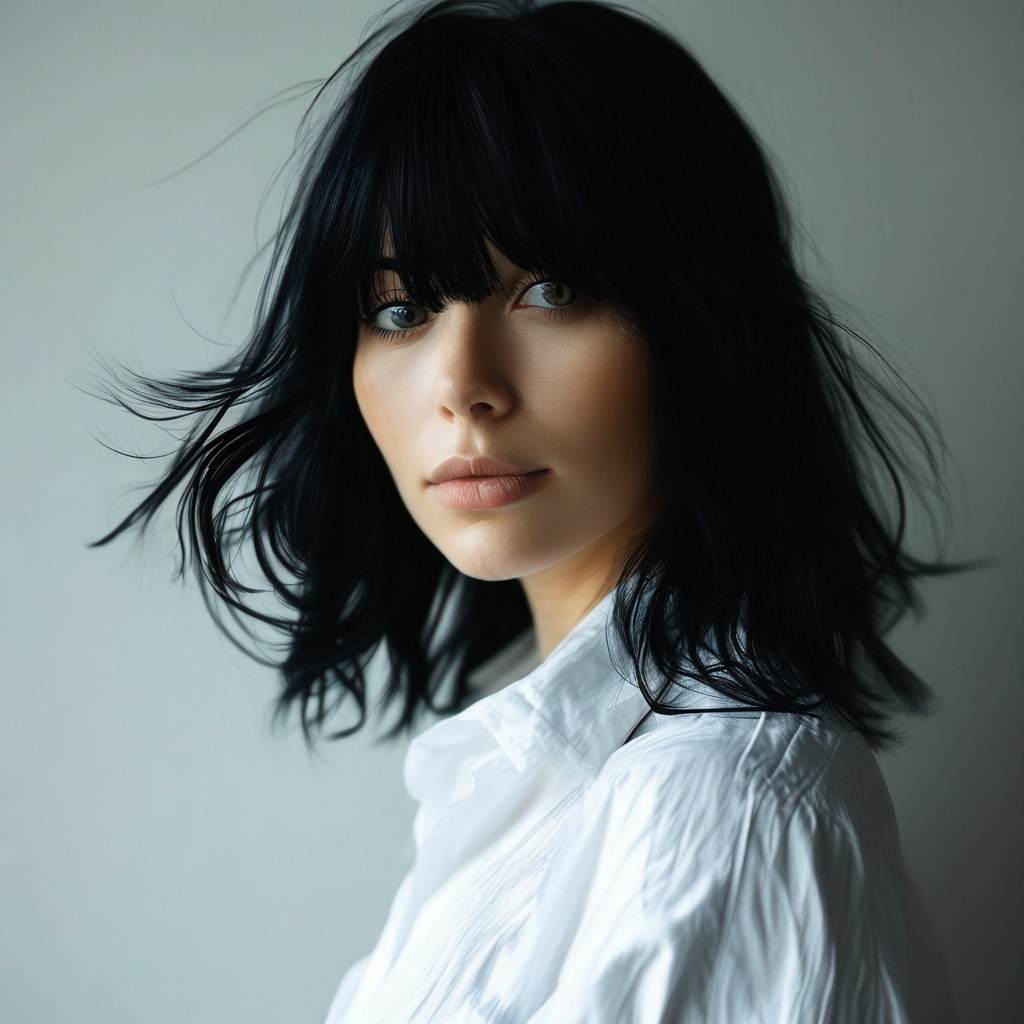} &
\includegraphics[width=0.10\textwidth]{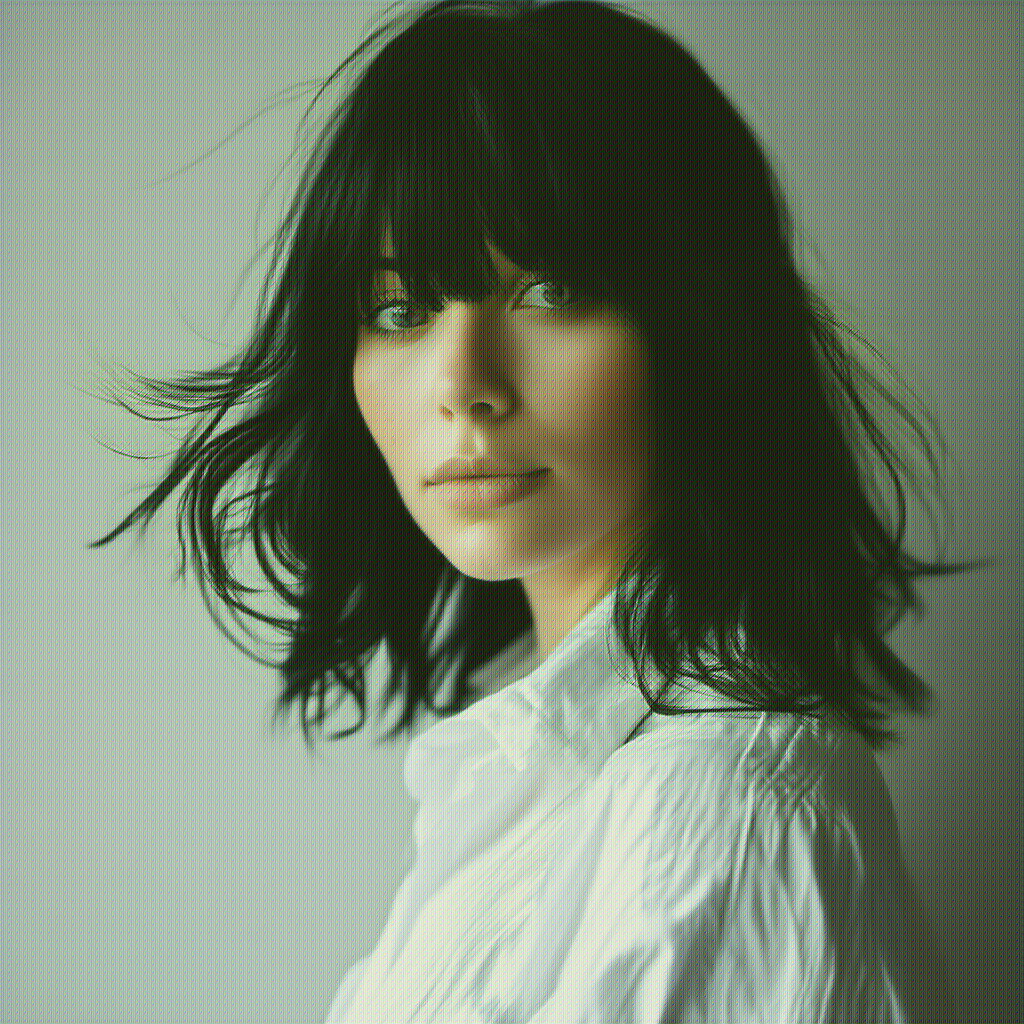} &
\includegraphics[width=0.10\textwidth]{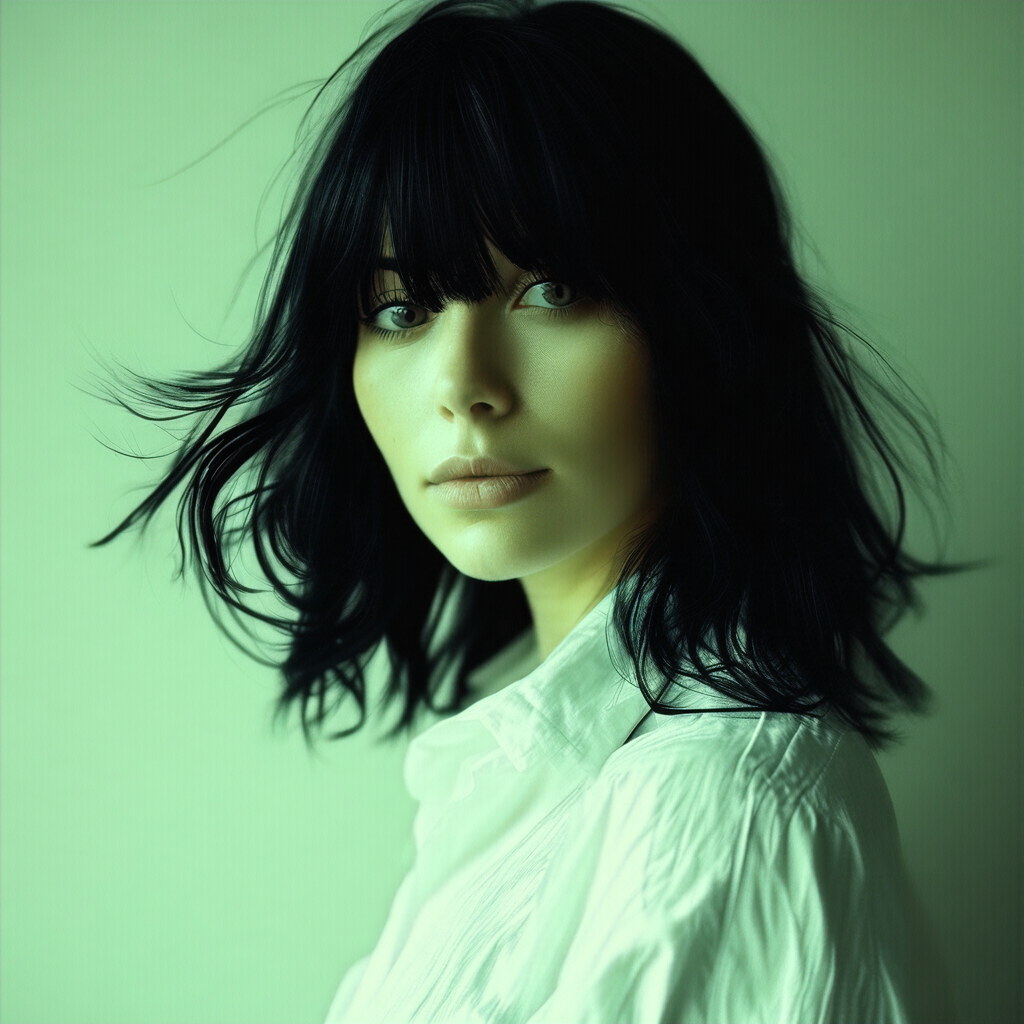} &
\includegraphics[width=0.10\textwidth]{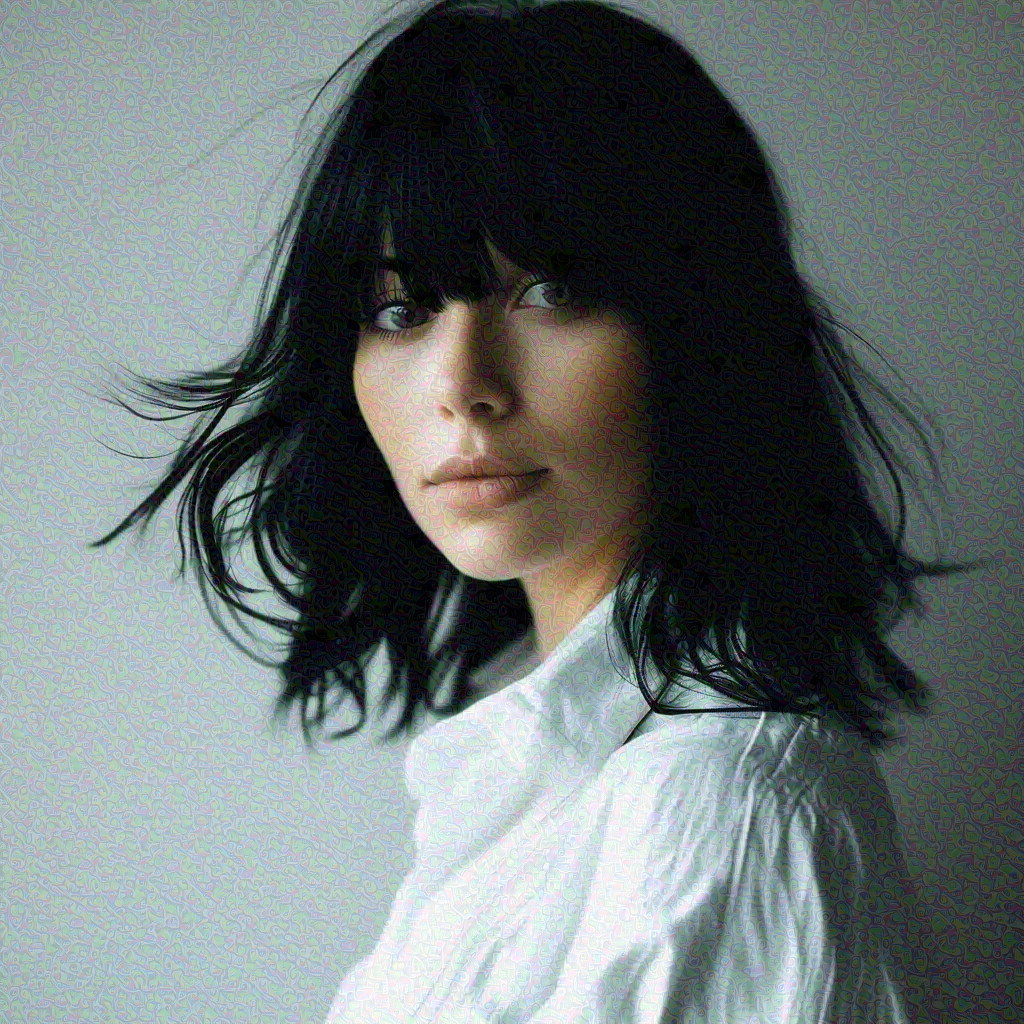} \\
\textbf{Reference} & \textbf{LPIPS} & \textbf{DSSIM} & \textbf{$l_\infty$} \\
\end{tabular}
}
\caption{Visual quality under different visual constraints in \method{} optimization on the CelebA dataset. Following prior work, we adopt the following thresholds: LPIPS = 0.05, DSSIM = 0.007, and $\epsilon$ = 32/255.
}
\label{fig:visual_quality_motivation}
\end{figure}

Traditional adversarial optimization generally uses a single visual constraint to bound the visual difference of adversarial examples and preserve their stealthiness~\cite{goodfellow2015explaining,madry2018pgd}.
Specifically, LPIPS-based constraints are used by Glaze and Nightshade, DSSIM-based constraints are used by Fawkes~\cite{shan2020fawkes}, and $l_\infty$-based constraints are widely used in prior image-protection systems~\cite{madry2018pgd}; these constraints have been verified effective in their original optimization settings~\cite{shan2023glaze,shan2024nightshade,shan2020fawkes,liang2023advdm,salman2023photoguard}.
However, when these losses are applied to our optimization setting, they fail to adequately preserve the visual quality of the generated images. 
The optimized images are shown in Figure~\ref{fig:visual_quality_motivation}.
We find that the optimized image may satisfy the single chosen constraint, such as LPIPS, DSSIM, or $l_\infty$, while still exhibiting visible color shifts, local artifacts, and high-frequency noise that would be unacceptable in a commercial T2I service.
We speculate that, under the challenging shared decoder optimization setting, relying on a single visual constraint leads to metric-specific overfitting. 
Consequently, the optimized images may still suffer from noticeable perceptual degradation relative to the original images, even when they satisfy the prescribed metric constraint.
Therefore, this observation motivates the use of multiple visual metrics in the optimization objective, with the visual fidelity term treated as the dominant serving constraint. \looseness=-1

\section{Methodology}
\method{} is designed around the three service-level challenges identified above.
First, it addresses the latency challenge by moving adversarial optimization offline into decoder protection training, which induces no sample-wise online optimization.
Second, \method{} analyzes the factors that limit disruption effectiveness after sample-wise objectives are transferred to a shared decoder, and uses this analysis to guide the method design.
Third, \method{} addresses the visual quality challenge with composite constraints that preserve color fidelity, perceptual structure, spatial smoothness, and local consistency. \looseness=-1

\subsection{Service-Internal Forward-Pass Protection}
\label{subsec:forward_pass_intrinsic_protection}

\method{} addresses the latency challenge by moving protection from online image optimization into the offline service-side VAE decoder optimization.
Inspired by prior work~\cite{fernandez2023stable}, the provider learns a protected VAE decoder offline and deploys it as part of the normal generation path.
Let $\theta_0$ be the parameters of the frozen reference decoder, with $D_0=D_{\theta_0}$, and let $\theta^\star$ be the protected decoder parameters learned offline.
During offline optimization, $D_\theta$ denotes the trainable protected decoder with current parameters $\theta$; after training, $\theta$ is set to $\theta^\star$ for deployment.
For a service latent $z$, the clean reference output and protected public output are \(x_r = D_{\theta_0}(z), x_p = D_{\theta^\star}(z)\).
Here, $z$ is produced by the denoiser, $x_r$ is the image the original decoder would have returned, and $x_p$ is the image returned by the protected decoder.
At deployment, the provider replaces $D_{\theta_0}$ with $D_{\theta^\star}$.
This design separates offline training cost from online serving cost, so that serving-time cost no longer scales with the number of protection optimization steps.\looseness=-1

\subsection{Reachable Shared Decoder Disruption}
\label{subsec:generalizable_economical_objective}
\mypara{Difficulty in the Shared Decoder Optimization.}
The motivating study in Section~\ref{subsec:motivationstudy} shows that sample-wise objectives cannot be directly transferred to decoder training, because moving protection from image-level perturbations to a shared decoder update changes the feasible optimization geometry. 
This setting imposes two coupled restrictions. First, decoder protection is subject to a \emph{parameter-sharing restriction}: instead of assigning an independent perturbation $\delta_i$ to each image, all protected outputs must be produced by a single shared update $\Delta\theta=\theta-\theta_0$. 
Second, it is subject to a \emph{visual-preservation restriction}: the update must remain within a small visual budget so that service outputs stay close to their clean references.
We therefore define the visually feasible update set as \looseness=-1
$\mathcal C_\epsilon={\Delta\theta:\mathcal L_{vis}(\Delta\theta)\le \epsilon},$
where $\mathcal L_{vis}$ measures the visual deviation between protected outputs and their clean references.
Because protected outputs for all inputs must remain visually close to their clean counterparts, the shared parameter update $\Delta\theta$ is constrained to be small. 
Our analysis therefore stays in the first-order regime around $\theta_0$ and omits higher-order terms.
Around the clean decoder, the image perturbation induced by $\Delta\theta$ can be locally approximated as\looseness=-1

\begin{equation}
\small
\Delta x(z)
=
D_{\theta_0+\Delta\theta}(z)-D_{\theta_0}(z)
\approx
\left.
\frac{\partial D_\theta(z)}{\partial \theta}
\right|_{\theta=\theta_0}[\Delta\theta]
=
\mathcal J_z[\Delta\theta].
\label{eq:shared_decoder_local_shift}
\end{equation}
where $\mathcal J_z=\left.\frac{\partial D_\theta(z)}{\partial \theta}\right|_{\theta=\theta_0}$ denotes the Jacobian of the decoder output with respect to its parameters. 
This follows from a first-order Taylor expansion around $\theta_0$, with higher-order terms omitted under the small-update regime enforced by visual preservation. 
To compare objectives under the same constraints, we move to the surrogate VAE latent space.
For training latents $\mathcal D=\{z_i\}_{i=1}^{N}$, let $x_r^i=D_{\theta_0}(z_i)$, $x_p^i=D_{\theta_0+\Delta\theta}(z_i)$, $h_r^i=E_s(x_r^i)$, and $h_p^i=E_s(x_p^i)$.
With $J_i\Delta\theta$ denoting the coordinate form of $\mathcal J_{z_i}[\Delta\theta]$ and $K_i=\left.\partial E_s(x)/\partial x\right|_{x=x_r^i}$, the local latent response is
\begin{equation}
    h_p^i-h_r^i=A_i\Delta\theta+O(\|\Delta\theta\|_2^2),\qquad A_i=K_iJ_i.
    \label{eq:latent_local_shift}
\end{equation}
Every $A_i$ is evaluated at the clean decoder and treated as fixed, rather than changing with the current decoder parameters, since $\Delta\theta$ is small.
We use $A_i\Delta\theta$ as the dominant reachable response throughout this local analysis.

We now characterize reachability under the shared decoder setting. 
Specifically, reachability is measured by the volume of decoder parameter updates $\Delta\theta$ that remain feasible and satisfy the current optimization objective within a prescribed loss threshold.

\begin{definition}[Reachable Set]
Given the local shared response map $A$ and the visual feasible set $\mathcal C_\epsilon$, the reachable set of a minimization objective $L$ at level $\tau$ is defined as
\begin{equation}
\Omega_L(\tau)
=
\{\Delta\theta\in\mathcal C_\epsilon: L(\Delta\theta)\le \tau\}.
\end{equation}
\end{definition}

The reachable set contains the visually feasible shared decoder updates that can satisfy the objective up to level $\tau$. Since different objectives may have different numerical scales and geometries, we use this definition to compare their reachability under the same shared response map $A$ and the same visual feasible set $\mathcal C_\epsilon$.

\begin{figure}[t]
    \centering
    \includegraphics[width=\columnwidth]{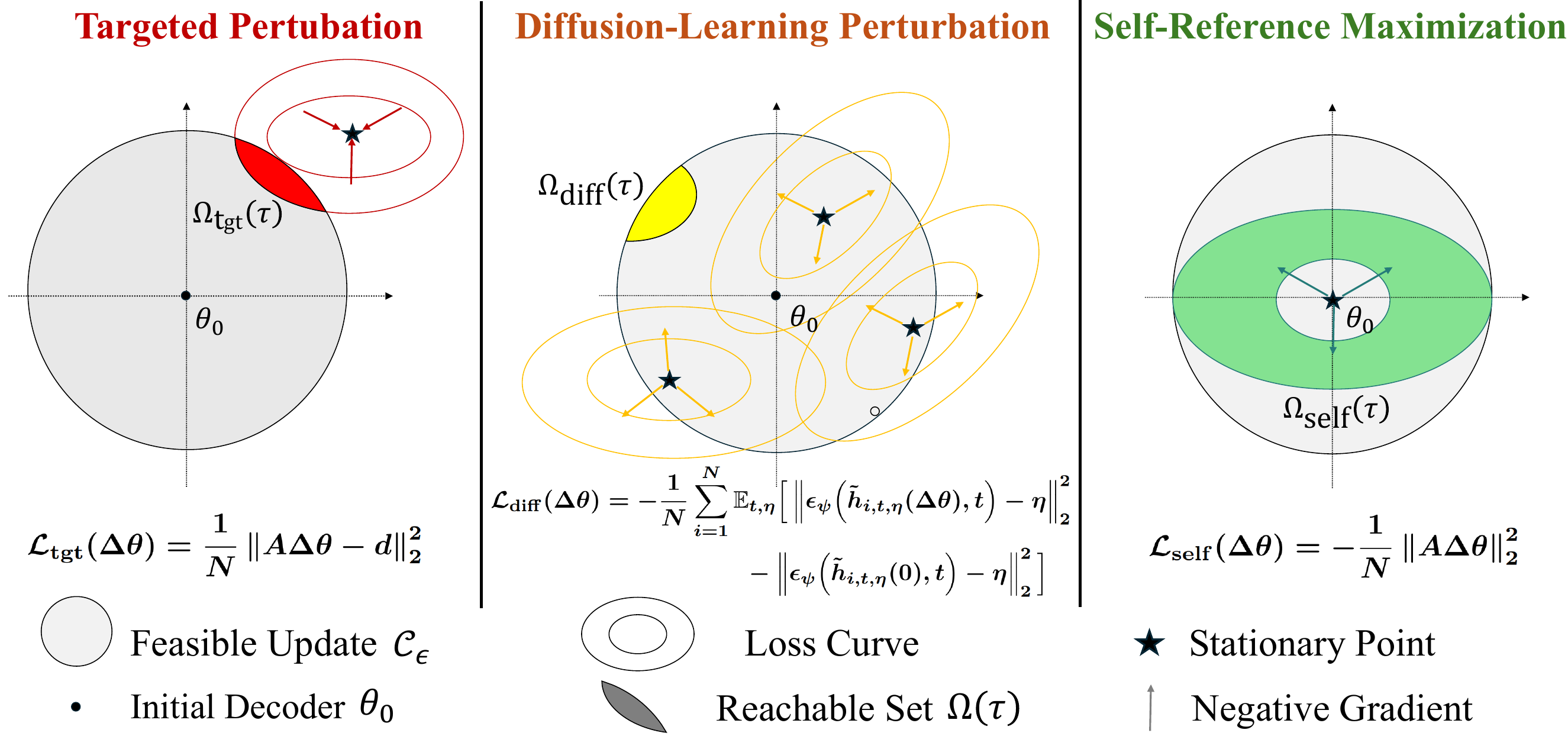}
    \caption{Conceptual comparison of effective reachable sets across three loss categories under the same visually feasible region. The gray circle denotes the visually feasible update set $\mathcal{C}_\epsilon$. 
    The colored region represents the reachable set, and its area indicates the size of this set. }
    \label{fig:motivation_reachable_sets}
\end{figure}

\mypara{Failure Modes of Existing Objectives.}
Existing optimization objectives can be broadly categorized into two types: \textit{targeted perturbation} objectives that operate in the VAE latent space~\cite{shan2023glaze,shan2024nightshade,salman2023photoguard}, and \textit{diffusion-learning perturbation} objectives that disrupt the diffusion learning stage~\cite{liang2023advdm,vanle2023antidreambooth}.
They are effective in their original sample-wise settings, but they become restrictive once all samples must share one decoder update.
For targeted perturbations with an external target, each image has its own perturbation $\delta_i$, so the optimizer can move sample $i$ toward its own target independently.
The target latent $t_i$ therefore only constrains a sample-wise variable.
In contrast, shared decoder optimization must fit all target residuals with one vector $\Delta\theta$.
Let $d_i=t_i-h_r^i$ and $d=[d_1;\ldots;d_N]$.
The local target-matching loss is
\begin{equation}
    \mathcal L_{\mathrm{tgt}}(\Delta\theta)
    =
    \frac{1}{N}\sum_{i=1}^{N}\|A_i\Delta\theta-d_i\|_2^2
    =
    \frac{1}{N}\|A\Delta\theta-d\|_2^2.
    \label{eq:target_local_loss}
\end{equation}
\begin{equation}
    \Omega_{\mathrm{tgt}}(\tau)=\left\{\Delta\theta\in\mathcal C_\epsilon:\|A\Delta\theta-d\|_2^2\le N\tau\right\}.
    \label{eq:tgt_effective_set}
\end{equation}
This reachable set is small because the shared response must land near a preassigned residual vector $d$, rather than merely become disruptive.
As shown in Figure~\ref{fig:motivation_reachable_sets}, the target often lies far from the original parameters that can be reached to achieve disruption effectiveness. 
The reachable set is given by the intersection between $\mathcal C_\epsilon$ and the set of parameters whose loss is below $\tau$, and occupies only a limited region.
Moreover, every response induced by the shared decoder lies in $\mathrm{Col}(A)$.
If $P_A$ is the orthogonal projection onto $\mathrm{Col}(A)$, then the component of $d$ outside this column space creates an unavoidable residual: \looseness=-1
\begin{equation}
    \inf_{\Delta\theta\in\mathcal C_\epsilon}
    \mathcal L_{\mathrm{tgt}}(\Delta\theta)
    \ge
    \frac{1}{N}\|(I-P_A)d\|_2^2.
    \label{eq:tgt_unavoidable_residual}
\end{equation}
Therefore, these objectives fail for a structural reason: they depend on an external residual $d$ that may be incompatible with the response space reachable by one visually constrained shared update.
We refer to this failure as \emph{target dependence}. \looseness=-1

Diffusion-learning perturbations avoid explicit external targets, but introduce a different failure mode.
For a single image, the perturbation can be repeatedly adapted to sampled timesteps and noise values until it increases the surrogate denoising error.
In shared decoder optimization, however, the same $\Delta\theta$ must work across all samples and across the stochastic diffusion process.
Let $\epsilon_\psi$ denote the surrogate denoiser, $t$ a sampled diffusion timestep, $\eta\sim\mathcal N(0,I)$ the forward noising seed, and $\bar\alpha_t$ the cumulative noise-schedule coefficient.
Using $h_p^i\approx h_r^i+A_i\Delta\theta$, the signed minimization loss expands as\looseness=-1


\begin{align}
\tilde h_{i,t,\eta}(\Delta\theta)
&=
\sqrt{\bar\alpha_t}(h_r^i+A_i\Delta\theta)
+\sqrt{1-\bar\alpha_t}\eta .
\label{eq:diff_noisy_latent}
\\[-1mm]
\mathcal L_{\mathrm{diff}}(\Delta\theta)
&=
-\frac{1}{N}\sum_{i=1}^{N}\mathbb E_{t,\eta}
\Big[
\left\|
\epsilon_\psi(\tilde h_{i,t,\eta}(\Delta\theta),t)-\eta
\right\|_2^2
\nonumber\\
&\qquad\qquad
-
\left\|
\epsilon_\psi(\tilde h_{i,t,\eta}(0),t)-\eta
\right\|_2^2
\Big].
\label{eq:advdm_signed_average}
\end{align}
\begin{equation}
    \Omega_{\mathrm{diff}}(\tau)=\left\{\Delta\theta\in\mathcal C_\epsilon:\mathcal L_{\mathrm{diff}}(\Delta\theta)\le \tau\right\}.
    \label{eq:diff_effective_set}
\end{equation}
Here, $\tilde h_{i,t,\eta}(\Delta\theta)$ is the noisy latent obtained by applying the diffusion forward process to the locally approximated protected latent, not a generated image.
The expectation over the sampled timestep $t$ and noise seed $\eta$ is critical.
A standard diffusion training schedule, such as DDPM, typically involves about $1000$ timesteps~\cite{ho2020ddpm}. Moreover, each optimization step samples both the timestep and the initial noise independently, so the objective is optimized only through stochastic timestep-noise pairs rather than a fixed deterministic target.
However, different timestep-noise pairs induce substantially different optimization loss curves.
As illustrated in Figure~\ref{fig:motivation_reachable_sets}, for clarity, we only plot the loss curves induced by three timestep-noise pairs. 
Since the objective requires each loss term to be minimized, namely moving away from each corresponding stationary point, the reachable set becomes the intersection of the reachable sets associated with multiple loss curves. 
This intersection further narrows the overall reachable region.
Due to the large variation among loss curves, this averaging can induce a concentration effect: a shared update may satisfy the reachable set by concentrating errors on a few samples or timestep-noise pairs, leaving most training outputs insufficient to disrupt distillation.
In practice, many more timestep-noise pairs are involved, making the intersection increasingly restrictive. As a result, the reachable region further shrinks, which leads to poor reachability.
We refer to this failure as \emph{stochasticity dependence}.\looseness=-1


\mypara{Self-Referenced Maximization Avoids Two Failure Modes.}
The two restrictive cases reveal two different failure sources: latent targeted perturbations depend on a target residual that may be unreachable, while diffusion-learning perturbations depend on a stochastic denoising proxy whose gradients may not align under one shared update.
We therefore optimize directly in the VAE latent space with a self-referenced objective.
Each clean output acts as its own reference, so the objective rewards the magnitude of the response produced by the shared decoder itself:
\begin{equation}
    \begin{aligned}
    \mathcal L_{\mathrm{self}}(\Delta\theta)
    &=
    -\frac{1}{N}\sum_{i=1}^{N}\|A_i\Delta\theta\|_2^2
    =
    -\frac{1}{N}\|A\Delta\theta\|_2^2.
    \end{aligned}
    \label{eq:self_batch_loss}
\end{equation}
\begin{equation}
    \Omega_{\mathrm{self}}(\tau)
    =
    \left\{
    \Delta\theta\in\mathcal C_\epsilon:
    \|A\Delta\theta\|_2^2\ge -N\tau
    \right\}.
    \label{eq:self_effective_set}
\end{equation}
This set is less restrictive because it imposes only an energy condition on the same feasible set $\mathcal C_\epsilon$.
As shown in Figure~\ref{fig:motivation_reachable_sets}, the optimization is directly anchored at the clean decoder parameters $\theta_0$. 
By contrast, \method{} adopts a self-referenced objective centered at the clean decoder $\theta_0$. 
Instead of pursuing an external target or optimizing against stochastic diffusion-proxy losses, both of which introduce additional optimization dependency, it directly encourages the decoder-induced latent perturbation to deviate from its original latent representation. 
This formulation removes unnecessary external constraints, reduces conflicts among sampled objectives, and better exploits the shared nature of decoder parameters.
As a result, \method{} induces a larger reachable effective region under the same visual feasibility constraint.
Note that independent reparameterized sampling in the surrogate VAE encoder introduces slight latent differences even when $x_p=x_r$, thereby providing a nonzero training signal at initialization.
\looseness=-1

\mypara{Avoiding the Latent Shortcut with Color Regularization.}
The preceding analysis motivates a self-referenced objective, with latent-discrepancy maximization as a natural choice.
However, as shown in Section~\ref{subsec:CH3}, latent-only maximization constrains only the output of the surrogate encoder, and its effect may not persist through the complete VAE bottleneck, thereby inducing latent shortcuts.
This behavior is consistent with the denoising and manifold-projection view of autoencoders~\cite{vincent2008denoising,alain2014regularized,rombach2022high}, where the decoder reconstruction tends to map corrupted or perturbed inputs back toward the learned data manifold, thereby attenuating perturbations that are not effectively supported by the decoder.
Consequently, large latent discrepancies can be suppressed or smoothed by the surrogate decoder, leading to visually similar reconstructions. \looseness=-1

To block this shortcut, \method{} adds a reconstruction-guided color regularization term while preserving the same self-referenced optimization structure.
The term still compares each protected sample with its own clean reference, rather than introducing an external target direction.
After the protected latent passes through the surrogate decoder, the regularization encourages the discrepancy induced by the same shared update $\Delta\theta$ in reconstructed image space.
Because the perturbation budget within $\mathcal C_\epsilon$ is limited, the reconstruction signal should allocate this constrained perturbation capacity to perceptually meaningful directions, rather than merely arbitrary pixel-level perturbations.
Color appearance is a salient component of human vision, and standardized color-difference metrics are specifically designed to approximate perceived color changes~\cite{wyszecki1982colorScience,luo2001ciede2000,sharma2005ciede2000}. We therefore instantiate this reconstruction term using CIEDE2000~\cite{luo2001ciede2000}:
\begin{equation}
    \mathcal{L}_{color}
    =
    \Delta E_{00}(\hat{x}_p,x_r)
    =
    \Delta E_{00}(D_s(h_p),x_r),
\end{equation}
where $\hat{x}_p=D_s(h_p)$ is the image reconstructed by the surrogate decoder.
Maximizing $\Delta E_{00}(\hat{x}_p,x_r)$ encourages the latent shift to appear in reconstructed image space to enhance disruption effectiveness: the delivered image $x_p$ remains visually close to $x_r$, while the downstream reconstruction $\hat{x}_p$ becomes inconsistent with the clean reference.
Note that the optimization remains self-referenced to achieve reachability, as the rewarded discrepancy is still measured between the protected output induced by the shared decoder and its corresponding clean reference.\looseness=-1

\mypara{Final disruption objective.}
The objective combines self-referenced latent maximization with reconstruction-guided color regularization:
\begin{equation}
    \mathcal{L}_{obj}
    =
    -\lambda_z\mathcal{L}_{latent}
    -
    \lambda_c\mathcal{L}_{color}.
\end{equation}
Here, $\mathcal{L}_{latent}=MSE(h_p,h_r)$ and $\mathcal{L}_{color}$ are weighted by $\lambda_z$ and $\lambda_c$.
The negative signs maximize downstream divergence under the global minimization objective.

\subsection{Composite Visual Fidelity Constraints}
As observed in Section~\ref{subsec:CH3}, the substantial optimization difficulty can cause the model to overfit to a single visual metric, leaving visible color shifts, local artifacts, and high-frequency noise despite satisfying that metric. 
To address this issue, \method{} introduces composite visual constraints that jointly regulate color consistency, perceptual similarity, pixel-level distortion, and structural fidelity.
These constraints regularize different aspects of visual quality, including overall color consistency with CIEDE2000~\cite{luo2001ciede2000}, perceptual structure with LPIPS~\cite{zhang2018lpips}, high-frequency smoothness with total variation~\cite{rudin1992tv}, and local color consistency with a patch-local CIEDE2000 term, inspired by local quality maps~\cite{wang2004ssim}:\looseness=-1
\begingroup
\setlength{\abovedisplayskip}{4pt plus 1pt minus 1pt}
\setlength{\belowdisplayskip}{4pt plus 1pt minus 1pt}
\setlength{\jot}{1.5pt}
\begin{equation}
\begin{aligned}
    \mathcal{L}_{cie}
    &=
    \max(\Delta E_{00}(x_p,x_r)-\tau_{cie},0),
    \\
    \mathcal{L}_{lpips}
    &=
    \max(LPIPS(x_p,x_r)-\tau_{lpips},0),
    \\
    \mathcal{L}_{tv}
    &=
    \max(TV(x_p)-TV(x_r)-\tau_{tv},0),
    \\
    \mathcal{L}_{patch}
    &=
    \mathbb{E}_{j}[
    \max(\Delta E_{00}(x_{p,j},x_{r,j})-\tau_{patch},0)].
\end{aligned}
\end{equation}
\endgroup

\begin{table*}[t]
    \centering
    \caption{Main results of \method{} compared with five baselines. Adv, Anti, Night, and PG denote AdvDM, Anti-DreamBooth, Nightshade, and PhotoGuard~\cite{salman2023photoguard}, respectively.}
    \label{tab:main_extraction_results}
    \footnotesize
    \setlength{\tabcolsep}{0pt}
    \renewcommand{\arraystretch}{1.10}
    \begin{minipage}[t]{0.495\linewidth}
    \begin{tabularx}{\linewidth}{@{}ll*{8}{>{\centering\arraybackslash}X}@{}}
    \toprule
    \rowcolor{tablepanel}\multicolumn{10}{c}{\textbf{SSD}} \\
    \midrule
    \rowcolor{tablehead}\textbf{Dataset} & \textbf{Metric} & \textbf{Ref.} & \textbf{Clean} & \textbf{Ours} & \textbf{Adv} & \textbf{Anti} & \textbf{Glaze} & \textbf{Night} & \textbf{PG} \\
    \midrule
    \multirow{4}{*}{\textbf{AFHQ}} & \mFID$\uparrow$ & -- & 40.14 & 48.67 & \best{86.72} & 40.50 & \second{49.34} & 48.12 & 46.64 \\
     & \mQA$\downarrow$ & 4.63 & 3.02 & \best{1.84} & 2.71 & 2.96 & \second{2.47} & 2.74 & 2.66 \\
     & \mPK$\downarrow$ & 22.50 & 21.38 & \best{20.60} & 21.41 & 21.42 & 20.93 & 20.95 & \second{20.88} \\
     & \mAES$\downarrow$ & 5.85 & 5.88 & \best{5.27} & 5.91 & 5.95 & 5.68 & 5.73 & \second{5.60} \\
    \addlinespace[0.2em]
    \multirow{4}{*}{\textbf{Pokemon}} & \mFID$\uparrow$ & -- & 64.02 & \best{80.81} & 66.27 & 64.68 & 66.21 & \second{72.13} & 71.90 \\
     & \mQA$\downarrow$ & 4.62 & 2.95 & \best{1.93} & 2.97 & 3.04 & 3.22 & 2.56 & \second{2.36} \\
     & \mPK$\downarrow$ & 23.31 & 22.60 & \best{21.88} & 22.62 & 22.67 & 22.46 & \second{22.17} & 22.24 \\
     & \mAES$\downarrow$ & 5.78 & 5.62 & \best{5.01} & 5.64 & 5.67 & 5.53 & 5.36 & \second{5.25} \\
    \addlinespace[0.2em]
    \multirow{4}{*}{\textbf{CelebA}} & \mFID$\uparrow$ & -- & 45.98 & \best{59.91} & 46.17 & 45.64 & 51.74 & \second{59.40} & 53.90 \\
     & \mQA$\downarrow$ & 4.91 & 3.97 & \best{1.95} & 4.21 & 4.05 & 3.15 & 2.91 & \second{2.67} \\
     & \mPK$\downarrow$ & 22.62 & 21.23 & \best{19.76} & 21.26 & 21.19 & 20.70 & \second{20.23} & 20.41 \\
     & \mAES$\downarrow$ & 6.04 & 5.94 & \best{5.43} & 6.07 & 6.02 & 5.87 & 5.63 & \second{5.50} \\
    \addlinespace[0.2em]
    \multirow{4}{*}{\textbf{Landscape}} & \mFID$\uparrow$ & -- & 93.07 & \best{113.66} & 91.60 & 89.46 & 103.70 & \second{112.39} & 97.15 \\
     & \mQA$\downarrow$ & 4.51 & 4.17 & \best{1.76} & 4.17 & 4.20 & 3.30 & 3.32 & \second{3.28} \\
     & \mPK$\downarrow$ & 22.70 & 22.12 & \best{20.84} & 22.23 & 22.29 & 21.61 & \second{21.37} & 21.82 \\
     & \mAES$\downarrow$ & 5.94 & 5.94 & \best{5.12} & 5.99 & 5.97 & 5.63 & \second{5.55} & 5.68 \\
    \bottomrule
    \end{tabularx}
    \end{minipage}\hfill
    \begin{minipage}[t]{0.495\linewidth}
    \begin{tabularx}{\linewidth}{@{}ll*{8}{>{\centering\arraybackslash}X}@{}}
    \toprule
    \rowcolor{tablepanel}\multicolumn{10}{c}{\textbf{PixArt}} \\
    \midrule
    \rowcolor{tablehead}\textbf{Dataset} & \textbf{Metric} & \textbf{Ref.} & \textbf{Clean} & \textbf{Ours} & \textbf{Adv} & \textbf{Anti} & \textbf{Glaze} & \textbf{Night} & \textbf{PG} \\
    \midrule
    \multirow{4}{*}{\textbf{AFHQ}} & \mFID$\uparrow$ & -- & 60.13 & \best{132.62} & 76.65 & 86.00 & 103.33 & 87.20 & \second{112.95} \\
     & \mQA$\downarrow$ & 4.63 & 3.10 & \best{1.59} & 2.98 & 3.16 & 2.83 & 3.11 & \second{2.30} \\
     & \mPK$\downarrow$ & 22.50 & 20.81 & \best{19.75} & 20.99 & 20.93 & 20.56 & 20.81 & \second{20.32} \\
     & \mAES$\downarrow$ & 5.85 & 5.97 & \best{5.25} & 6.06 & 5.98 & 5.88 & 5.95 & \second{5.62} \\
    \addlinespace[0.2em]
    \multirow{4}{*}{\textbf{Pokemon}} & \mFID$\uparrow$ & -- & 76.94 & \best{101.27} & 80.47 & 78.82 & 80.36 & 89.23 & \second{91.50} \\
     & \mQA$\downarrow$ & 4.62 & 3.68 & \best{1.45} & 3.39 & 3.48 & 3.37 & 2.43 & \second{2.22} \\
     & \mPK$\downarrow$ & 23.31 & 22.29 & \best{21.38} & 22.17 & 22.23 & 22.17 & \second{21.69} & 21.70 \\
     & \mAES$\downarrow$ & 5.78 & 5.73 & \best{4.82} & 5.73 & 5.72 & 5.60 & \second{5.22} & 5.30 \\
    \addlinespace[0.2em]
    \multirow{4}{*}{\textbf{CelebA}} & \mFID$\uparrow$ & -- & 67.61 & \best{102.42} & 60.75 & 70.69 & \second{77.82} & 76.34 & 76.91 \\
     & \mQA$\downarrow$ & 4.91 & 4.18 & \best{1.91} & 4.36 & 4.31 & \second{3.13} & 3.22 & 3.80 \\
     & \mPK$\downarrow$ & 22.62 & 20.90 & \best{19.31} & 21.03 & 20.87 & 20.32 & \second{20.06} & 20.46 \\
     & \mAES$\downarrow$ & 6.04 & 6.16 & \best{5.19} & 6.12 & 6.23 & 5.74 & \second{5.71} & 5.99 \\
    \addlinespace[0.2em]
    \multirow{4}{*}{\textbf{Landscape}} & \mFID$\uparrow$ & -- & 106.37 & \best{146.86} & 110.50 & 109.80 & 117.26 & \second{128.78} & 112.14 \\
     & \mQA$\downarrow$ & 4.51 & 4.32 & \best{1.74} & 4.23 & 4.39 & 4.09 & 3.27 & \second{3.13} \\
     & \mPK$\downarrow$ & 22.70 & 21.78 & \best{20.31} & 21.67 & 21.83 & 21.54 & \second{21.08} & 21.34 \\
     & \mAES$\downarrow$ & 5.94 & 5.92 & \best{5.06} & 5.94 & 5.98 & 5.79 & \second{5.57} & 5.66 \\
    \bottomrule
    \end{tabularx}
    \end{minipage}
    
    \vspace{1.0em}
    \begin{minipage}[t]{0.495\linewidth}
    \begin{tabularx}{\linewidth}{@{}ll*{8}{>{\centering\arraybackslash}X}@{}}
    \toprule
    \rowcolor{tablepanel}\multicolumn{10}{c}{\textbf{SDXL}} \\
    \midrule
    \rowcolor{tablehead}\textbf{Dataset} & \textbf{Metric} & \textbf{Ref.} & \textbf{Clean} & \textbf{Ours} & \textbf{Adv} & \textbf{Anti} & \textbf{Glaze} & \textbf{Night} & \textbf{PG} \\
    \midrule
    \multirow{4}{*}{\textbf{AFHQ}} & \mFID$\uparrow$ & -- & 50.45 & \best{75.09} & 38.84 & 43.56 & 47.12 & \second{51.65} & 39.11 \\
     & \mQA$\downarrow$ & 4.63 & 3.61 & \best{1.64} & 3.88 & 3.91 & 3.45 & \second{3.14} & 3.49 \\
     & \mPK$\downarrow$ & 22.50 & 21.66 & \best{19.72} & 21.90 & 21.73 & 21.47 & \second{21.12} & 21.69 \\
     & \mAES$\downarrow$ & 5.85 & 5.96 & \best{5.16} & 5.93 & 6.04 & 6.08 & 5.89 & \second{5.84} \\
    \addlinespace[0.2em]
    \multirow{4}{*}{\textbf{Pokemon}} & \mFID$\uparrow$ & -- & 69.81 & \second{76.50} & 64.76 & 71.31 & \best{77.67} & 72.91 & 72.91 \\
     & \mQA$\downarrow$ & 4.62 & 4.22 & \best{2.78} & 4.30 & 4.18 & 3.95 & 3.68 & \second{3.35} \\
     & \mPK$\downarrow$ & 23.31 & 22.64 & \best{21.81} & 22.72 & 22.59 & 22.35 & \second{22.21} & 22.25 \\
     & \mAES$\downarrow$ & 5.78 & 5.71 & \best{5.25} & 5.71 & 5.68 & 5.62 & 5.61 & \second{5.45} \\
    \addlinespace[0.2em]
    \multirow{4}{*}{\textbf{CelebA}} & \mFID$\uparrow$ & -- & 49.60 & \best{72.49} & 49.17 & 50.07 & \second{69.83} & 68.13 & 62.02 \\
     & \mQA$\downarrow$ & 4.91 & 4.26 & \best{2.03} & 4.25 & 4.18 & 2.98 & 2.91 & \second{2.88} \\
     & \mPK$\downarrow$ & 22.62 & 21.68 & \best{19.30} & 21.62 & 21.54 & 20.73 & 20.52 & \second{20.27} \\
     & \mAES$\downarrow$ & 6.04 & 6.03 & \best{5.26} & 6.08 & 6.02 & 5.80 & 5.62 & \second{5.46} \\
    \addlinespace[0.2em]
    \multirow{4}{*}{\textbf{Landscape}} & \mFID$\uparrow$ & -- & 98.30 & \best{120.48} & 99.30 & 99.96 & \second{110.71} & 109.17 & 100.82 \\
     & \mQA$\downarrow$ & 4.51 & 4.21 & \best{2.59} & 4.27 & 4.23 & \second{3.05} & 3.52 & 3.34 \\
     & \mPK$\downarrow$ & 22.70 & 22.10 & \best{20.87} & 22.02 & 21.95 & \second{21.41} & 21.46 & 21.67 \\
     & \mAES$\downarrow$ & 5.94 & 5.82 & \best{5.22} & 5.84 & 5.79 & \second{5.57} & 5.60 & \second{5.57} \\
    \bottomrule
    \end{tabularx}
    \end{minipage}\hfill
    \begin{minipage}[t]{0.495\linewidth}
    \begin{tabularx}{\linewidth}{@{}ll*{8}{>{\centering\arraybackslash}X}@{}}
    \toprule
    \rowcolor{tablepanel}\multicolumn{10}{c}{\textbf{Shuttle3}} \\
    \midrule
    \rowcolor{tablehead}\textbf{Dataset} & \textbf{Metric} & \textbf{Ref.} & \textbf{Clean} & \textbf{Ours} & \textbf{Adv} & \textbf{Anti} & \textbf{Glaze} & \textbf{Night} & \textbf{PG} \\
    \midrule
    \multirow{4}{*}{\textbf{AFHQ}} & \mFID$\uparrow$ & -- & 62.59 & \best{120.37} & 69.40 & 67.23 & 79.11 & \second{80.77} & 79.03 \\
     & \mQA$\downarrow$ & 4.63 & 3.56 & \best{1.50} & 2.98 & 3.40 & 2.61 & \second{2.42} & 3.05 \\
     & \mPK$\downarrow$ & 22.50 & 21.38 & \best{19.79} & 21.24 & 21.37 & 20.74 & \second{20.56} & 20.97 \\
     & \mAES$\downarrow$ & 5.85 & 6.25 & \best{5.36} & 6.20 & 6.23 & 6.04 & \second{5.68} & 5.98 \\
    \addlinespace[0.2em]
    \multirow{4}{*}{\textbf{Pokemon}} & \mFID$\uparrow$ & -- & 73.88 & \best{102.17} & 76.97 & 79.94 & 78.09 & 86.90 & \second{87.74} \\
     & \mQA$\downarrow$ & 4.62 & 4.33 & \best{1.77} & 4.01 & 4.07 & 3.56 & \second{2.54} & 2.72 \\
     & \mPK$\downarrow$ & 23.31 & 22.62 & \best{21.27} & 22.47 & 22.50 & 22.32 & \second{21.97} & 22.07 \\
     & \mAES$\downarrow$ & 5.78 & 5.93 & \best{5.00} & 5.89 & 5.90 & 5.66 & \second{5.31} & 5.43 \\
    \addlinespace[0.2em]
    \multirow{4}{*}{\textbf{CelebA}} & \mFID$\uparrow$ & -- & 76.75 & \best{105.59} & 73.41 & 68.60 & 69.78 & \second{93.34} & 85.49 \\
     & \mQA$\downarrow$ & 4.91 & 4.57 & \best{1.81} & 4.67 & 4.73 & 4.50 & 2.99 & \second{2.77} \\
     & \mPK$\downarrow$ & 22.62 & 20.94 & \best{19.10} & 21.13 & 21.16 & 21.00 & \second{20.11} & 20.19 \\
     & \mAES$\downarrow$ & 6.04 & 6.36 & \best{5.28} & 6.46 & 6.39 & 6.28 & \second{5.67} & 5.74 \\
    \addlinespace[0.2em]
    \multirow{4}{*}{\textbf{Landscape}} & \mFID$\uparrow$ & -- & 102.43 & \best{115.87} & 101.47 & 106.01 & 109.77 & \second{114.51} & 113.33 \\
     & \mQA$\downarrow$ & 4.51 & 4.61 & \best{2.18} & 4.59 & 4.61 & 3.85 & 3.27 & \second{3.00} \\
     & \mPK$\downarrow$ & 22.70 & 22.15 & \best{21.13} & 22.08 & 22.10 & 21.69 & \second{21.31} & 21.44 \\
     & \mAES$\downarrow$ & 5.94 & 6.09 & \best{5.39} & 6.09 & 6.13 & 5.82 & \second{5.60} & 5.68 \\
    \bottomrule
    \end{tabularx}
    \end{minipage}
\end{table*}
Here, $x_p$ and $x_r$ denote the protected and reference images, $\tau_{\cdot}$ denotes the corresponding threshold, and $x_{p,j}$ and $x_{r,j}$ are the protected and reference patches indexed by $j$, with $\mathbb{E}_j$ averaging over patch indices.
The visual loss is defined as: \looseness=-1
\begin{equation}
    \mathcal{L}_{vis}
    =
    \mathcal{L}_{lpips}
    +
    w_{cie}\mathcal{L}_{cie}
    +
    w_{tv}\mathcal{L}_{tv}
    +
    w_{patch}\mathcal{L}_{patch}.
\end{equation}
Here, $w_{cie}$, $w_{tv}$, and $w_{patch}$ denote the weights of the corresponding term in this formulation.
Together, these complementary hinge penalties reduce reliance on any single visual metric and keep the delivered output close to the reference in color, structure, smoothness, and local consistency.

\subsection{Full Objective and Training Procedure}


\begin{algorithm}[t]
\caption{\method{} defensive training}
\label{alg:inguard_training}
\small
\begin{algorithmic}[1]
\REQUIRE Generated training images $\mathcal{X}$; modules and weights defined above.
\ENSURE Trained protected decoder $D_{\theta^\star}$.
\STATE Initialize trainable $D_\theta$ from $D_0$ and freeze $E,D_0,E_s,D_s$.
\STATE Build training latents $\mathcal{Z}\leftarrow\{E(x):x\in\mathcal{X}\}$.
\FOR{each training step with latent minibatch $\mathcal{B}_z\subset\mathcal{Z}$}
    \STATE $x_r\leftarrow D_0(\mathcal{B}_z)$,\quad $x_p\leftarrow D_\theta(\mathcal{B}_z)$.
    \STATE Compute thresholded $\mathcal{L}_{lpips}$, $\mathcal{L}_{cie}$, $\mathcal{L}_{tv}$, and $\mathcal{L}_{patch}$.
    \STATE $\mathcal{L}_{vis}\leftarrow\mathcal{L}_{lpips}+w_{cie}\mathcal{L}_{cie}+w_{tv}\mathcal{L}_{tv}+w_{patch}\mathcal{L}_{patch}$.
    \STATE Sample $h_p\sim E_s(x_p)$ and $h_r\sim E_s(x_r)$.
    \STATE $\hat{x}_p\leftarrow D_s(h_p)$.
    \STATE $\mathcal{L}_{latent}\leftarrow\operatorname{MSE}(h_p,h_r)$,\quad $\mathcal{L}_{color}\leftarrow\Delta E_{00}(\hat{x}_p,x_r)$.
    
    \STATE $\mathcal{L}_{total}\leftarrow\lambda_v\mathcal{L}_{vis}-\lambda_z\mathcal{L}_{latent}-\lambda_c\mathcal{L}_{color}$.
    \STATE Update $D_\theta$ by descending $\nabla_\theta\mathcal{L}_{total}$.
\ENDFOR
\STATE \textbf{return} $D_{\theta^\star}\leftarrow D_\theta$.
\end{algorithmic}
\end{algorithm}

The final training loss combines the composite visual loss with the two downstream-divergence terms introduced above: \looseness=-1
\begin{align}
    \mathcal{L}_{total}
    =
    \lambda_v\mathcal{L}_{vis}
    -\lambda_z\mathcal{L}_{latent}
    -
    \lambda_c\mathcal{L}_{color}.
    \label{eq:total_loss}
\end{align}
We set $\lambda_v \gg \lambda_z,\lambda_c$ to ensure that visual fidelity remains the dominant constraint under the prescribed threshold.
Training updates only $D_\theta$ while freezing $E,D_0,E_s,D_s$.
The training images are generated by the same source T2I model used by the service before being encoded by $E$.
As a result, the training latents are aligned with the service generation distribution, rather than being drawn from an unrelated image-encoding distribution.
Algorithm~\ref{alg:inguard_training} summarizes the procedure.

\begin{figure*}[tbp]
    \centering
    \footnotesize
    \setlength{\tabcolsep}{1.3pt}
    \renewcommand{\arraystretch}{1.08}
    \begin{tabularx}{0.86\textwidth}{@{}l*{7}{>{\centering\arraybackslash}X}@{}}
    \toprule
    \rowcolor{tablepanel}\multicolumn{8}{c}{\textbf{Disruption Effectiveness Comparison with Five Baselines}} \\
    \midrule
    \rowcolor{tablehead}\textbf{Dataset} & \textbf{Clean} & \textbf{\method{} (Ours)} & \textbf{AdvDM} & \textbf{Anti-DB} & \textbf{Glaze} & \textbf{Nightshade} & \textbf{PhotoGuard} \\
    \textbf{AFHQ} &
    \raisebox{-.5\height}{\includegraphics[width=\linewidth]{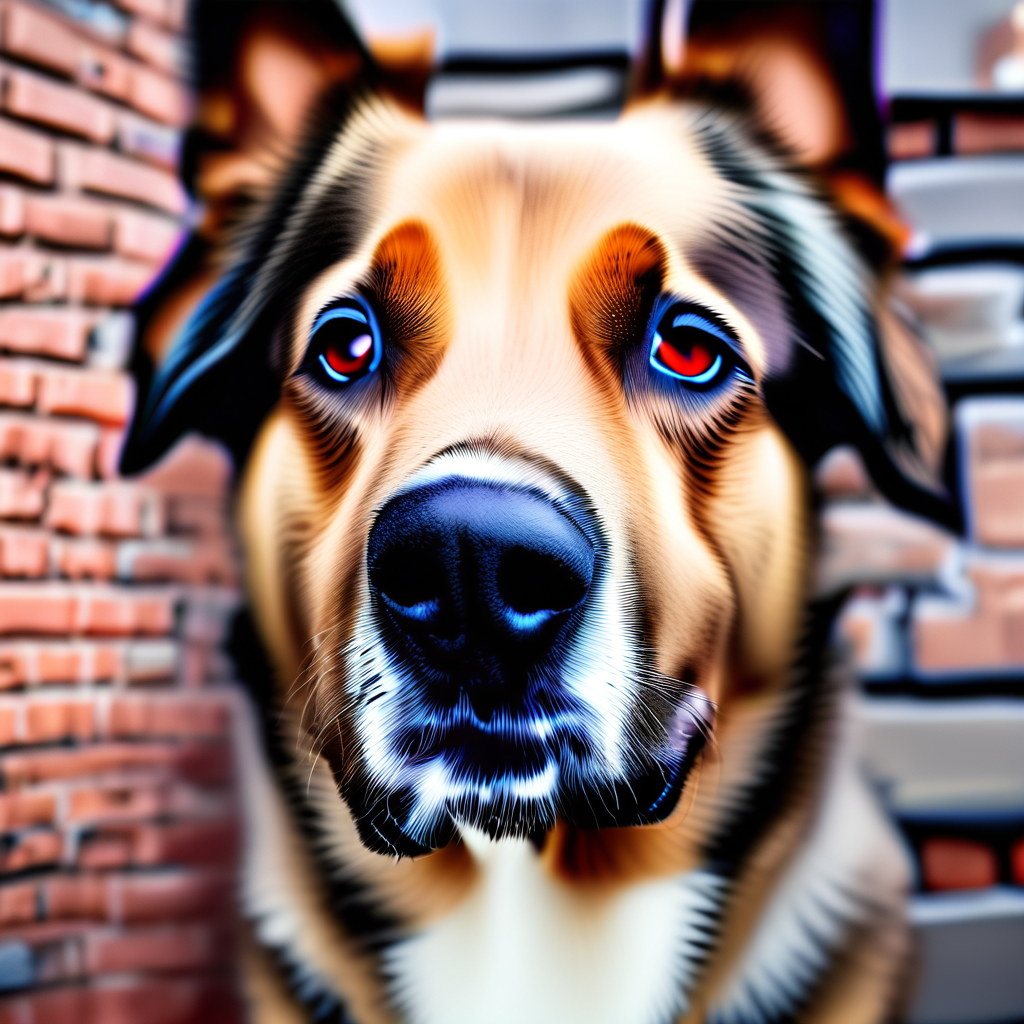}} &
    \raisebox{-.5\height}{\includegraphics[width=\linewidth]{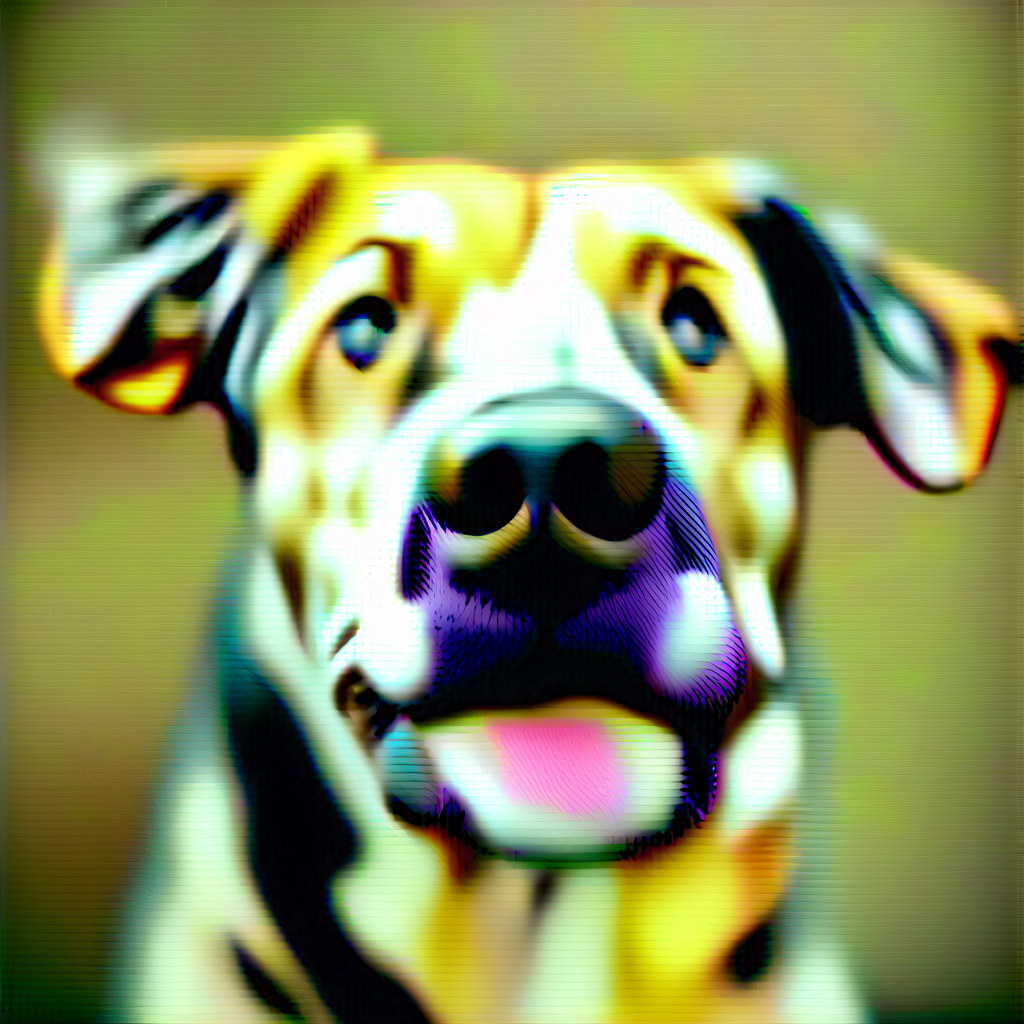}} &
    \raisebox{-.5\height}{\includegraphics[width=\linewidth]{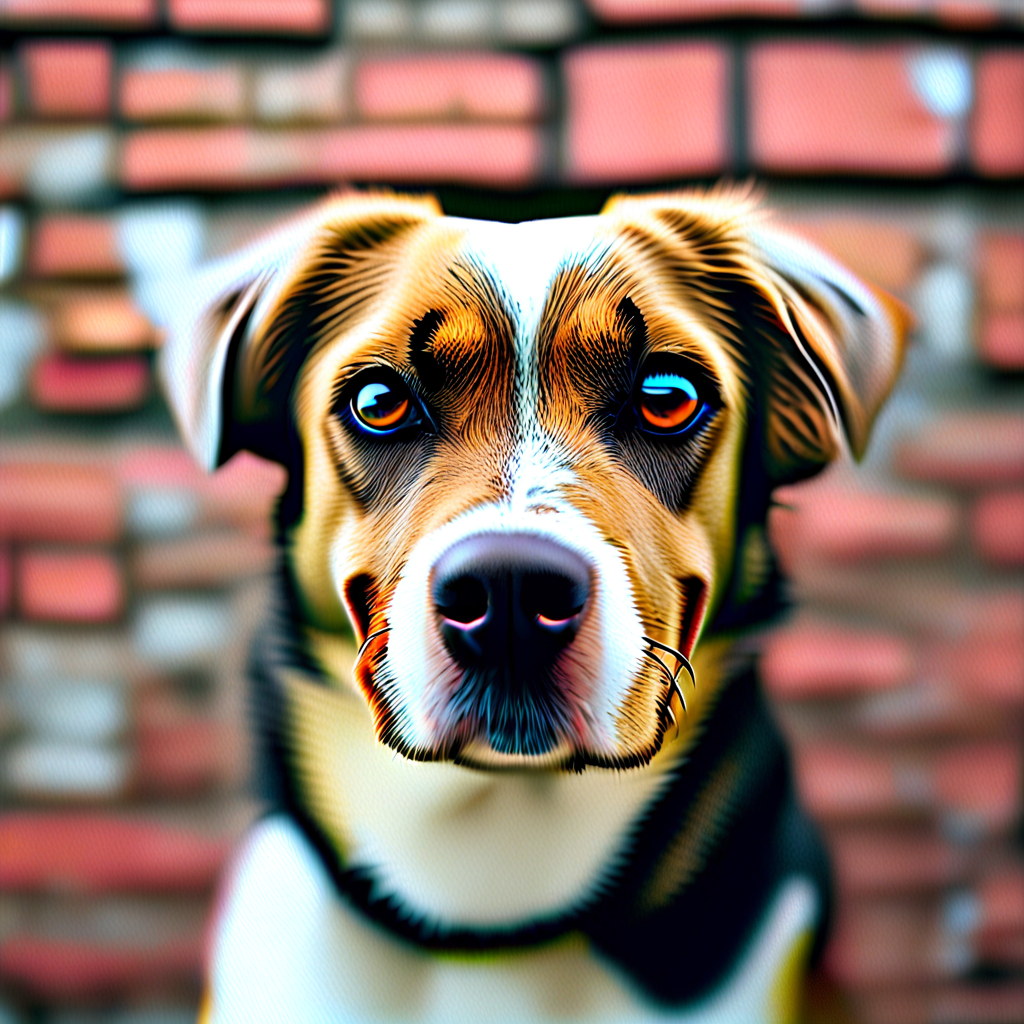}} &
    \raisebox{-.5\height}{\includegraphics[width=\linewidth]{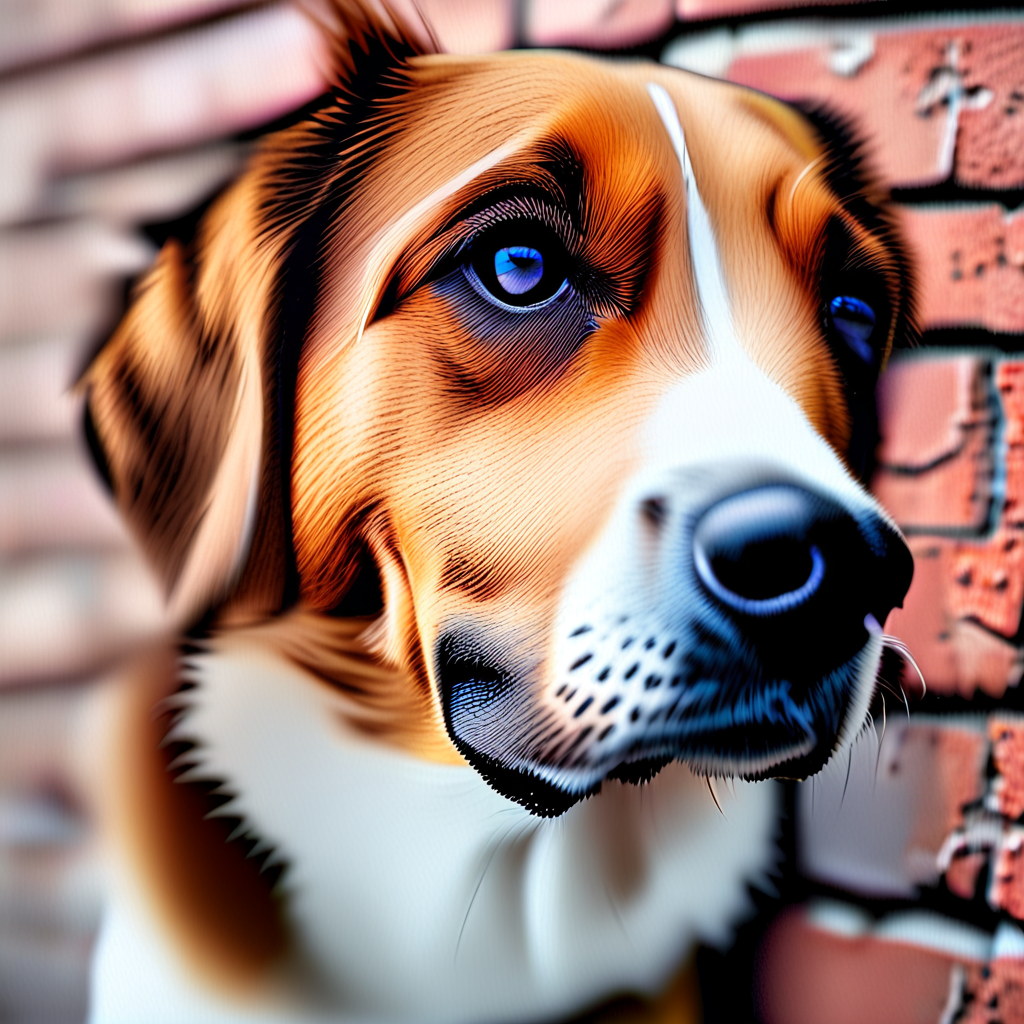}} &
    \raisebox{-.5\height}{\includegraphics[width=\linewidth]{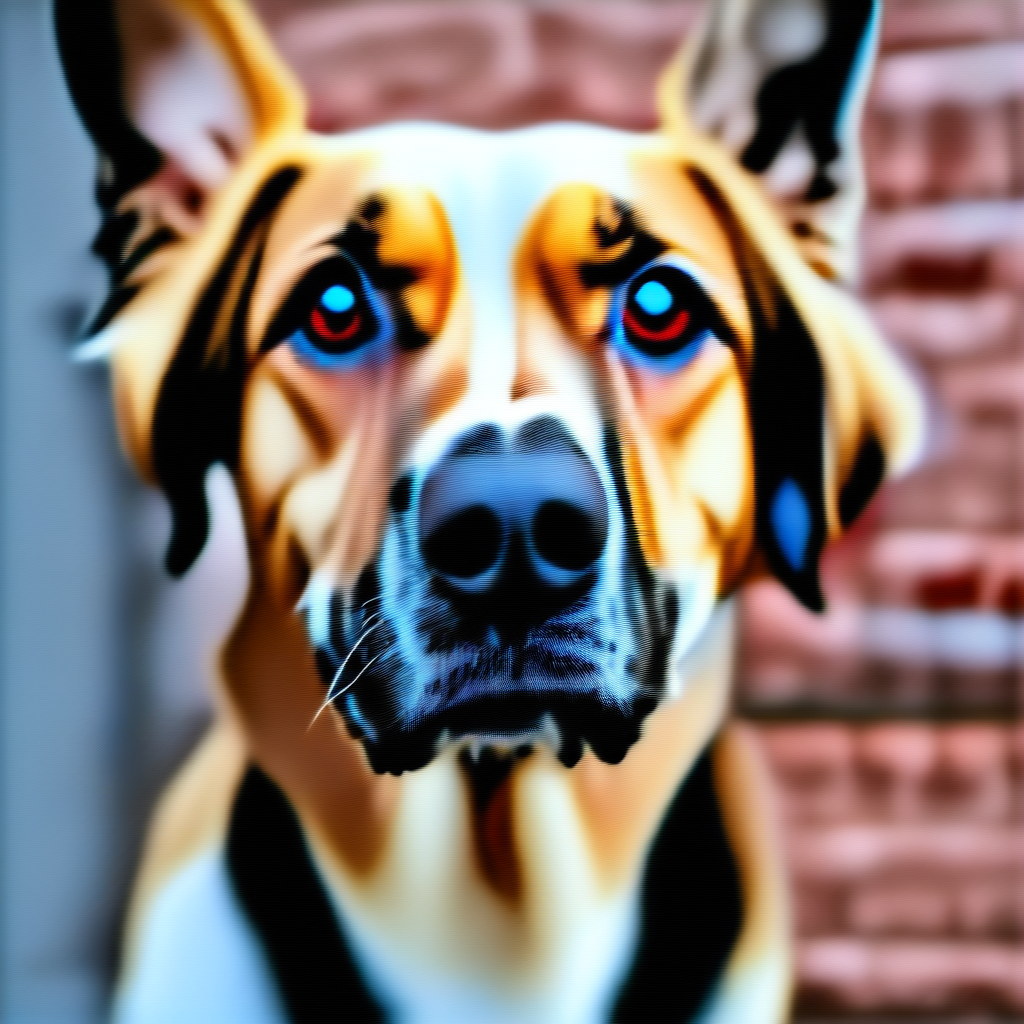}} &
    \raisebox{-.5\height}{\includegraphics[width=\linewidth]{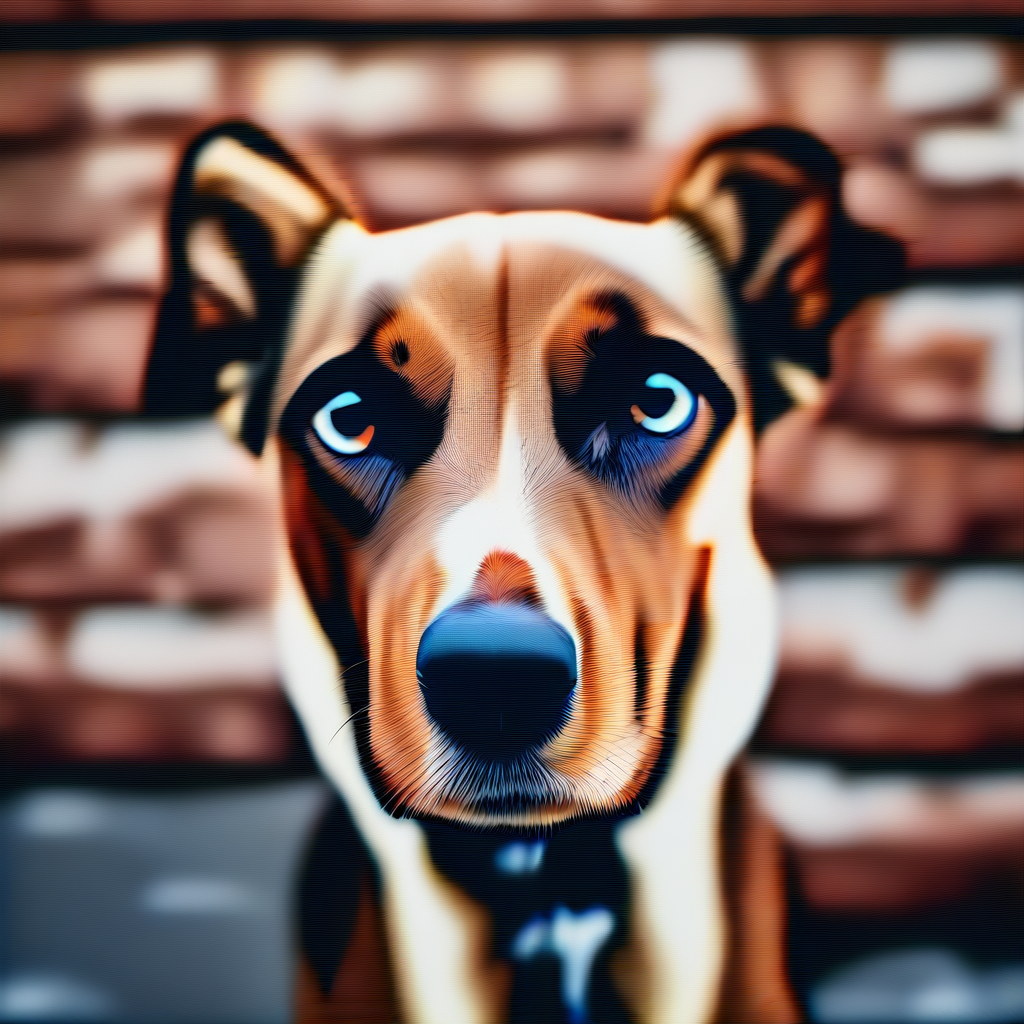}} &
    \raisebox{-.5\height}{\includegraphics[width=\linewidth]{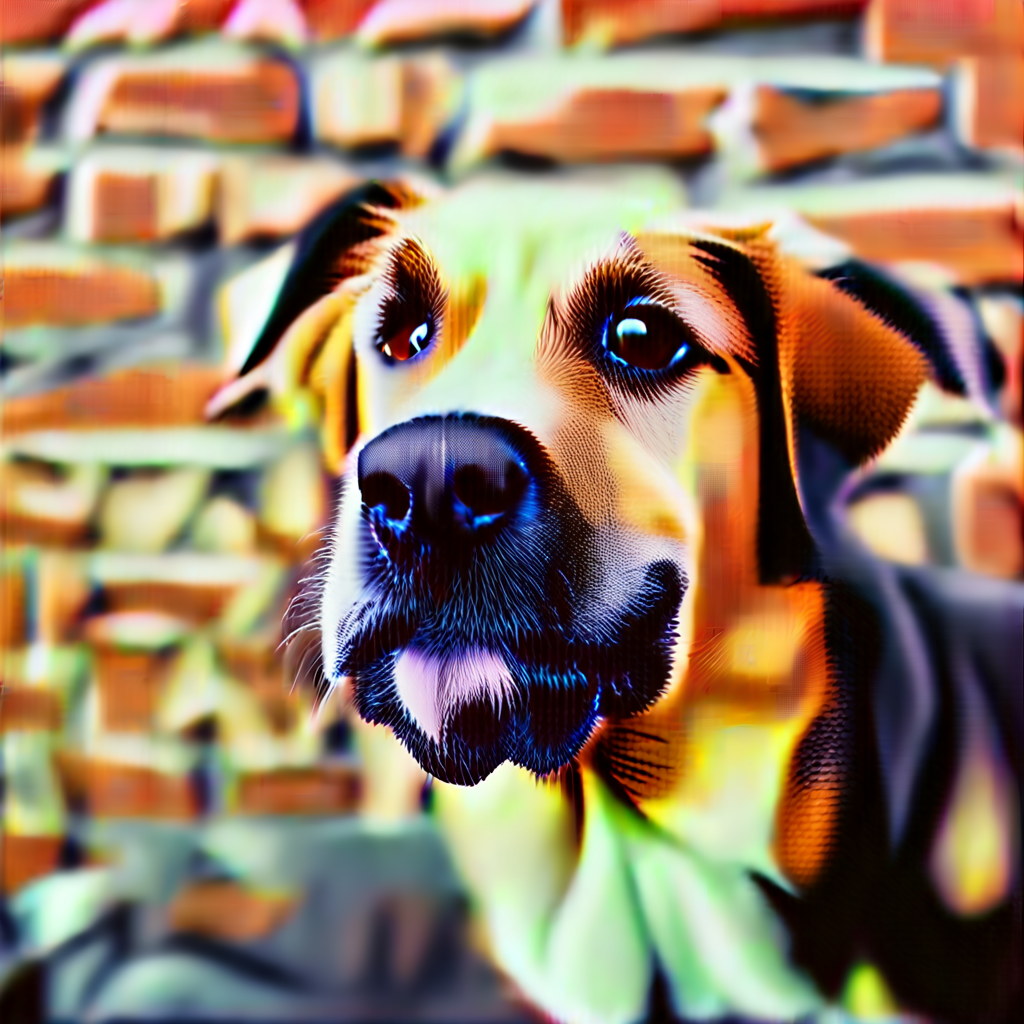}} \\
    \addlinespace[0.35em]
    \textbf{CelebA} &
    \raisebox{-.5\height}{\includegraphics[width=\linewidth]{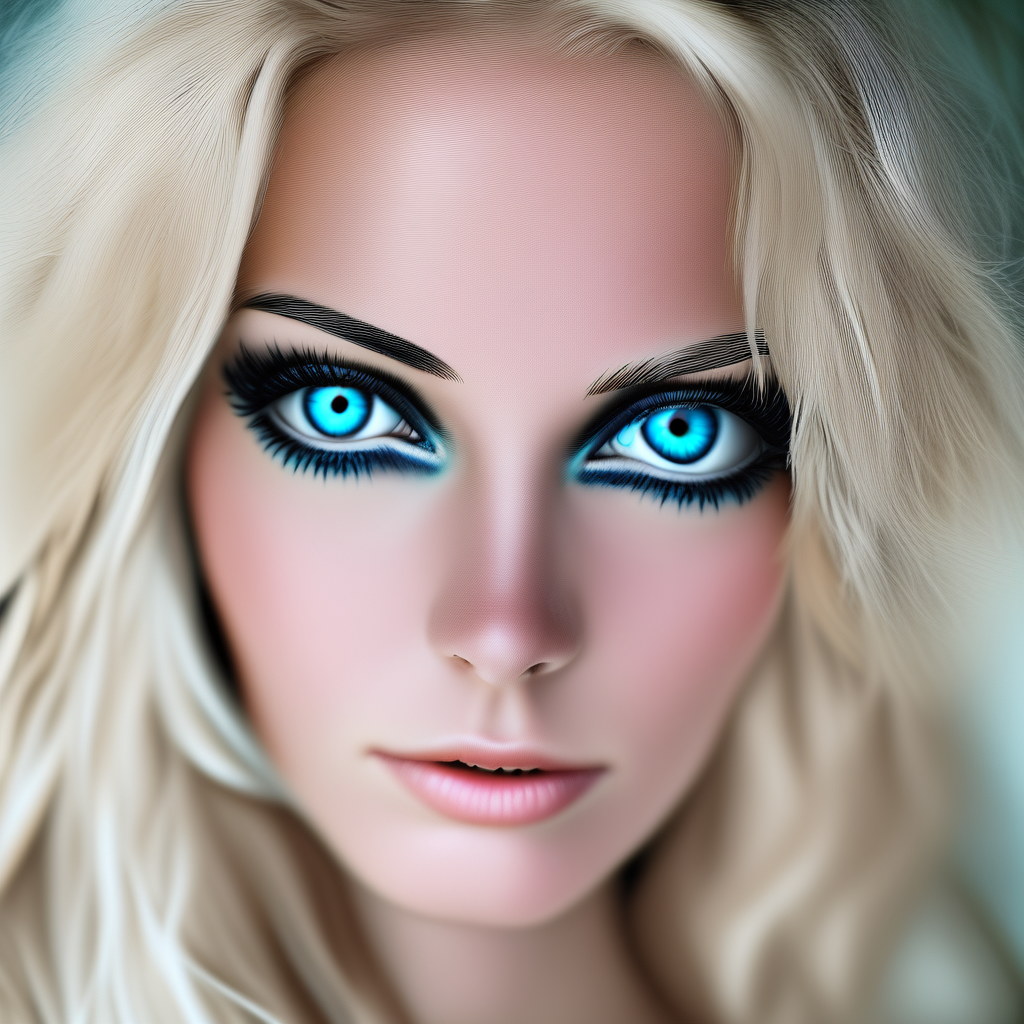}} &
    \raisebox{-.5\height}{\includegraphics[width=\linewidth]{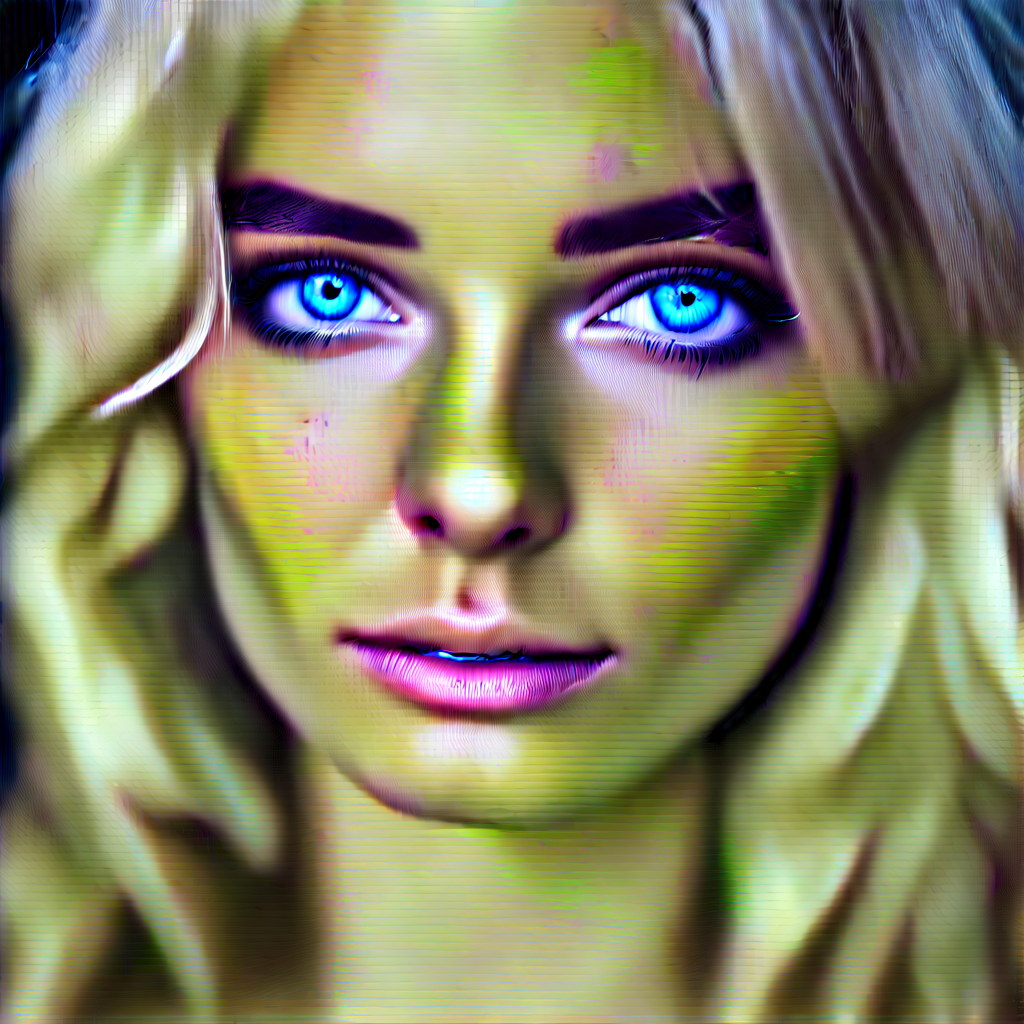}} &
    \raisebox{-.5\height}{\includegraphics[width=\linewidth]{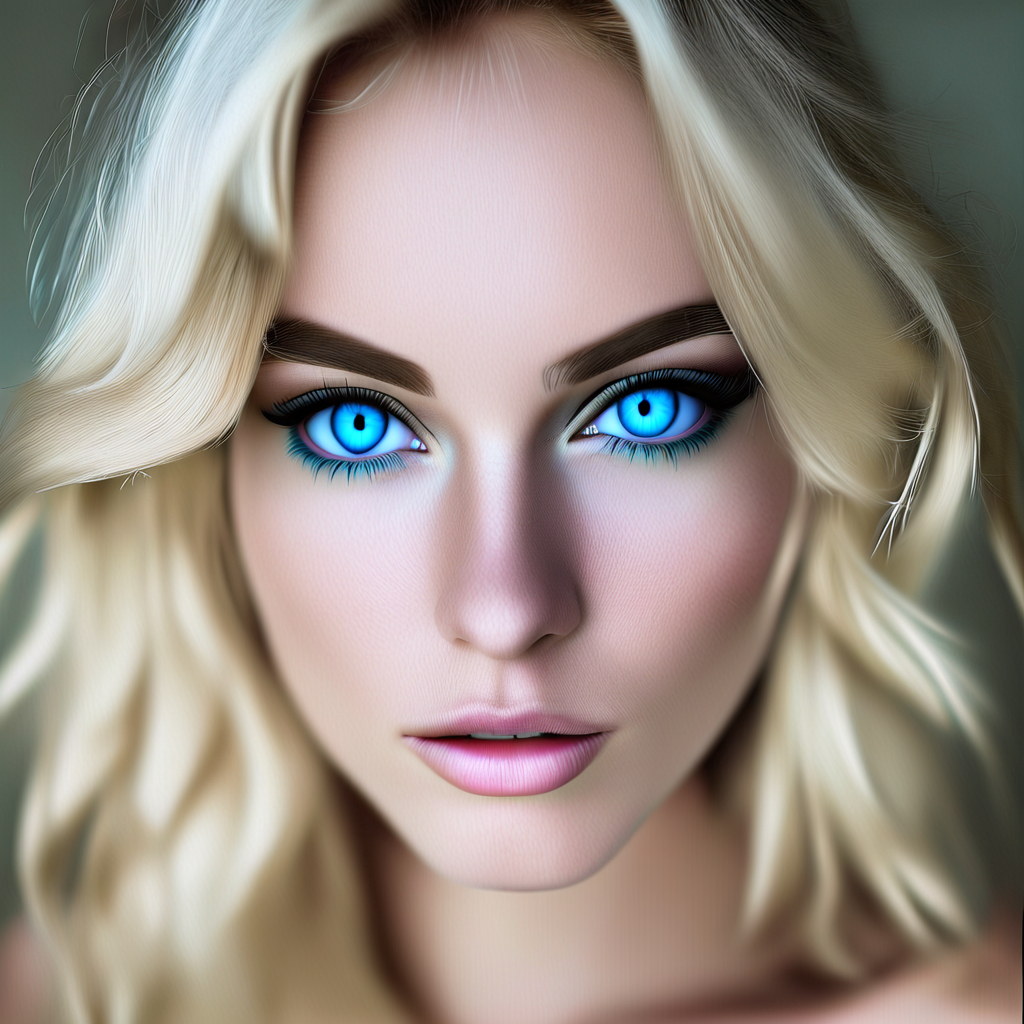}} &
    \raisebox{-.5\height}{\includegraphics[width=\linewidth]{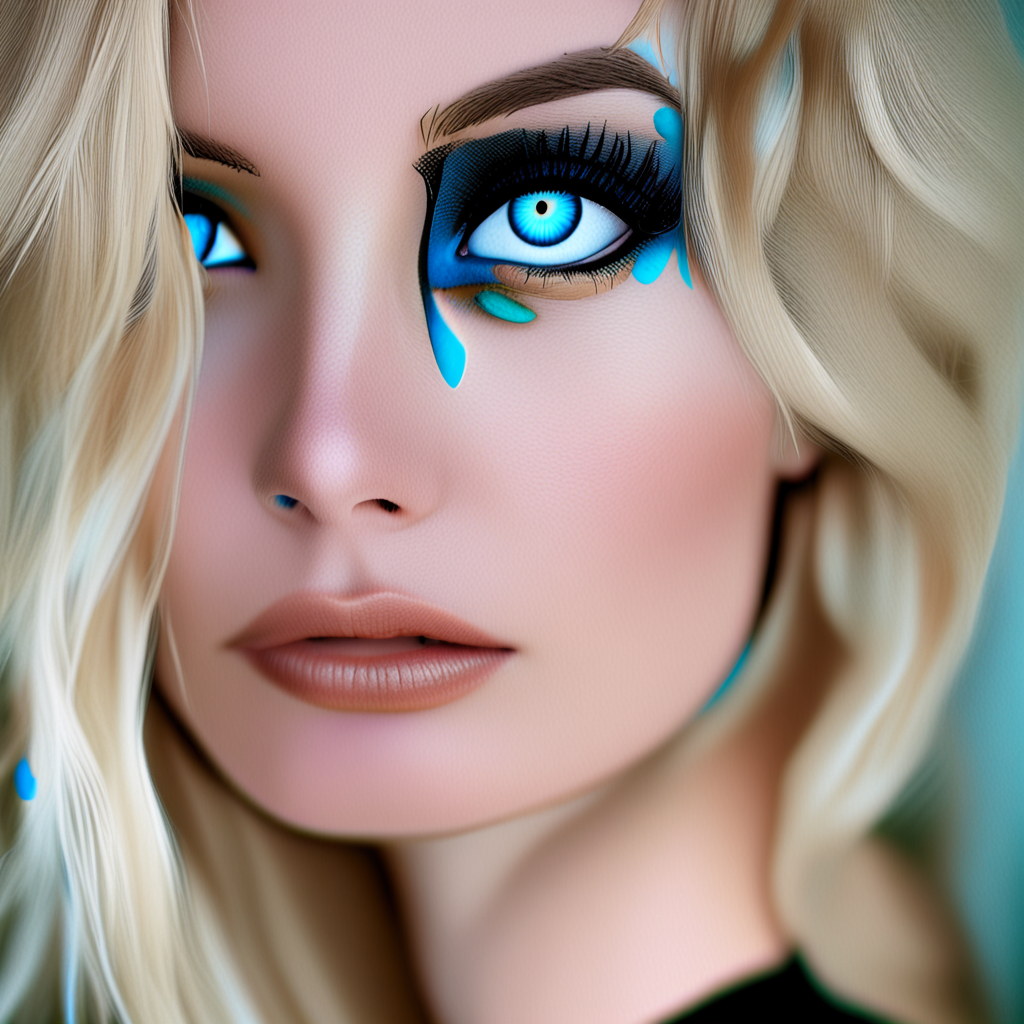}} &
    \raisebox{-.5\height}{\includegraphics[width=\linewidth]{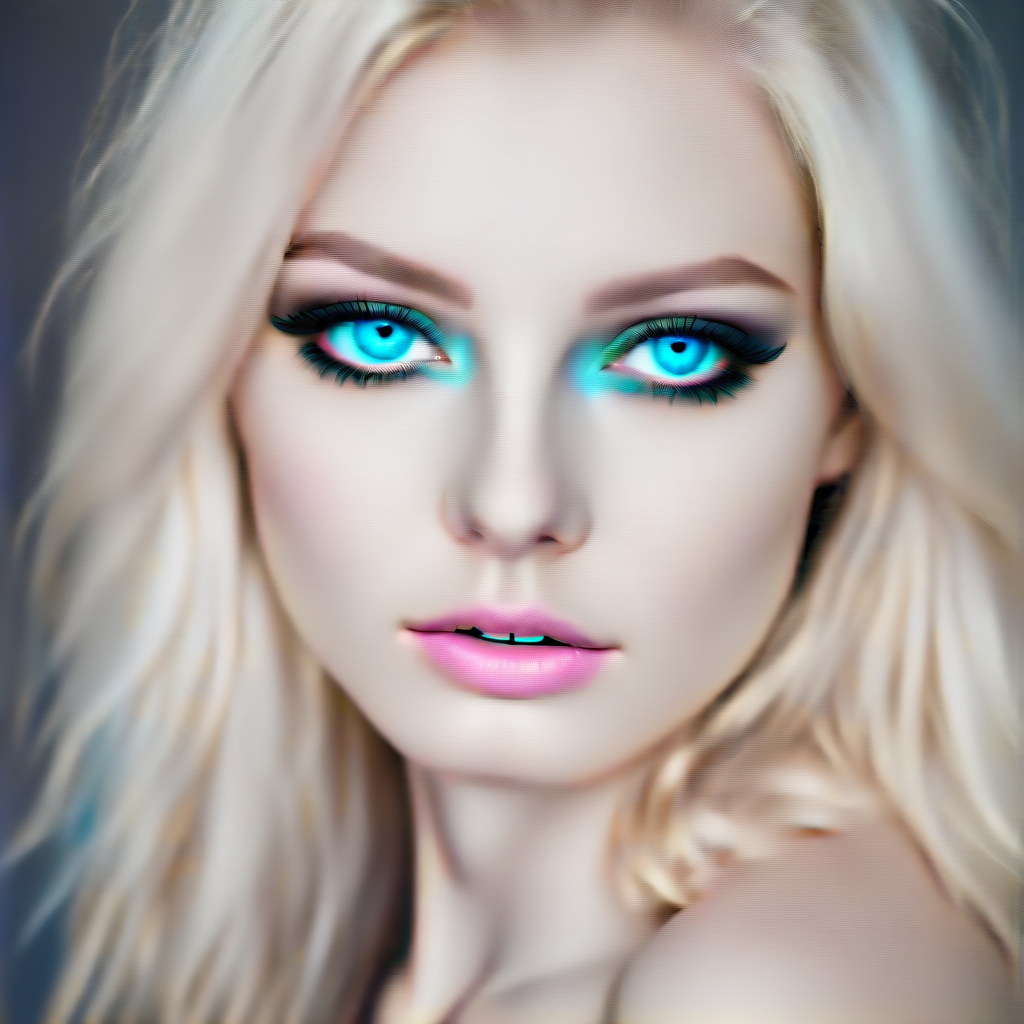}} &
    \raisebox{-.5\height}{\includegraphics[width=\linewidth]{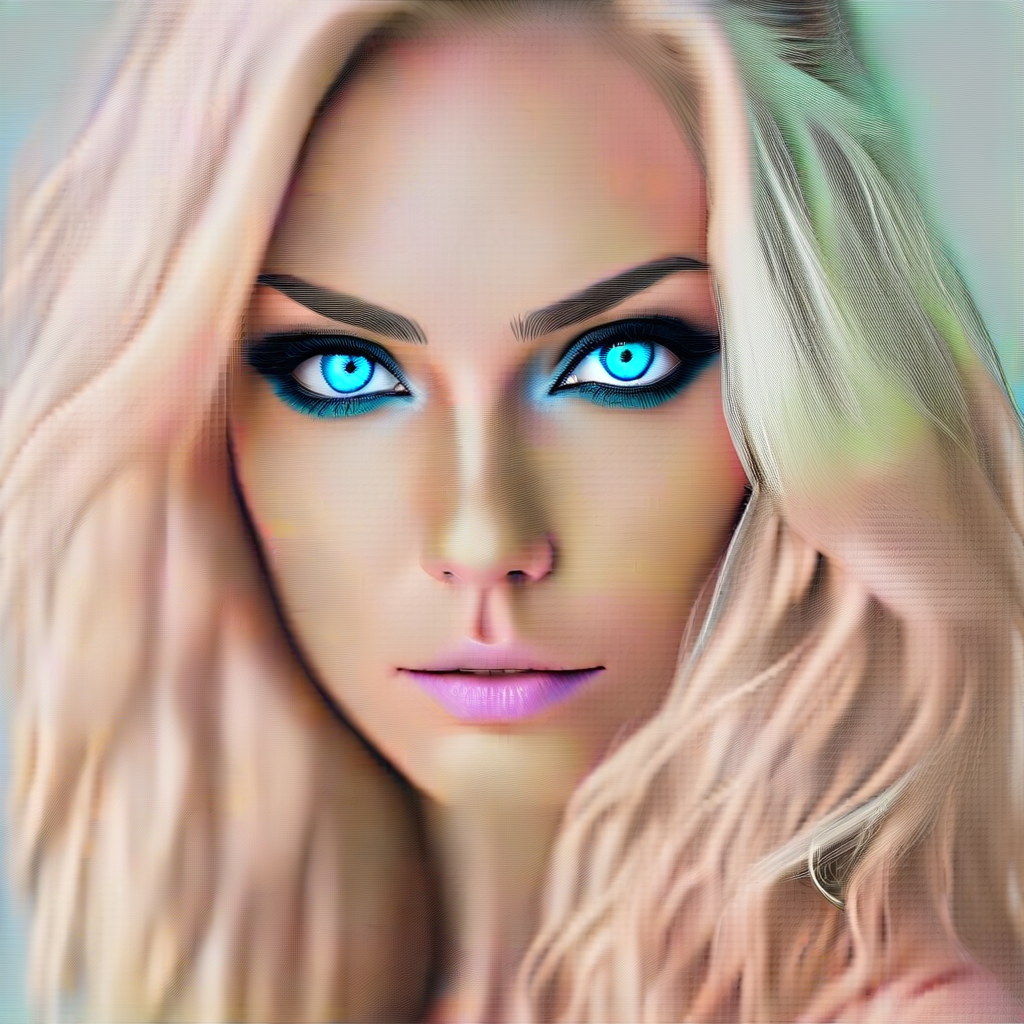}} &
    \raisebox{-.5\height}{\includegraphics[width=\linewidth]{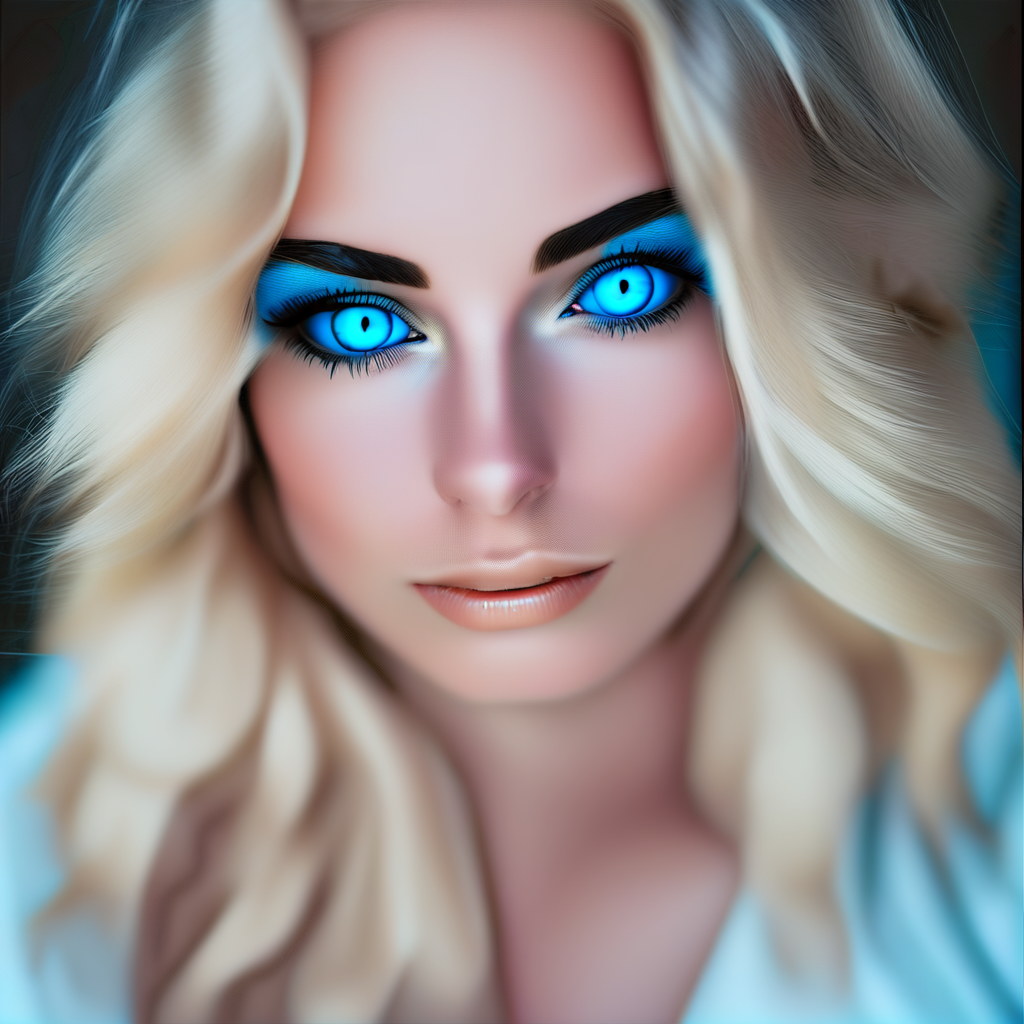}} \\
    \addlinespace[0.35em]
    \textbf{Landscape} &
    \raisebox{-.5\height}{\includegraphics[width=\linewidth]{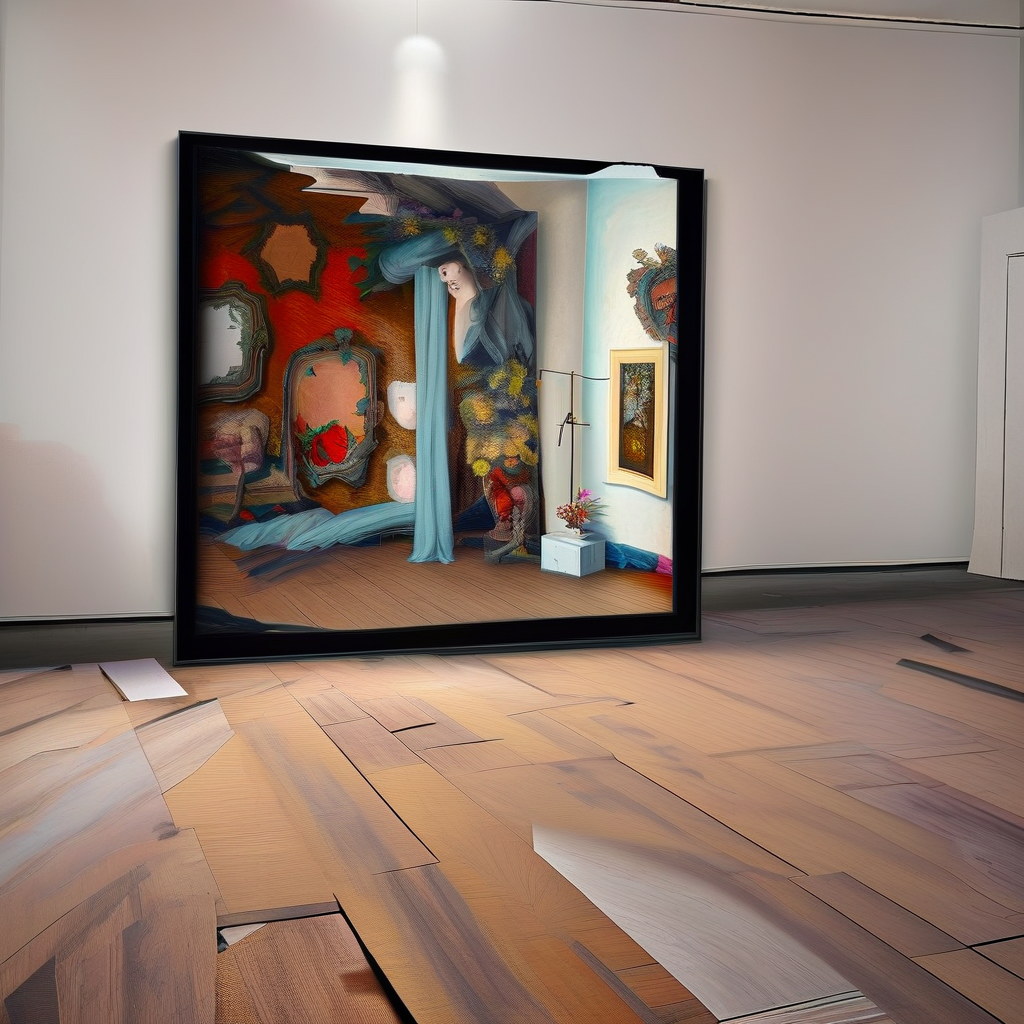}} &
    \raisebox{-.5\height}{\includegraphics[width=\linewidth]{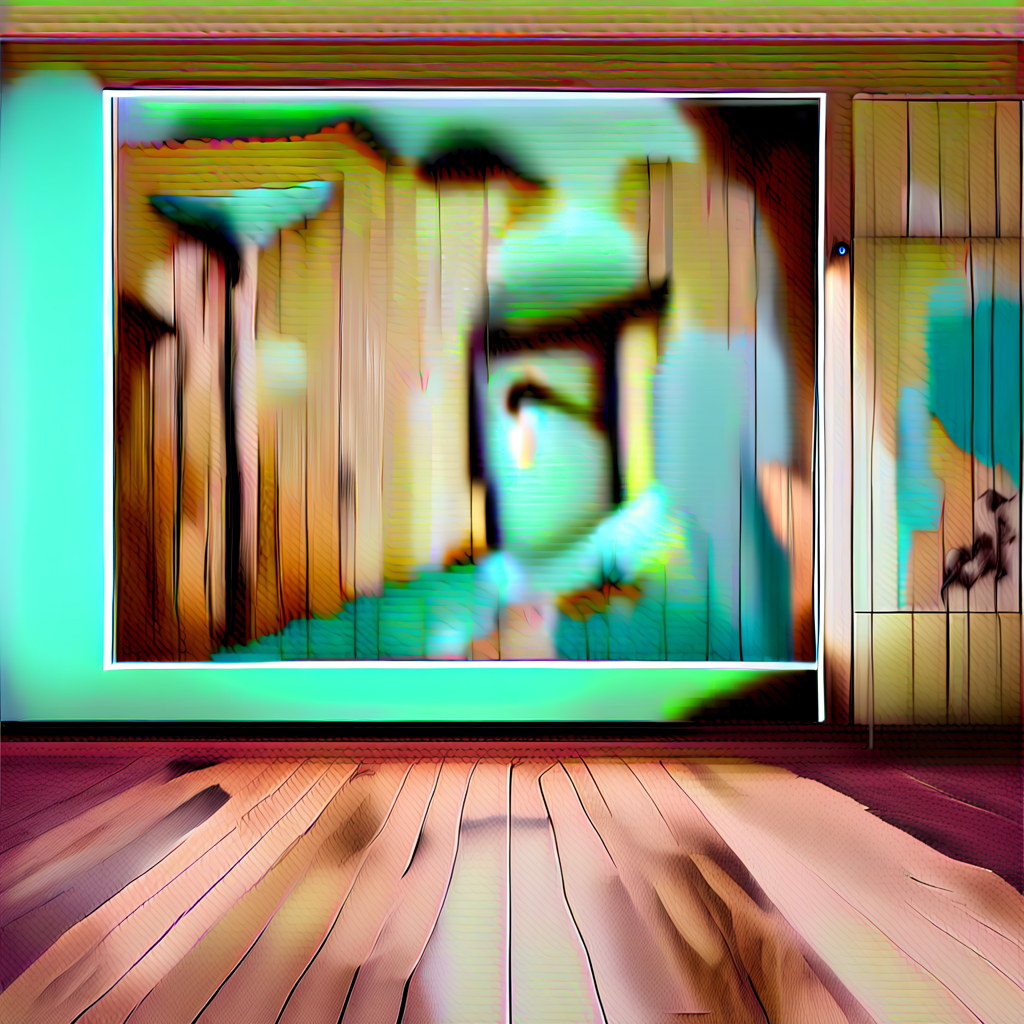}} &
    \raisebox{-.5\height}{\includegraphics[width=\linewidth]{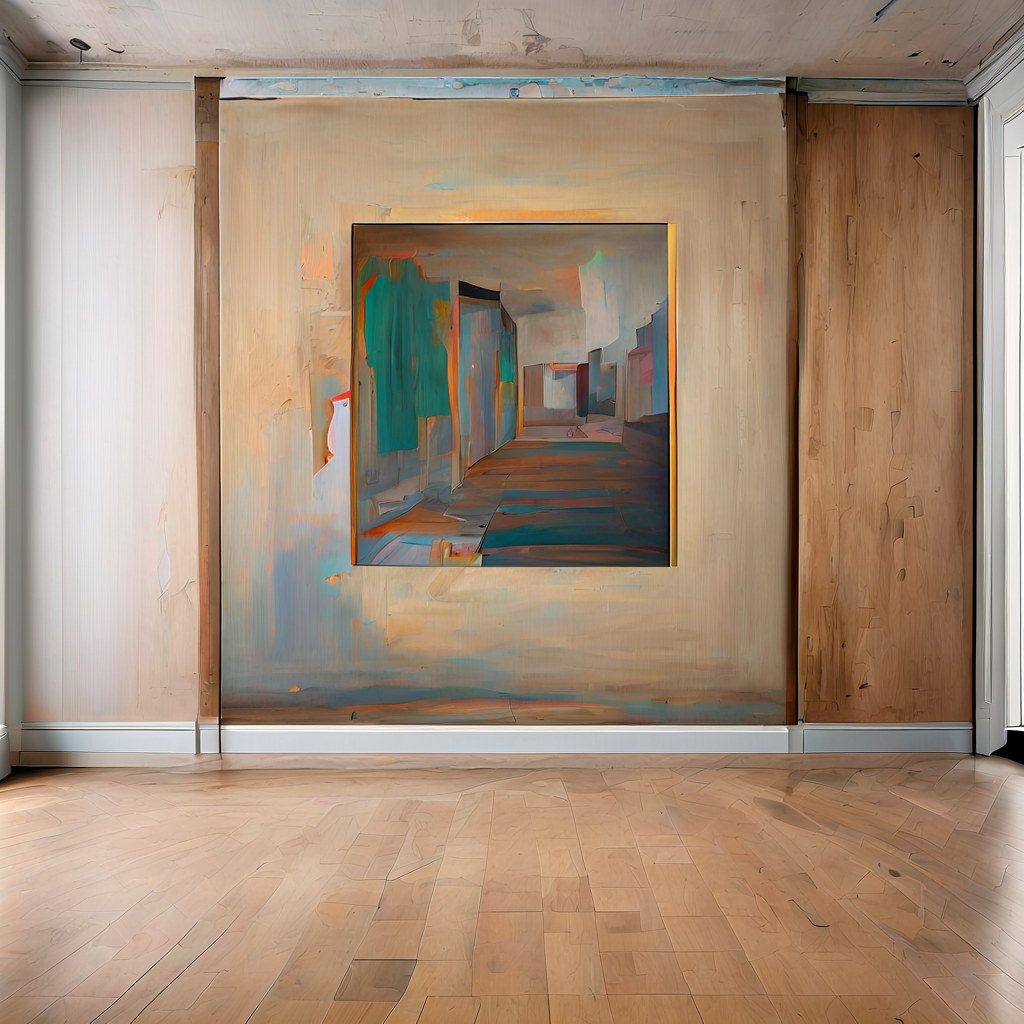}} &
    \raisebox{-.5\height}{\includegraphics[width=\linewidth]{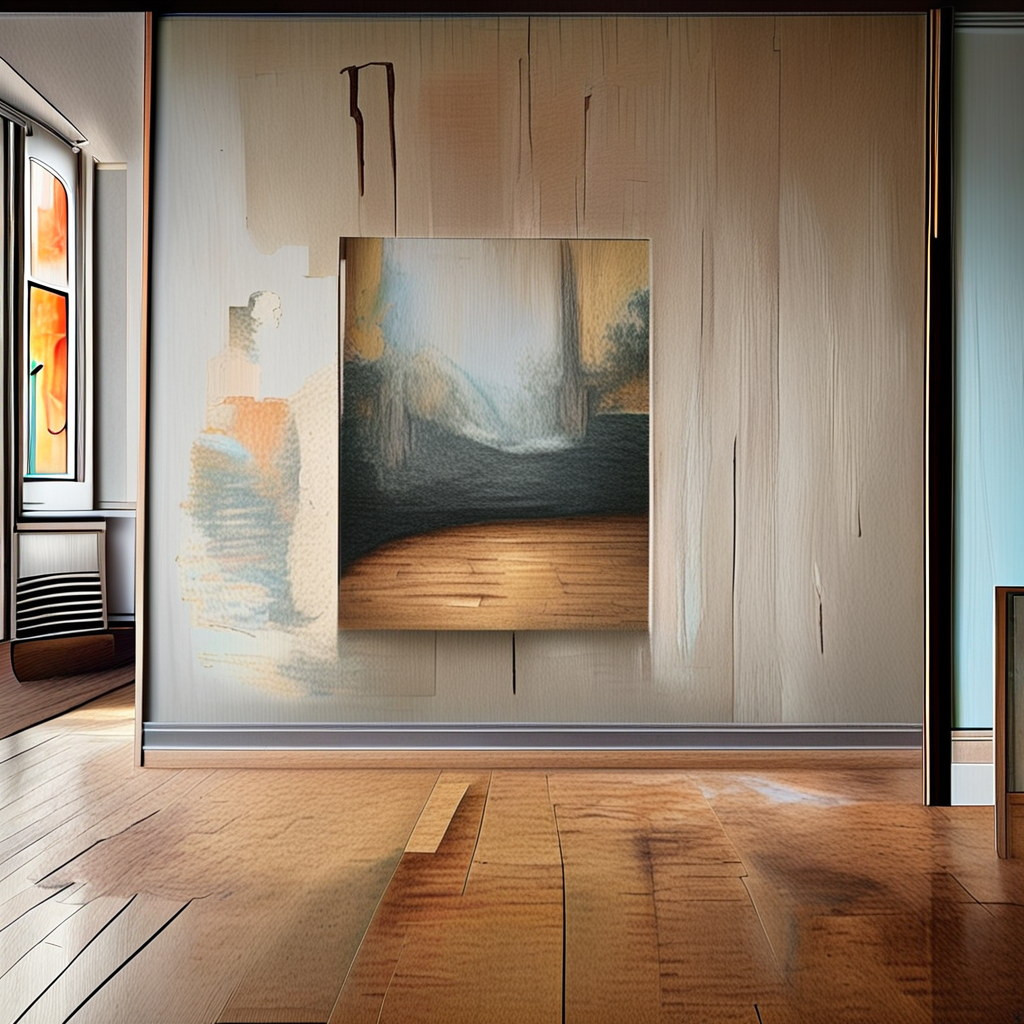}} &
    \raisebox{-.5\height}{\includegraphics[width=\linewidth]{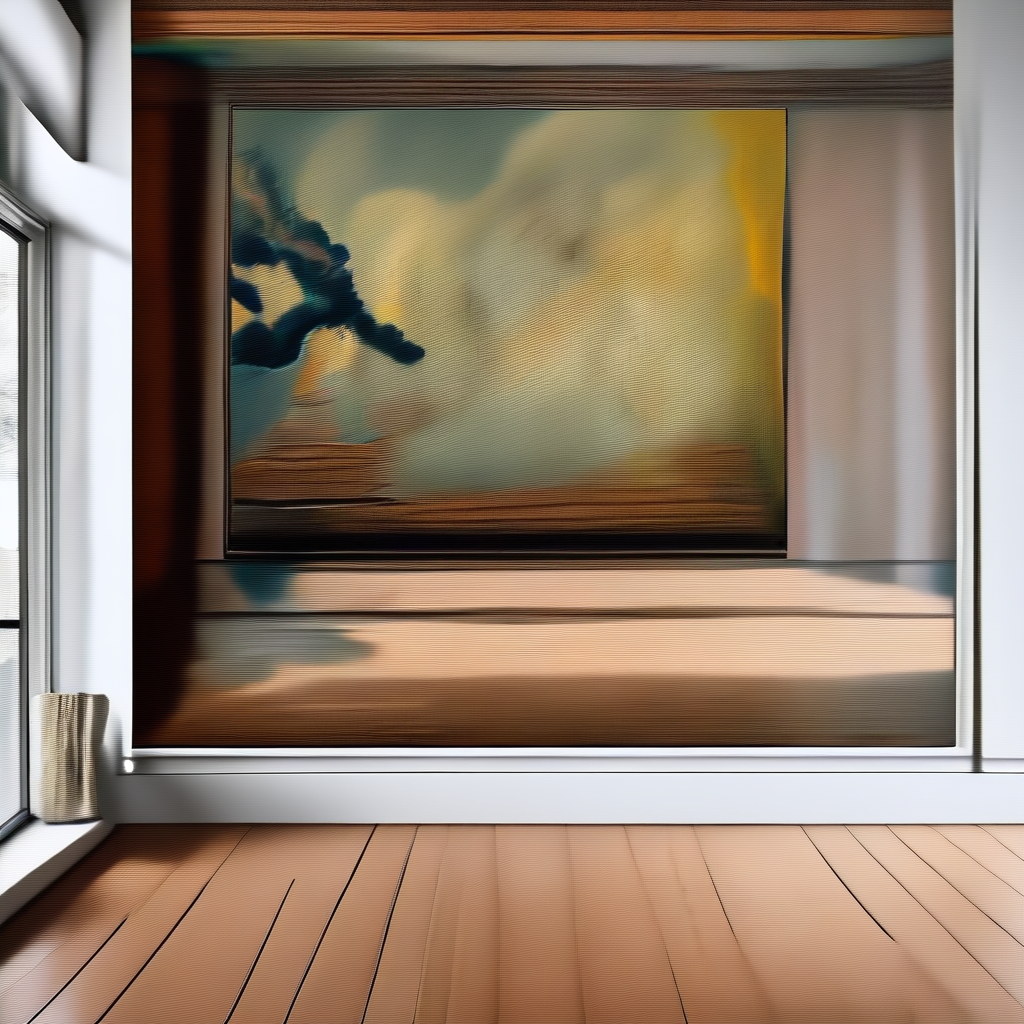}} &
    \raisebox{-.5\height}{\includegraphics[width=\linewidth]{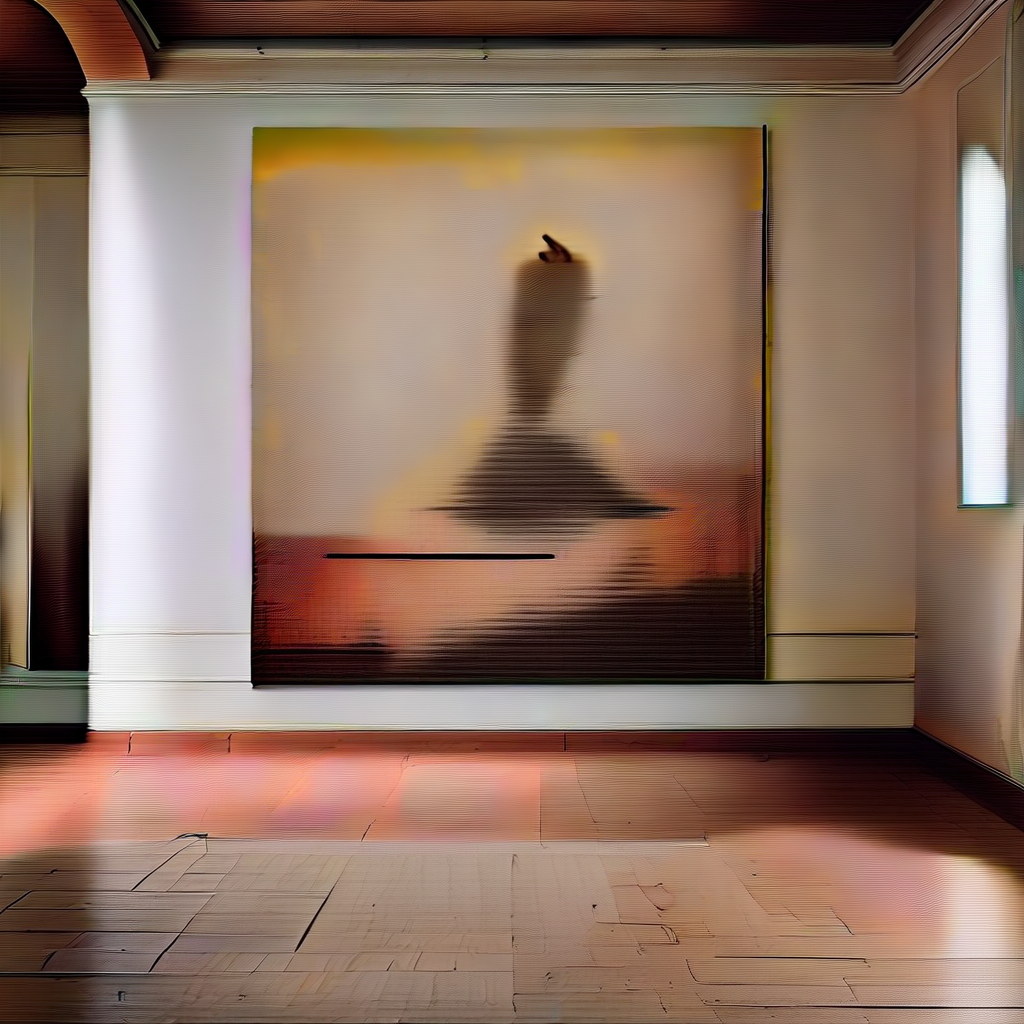}} &
    \raisebox{-.5\height}{\includegraphics[width=\linewidth]{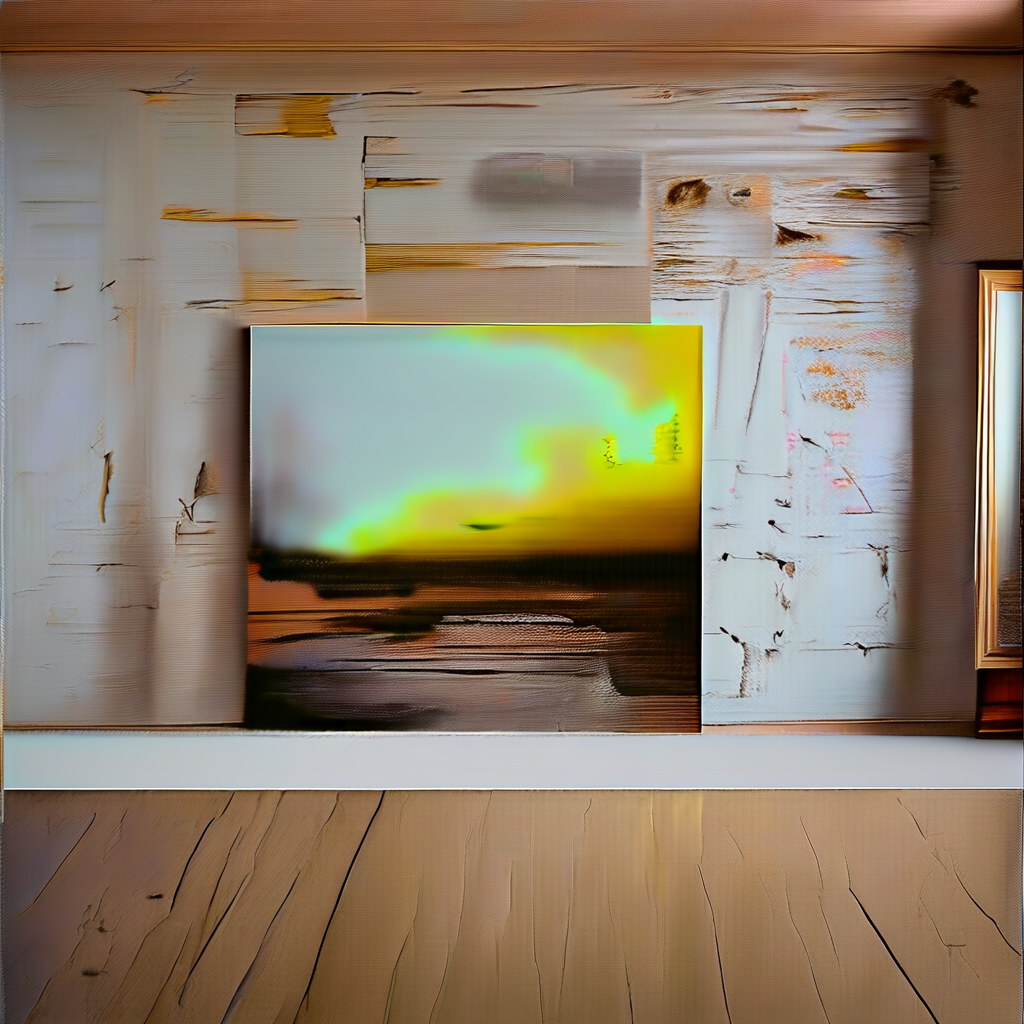}} \\
    \addlinespace[0.35em]
    \textbf{Pokemon} &
    \raisebox{-.5\height}{\includegraphics[width=\linewidth]{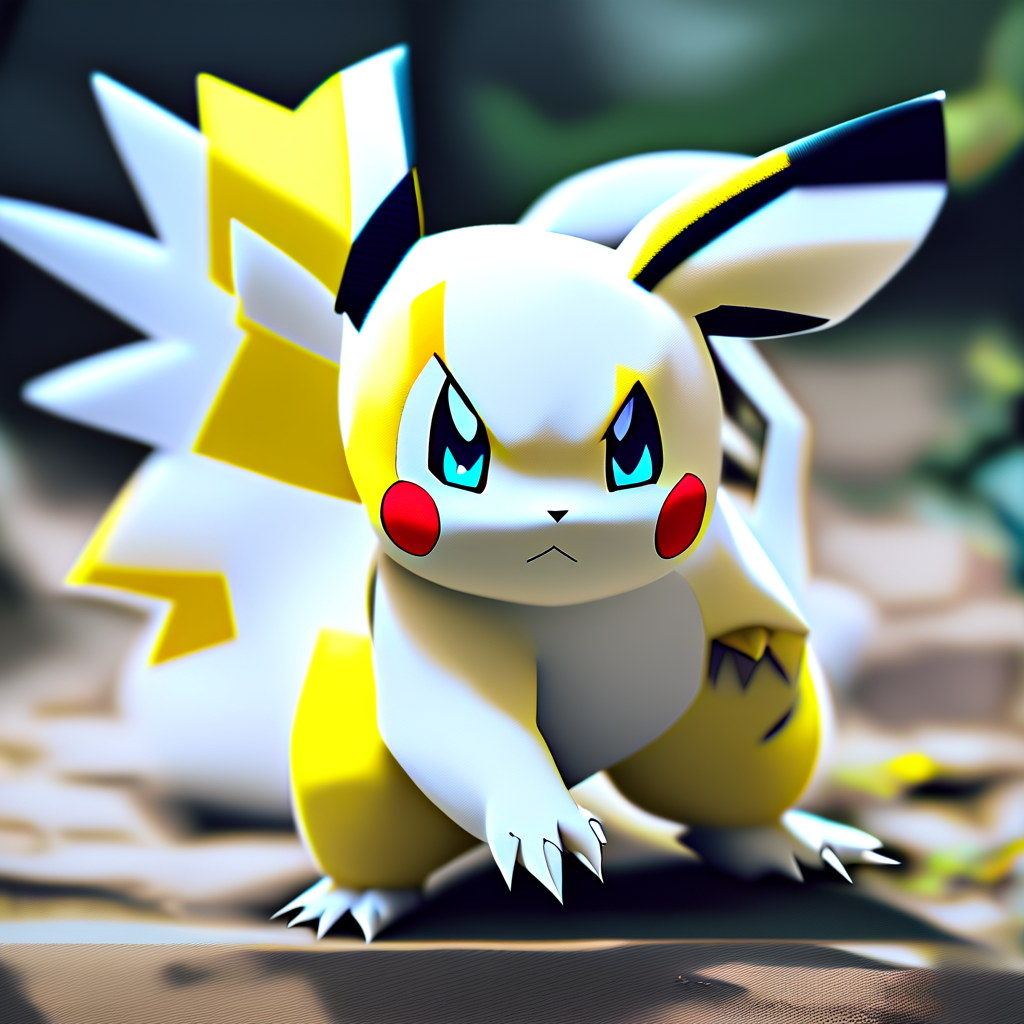}} &
    \raisebox{-.5\height}{\includegraphics[width=\linewidth]{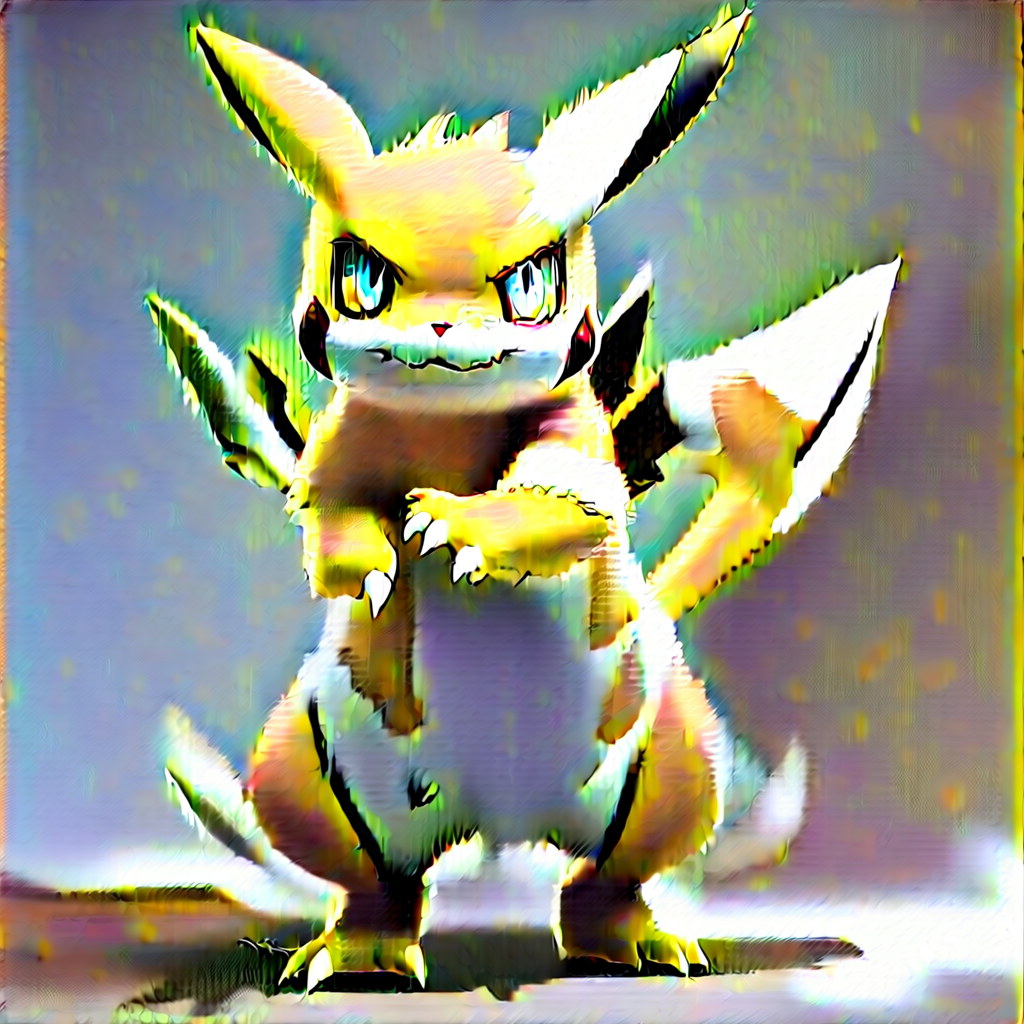}} &
    \raisebox{-.5\height}{\includegraphics[width=\linewidth]{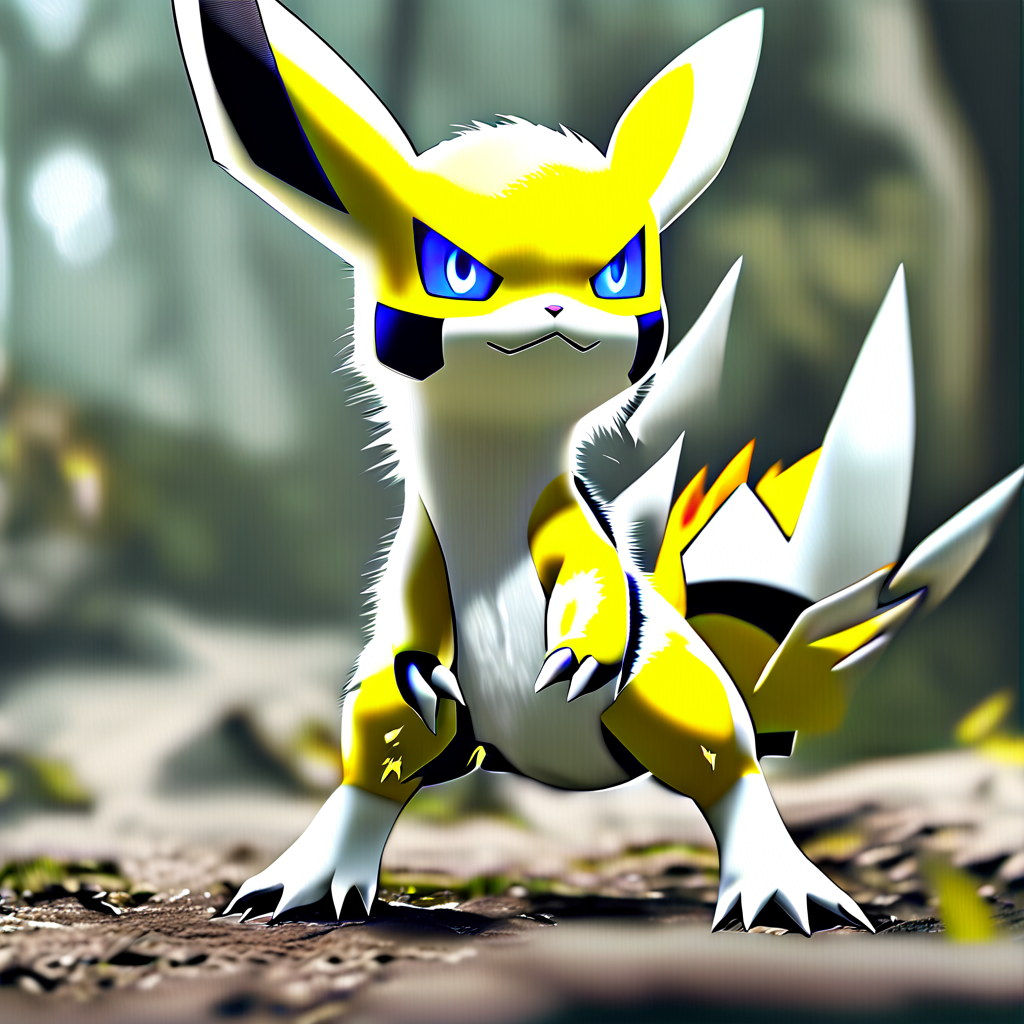}} &
    \raisebox{-.5\height}{\includegraphics[width=\linewidth]{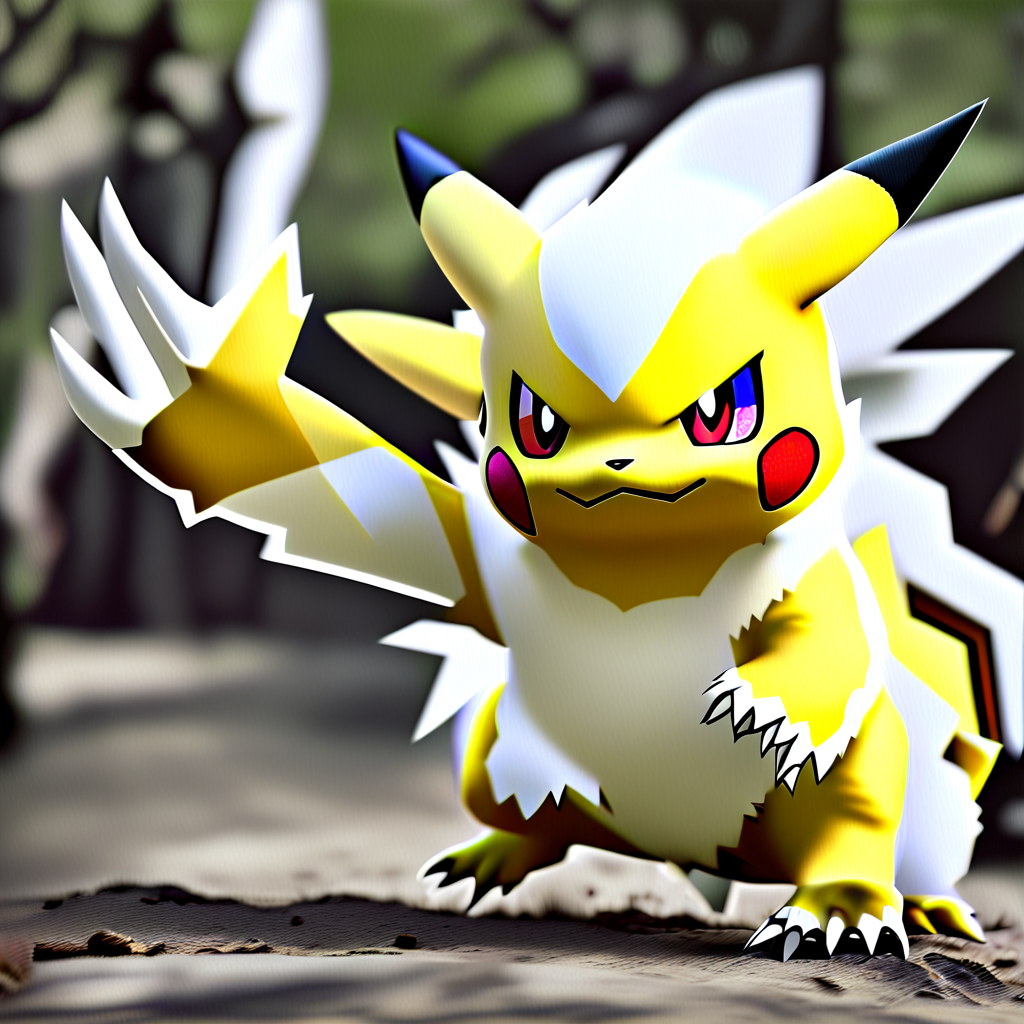}} &
    \raisebox{-.5\height}{\includegraphics[width=\linewidth]{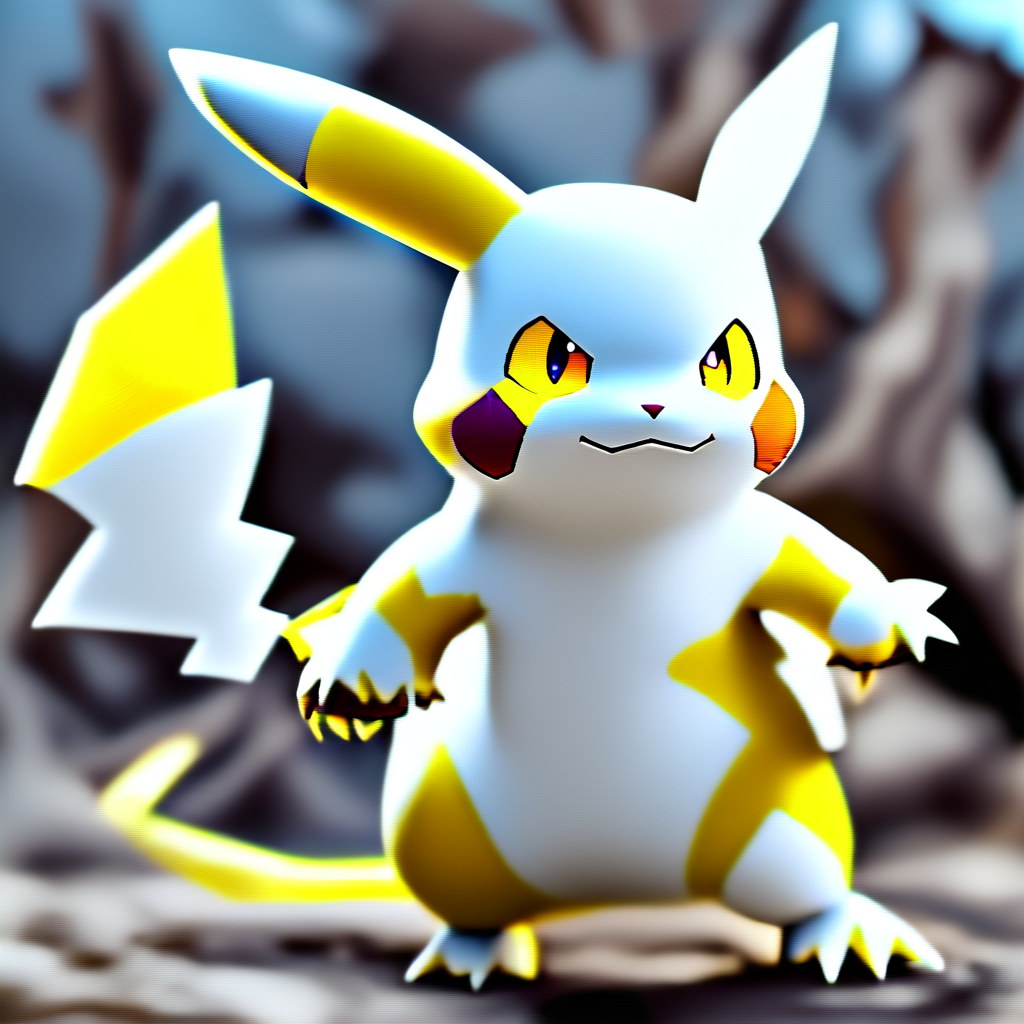}} &
    \raisebox{-.5\height}{\includegraphics[width=\linewidth]{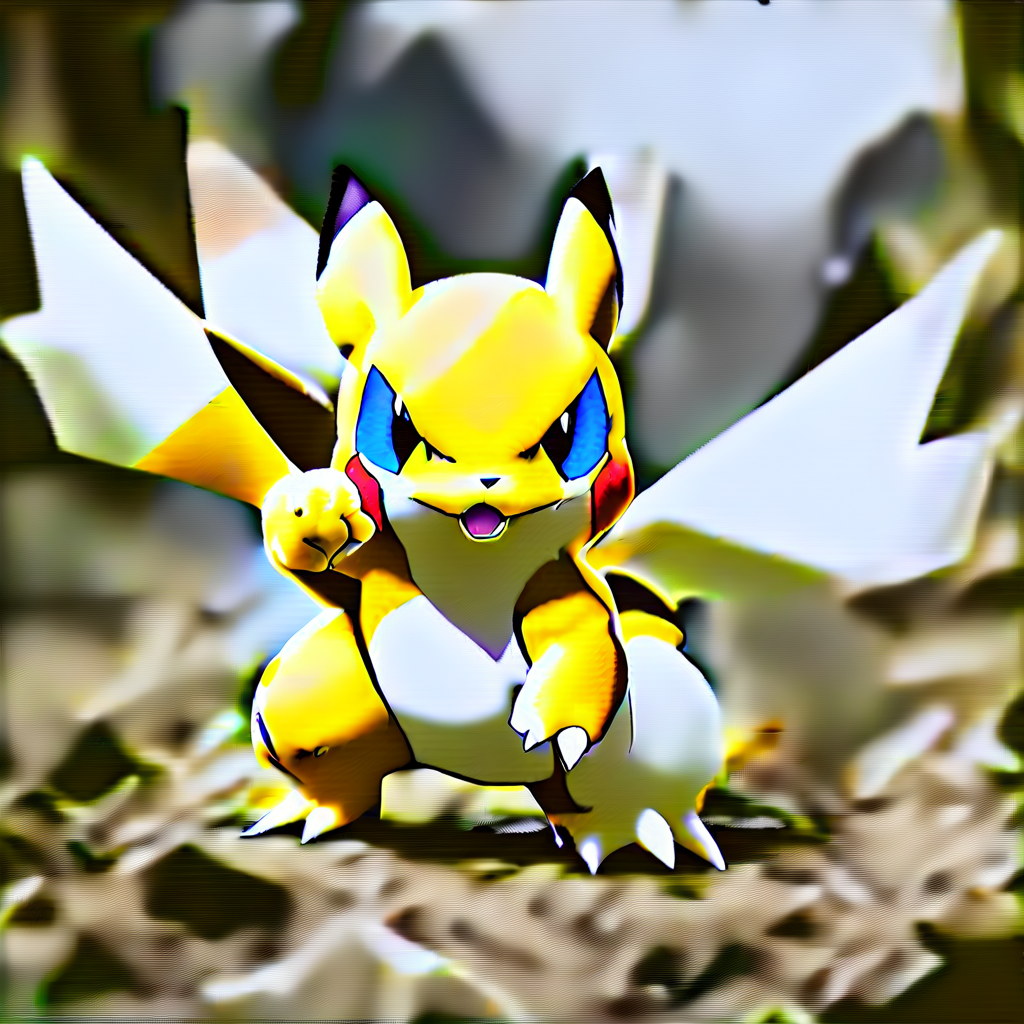}} &
    \raisebox{-.5\height}{\includegraphics[width=\linewidth]{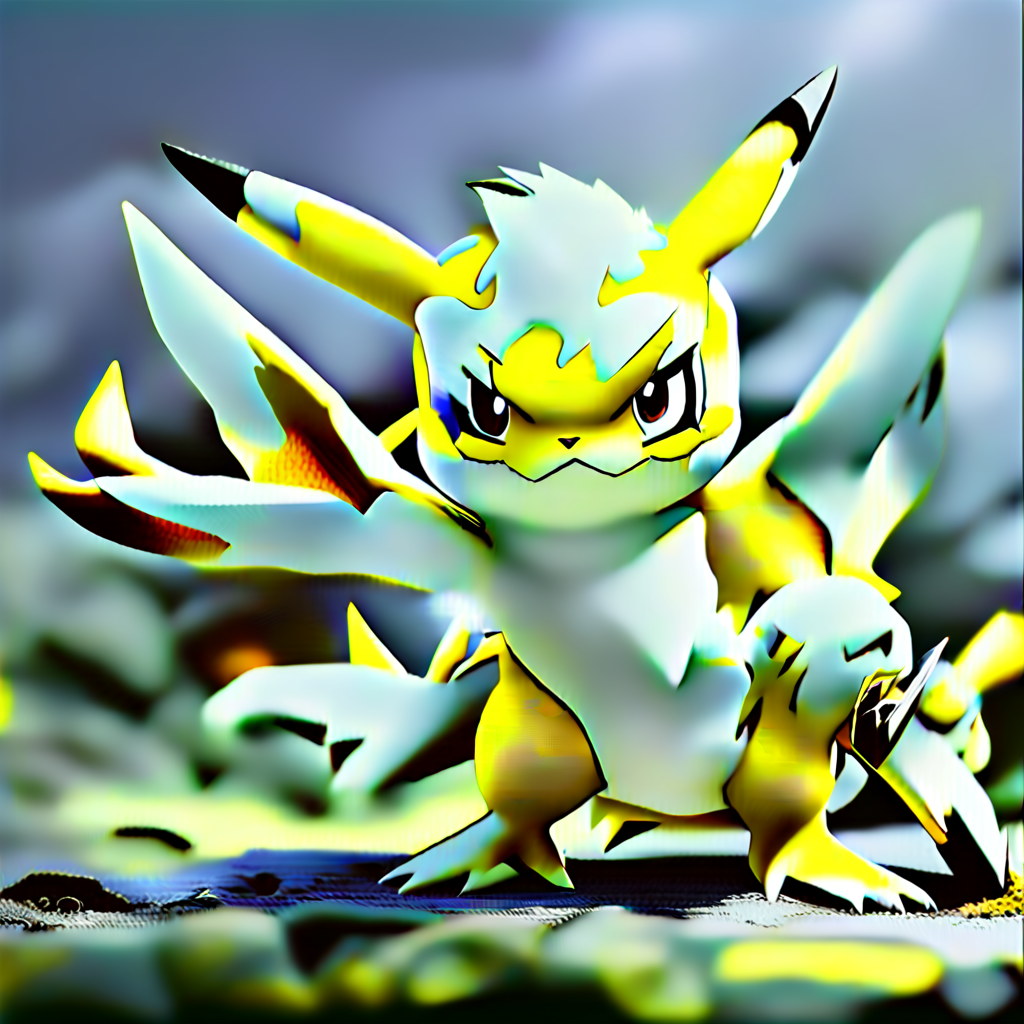}} \\
    \bottomrule
    \end{tabularx}
    \caption{Disruption effectiveness against model distillation using PixArt as the source model and SSD as the substitute model across four datasets.}
    \label{fig:generation_quality_comparison}
\end{figure*}

\section{Evaluation}
Due to space limitations, we provide the detailed experimental setup and results in Appendix~\ref{Appendix:expsetup} and Appendix~\ref{Appendix:expresult}.\looseness=-1

\subsection{Experimental Setup}
\label{sec:evaluation_setup}
\mypara{Model.}
We report results across four representative models: SSD, PixArt, SDXL, and Shuttle3~\cite{segmind2023ssd1b,chen2024pixartsigma,podell2023sdxl,shuttleai2024shuttle3}.
These models cover Stable-Diffusion-style latent diffusion backbones and newer high-resolution T2I generators. 

\mypara{Dataset.}
The evaluation uses four image datasets covering animal faces, human faces, natural landscapes, and Pokemon-style images: AFHQ, CelebA, Landscape, and Pokemon~\cite{choi2020starganv2,liu2015faceattributes,tabularisai2025artisticlandscape,djilax2024pokemon}.
For each dataset, we randomly sample 5{,}000 images to obtain decoder-training prompts, 500 disjoint images to obtain prompts for constructing the generated protected dataset for model distillation, and 500 disjoint images to obtain prompts for evaluating the distilled substitute model.
We caption each sampled image with BLIP-Large~\cite{blip} and use the resulting caption as its associated prompt.

\mypara{Metrics.}
We organize evaluation metrics into two groups: substitute model distillation metrics and visual fidelity metrics to evaluate the disruption effectiveness and visual-quality preservation.
For the compact table, we introduce a short name for each metric.\looseness=-1
\looseness=-1

\noindent\textit{Substitute distillation metrics.}
These metrics evaluate the quality of the substitute model trained on collected service outputs, where stronger defense corresponds to more severe degradation of the distilled model.
We report Fr\'echet Inception Distance (\mFID)~\cite{heusel2017fid}, QAlign score (\mQA)~\cite{wu2023qalign}, PickScore (\mPK)~\cite{kirstain2023pickapic}, and Aesthetic score (\mAES)~\cite{laion2022aesthetics}.
Higher \mFID{} indicates a larger distributional mismatch from the target reference, while lower \mQA{}, \mPK{}, and \mAES{} indicate lower human-aligned quality, preference, and visual appeal, respectively. \looseness=-1

\noindent\textit{Visual fidelity metrics.}
These metrics evaluate whether the delivered protected image $x_p$ remains visually close to the clean service reference $x_r$ produced by the original decoder.
We report LPIPS (\mLP)~\cite{zhang2018lpips}, SSIM (\mSS)~\cite{wang2004ssim}, CIEDE2000 (\mDE)~\cite{luo2001ciede2000}, and PSNR (\mPN)~\cite{hore2010psnr}.
Better visual fidelity corresponds to lower LPIPS and CIEDE2000, and higher SSIM and PSNR.
The same visual metrics are also used in the motivating study in Table~\ref{tab:motivation_gap}.\looseness=-1

\mypara{Baselines.}
We compare \method{} with five representative preventive image-protection baselines: AdvDM~\cite{liang2023advdm}, Anti-DreamBooth~\cite{vanle2023antidreambooth}, Glaze~\cite{shan2023glaze}, Nightshade~\cite{shan2024nightshade}, and PhotoGuard~\cite{salman2023photoguard}.
In their original form, these methods optimize sample-specific perturbations; in our comparison, we adapt each objective to the same decoder-training and deployment protocol as \method{}.
All baseline decoders use the same dataset splits, prompts, source model, substitute model, image resolution, and visual fidelity constraints to ensure a matched decoder-level comparison; this comparison evaluates whether their objectives remain effective in our service-internal setting, not whether the original methods fail in their intended deployment settings.
Detailed baseline adaptations are provided in Appendix~\ref{Appendix:expsetup}.
\looseness=-1

\mypara{Implementation Details.}
In the main experiments, 
Ref. denotes samples from the initial SSD substitute model before distillation, Clean denotes samples after distillation on clean service outputs, and defense columns denote samples after distillation on protected outputs.
We train the decoder protection stage for $5$ epochs with full parameters and then perform downstream distillation training for $2$ epochs using LORA~\cite{hu2021lora}. All generated images are produced at a resolution of $1024\times1024$.
For distillation evaluation, the attacker collects prompt-image pairs from the protected service and trains the SSD substitute model on this data.\looseness=-1

\subsection{Main Results}

We evaluate black-box T2I distillation across four source models and four datasets; Table~\ref{tab:main_extraction_results} reports the quantitative results.
Figure~\ref{fig:generation_quality_comparison} further shows representative substitute model generations after distillation and supports three key findings. First, the degradation is dataset-wide: across AFHQ, Pokemon, CelebA, and Landscape, \method{} consistently yields much worse distilled substitutes than Clean and is usually stronger than the adapted protection baselines. 
On PixArt, for instance, QAlign drops from 3.10 under Clean to 1.59 under \method{}, 3.68 to 1.45, 4.18 to 1.91, and 4.32 to 1.74 on the four datasets, respectively. 
Second, the effect is not tied to a single source model. 
\method{} is the strongest defense in most evaluated source-model and metric settings, including nearly all SSD settings and across all listed dataset-metric settings on PixArt. 
On Landscape, for example, it reduces QAlign from 4.17 under Clean to 1.76 under \method{} on SSD, 4.32 to 1.74 on PixArt, 4.21 to 2.59 on SDXL, and 4.61 to 2.18 on Shuttle3. 
Third, \method{} substantially outperforms the five adapted baselines. On PixArt with AFHQ, \method{} achieves a QAlign score of 1.59, far below AdvDM 2.98, Anti-DreamBooth 3.16, Glaze 2.83, Nightshade 3.11, and PhotoGuard 2.30.
Moreover, as shown in the generated images, \method{} introduces not only pronounced structural perturbations, such as noticeable background stripes, which we attribute to the induced latent shift produced by self-referenced latent maximization, but also clear color fluctuations, which are caused by the color regularization.
In contrast, other defenses may distort individual examples but often remain visually closer to the Clean column.
Together, the results show that \method{} delivers the most consistent degradation among the evaluated defenses of the attacker's distilled substitute models among the evaluated defenses. \looseness=-1

\begin{figure}[tbp]
\centering
\footnotesize
\setlength{\tabcolsep}{1.4pt}
\renewcommand{\arraystretch}{1.08}
\begin{tabularx}{\columnwidth}{@{}l*{4}{>{\centering\arraybackslash}X}@{}}
\toprule
\rowcolor{tablepanel}\multicolumn{5}{c}{\textbf{Qualitative Results of Benign Visual Fidelity}} \\
\midrule
\rowcolor{tablehead}\textbf{Output} & \textbf{AFHQ} & \textbf{CelebA} & \textbf{Landscape} & \textbf{Pokemon} \\
\textbf{Clean} &
\raisebox{-.5\height}{\includegraphics[width=\linewidth]{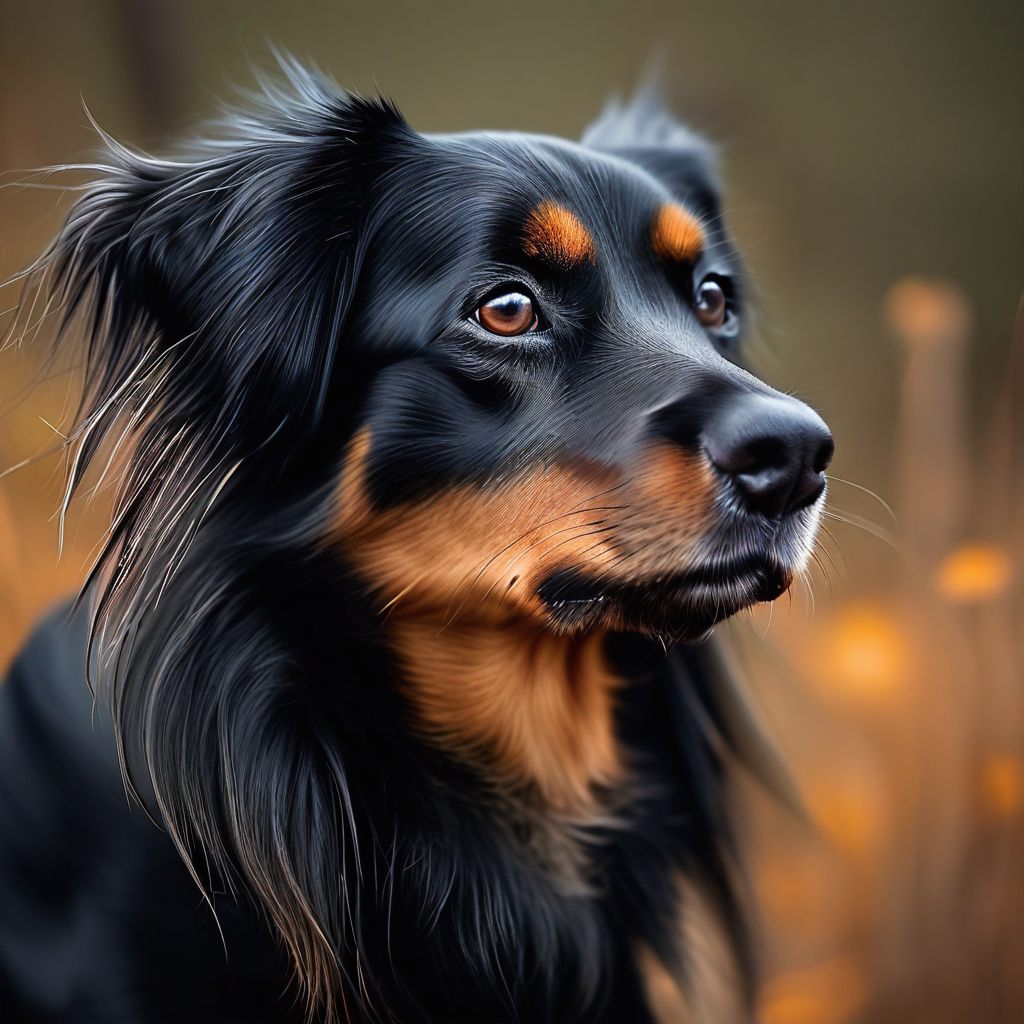}} &
\raisebox{-.5\height}{\includegraphics[width=\linewidth]{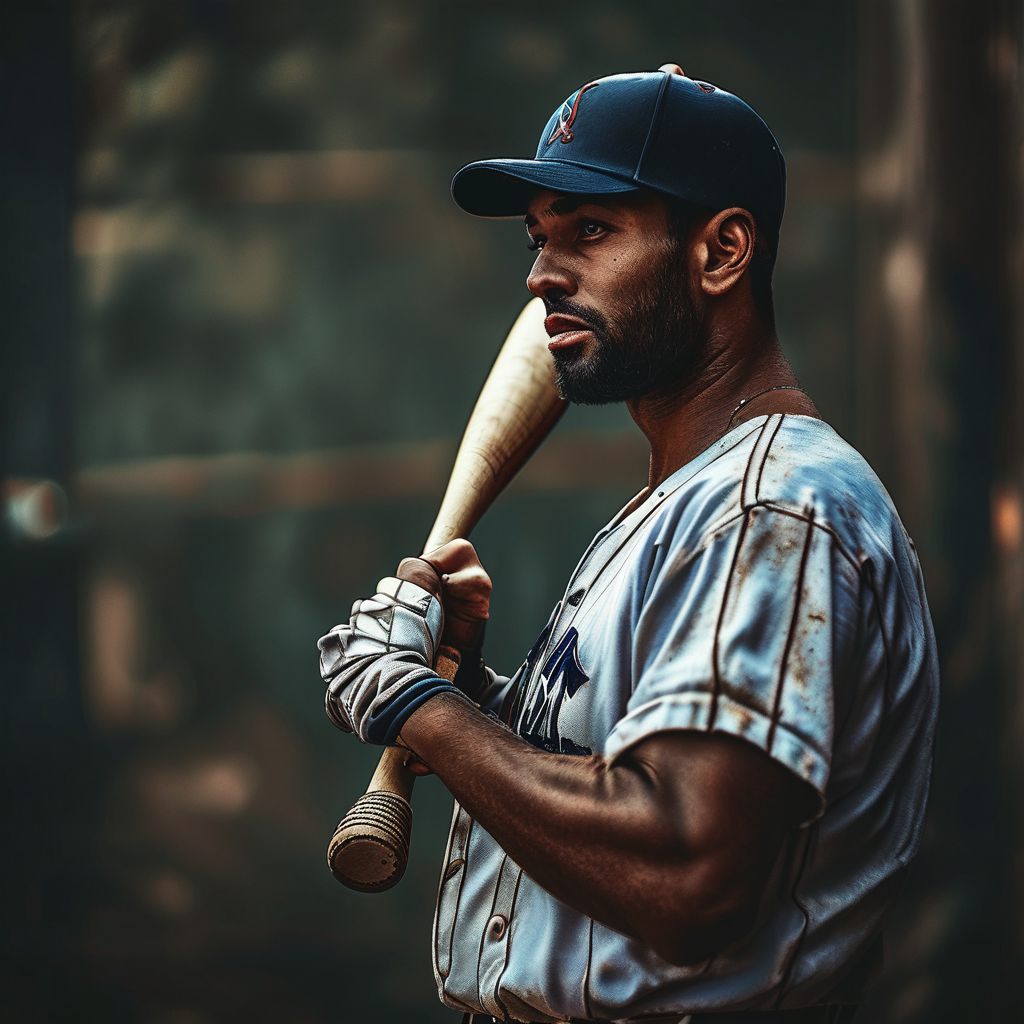}} &
\raisebox{-.5\height}{\includegraphics[width=\linewidth]{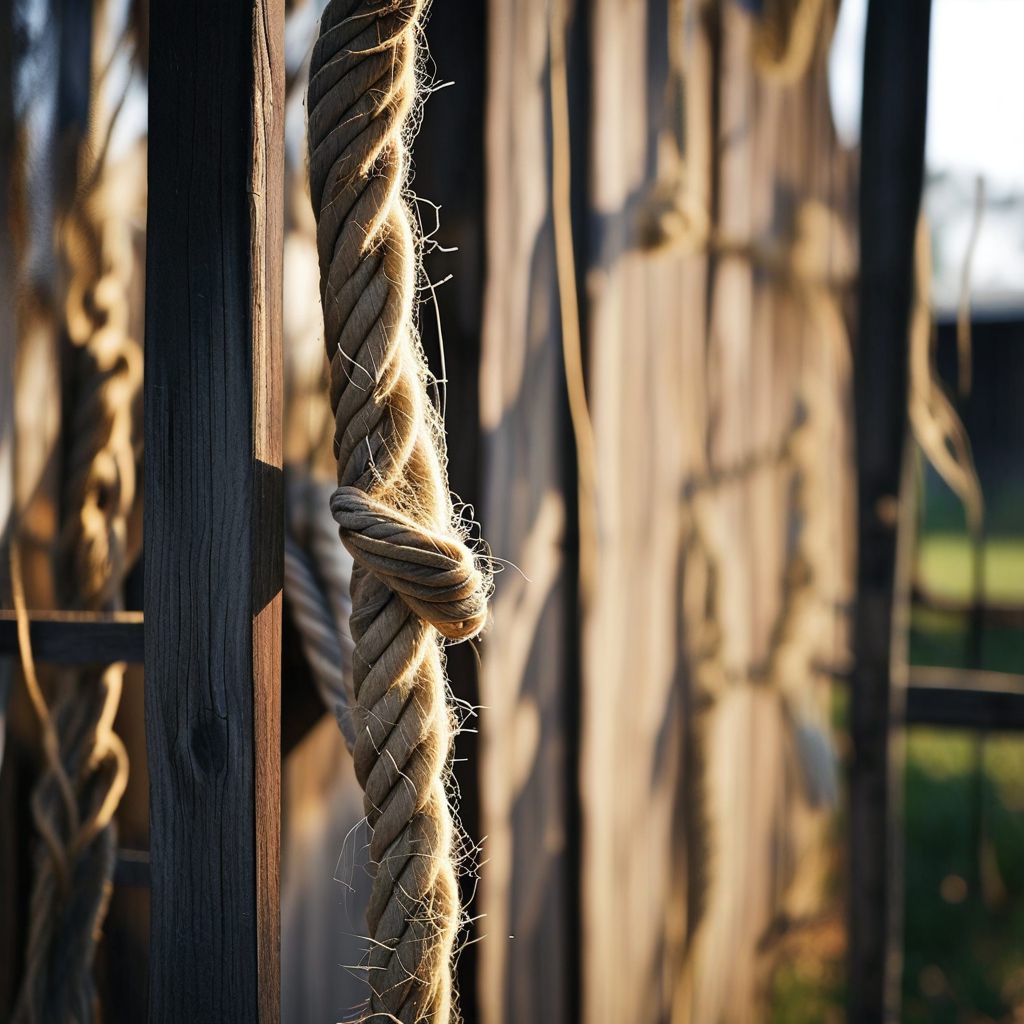}} &
\raisebox{-.5\height}{\includegraphics[width=\linewidth]{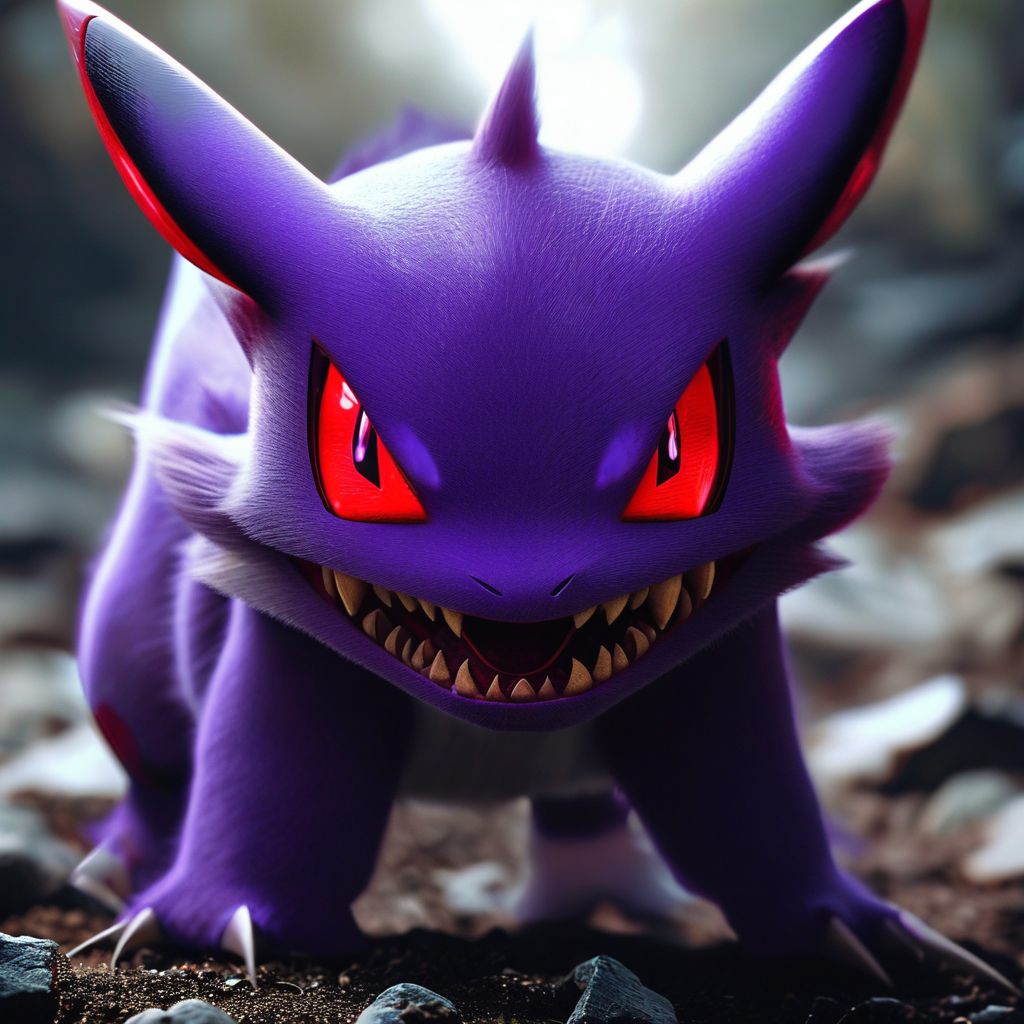}} \\
\addlinespace[0.3em]
\textbf{\method} &
\raisebox{-.5\height}{\includegraphics[width=\linewidth]{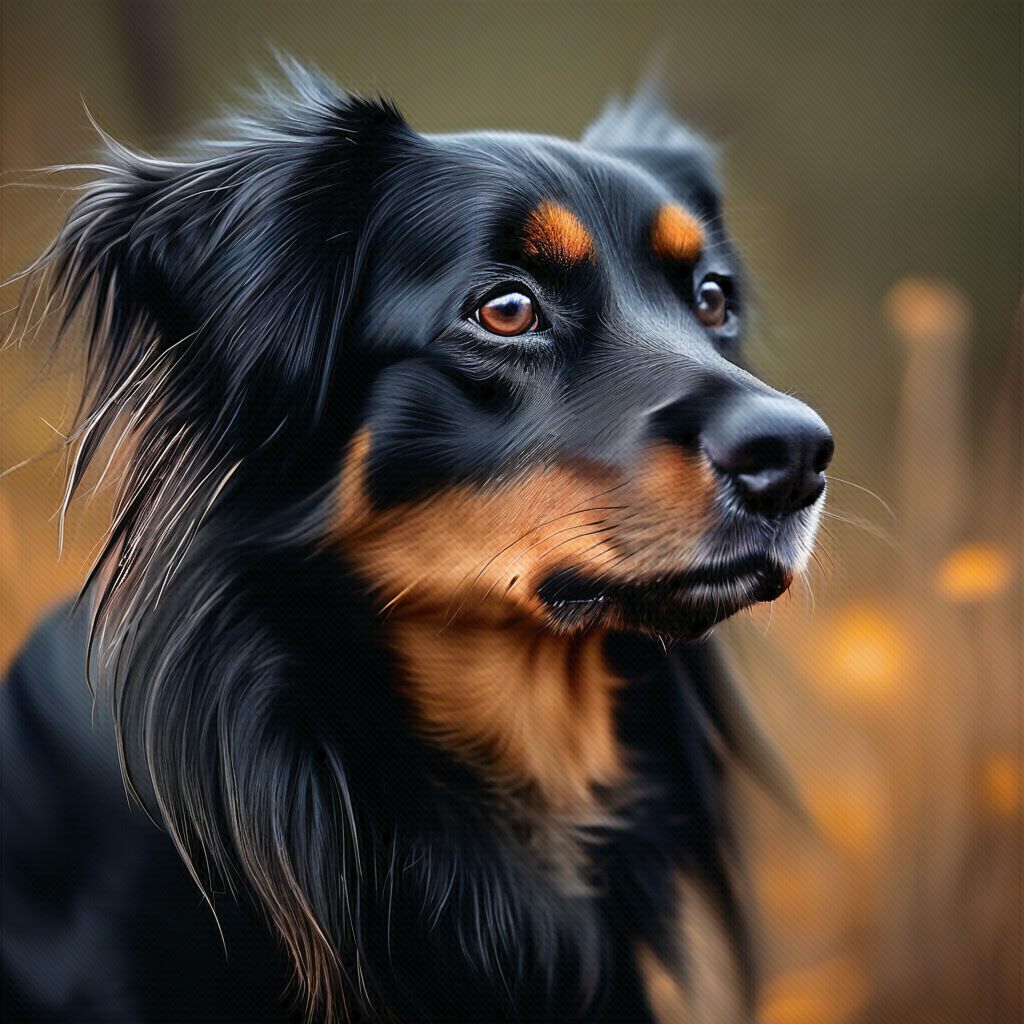}} &
\raisebox{-.5\height}{\includegraphics[width=\linewidth]{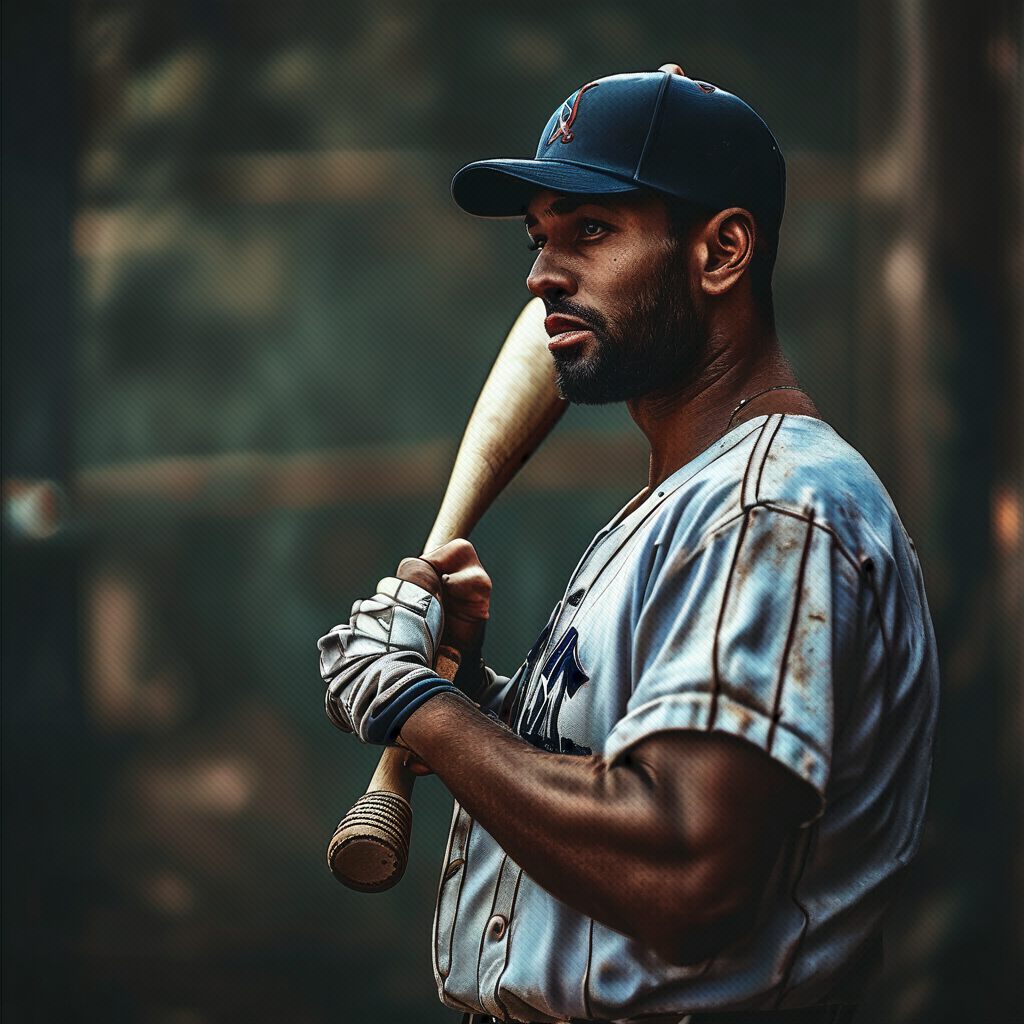}} &
\raisebox{-.5\height}{\includegraphics[width=\linewidth]{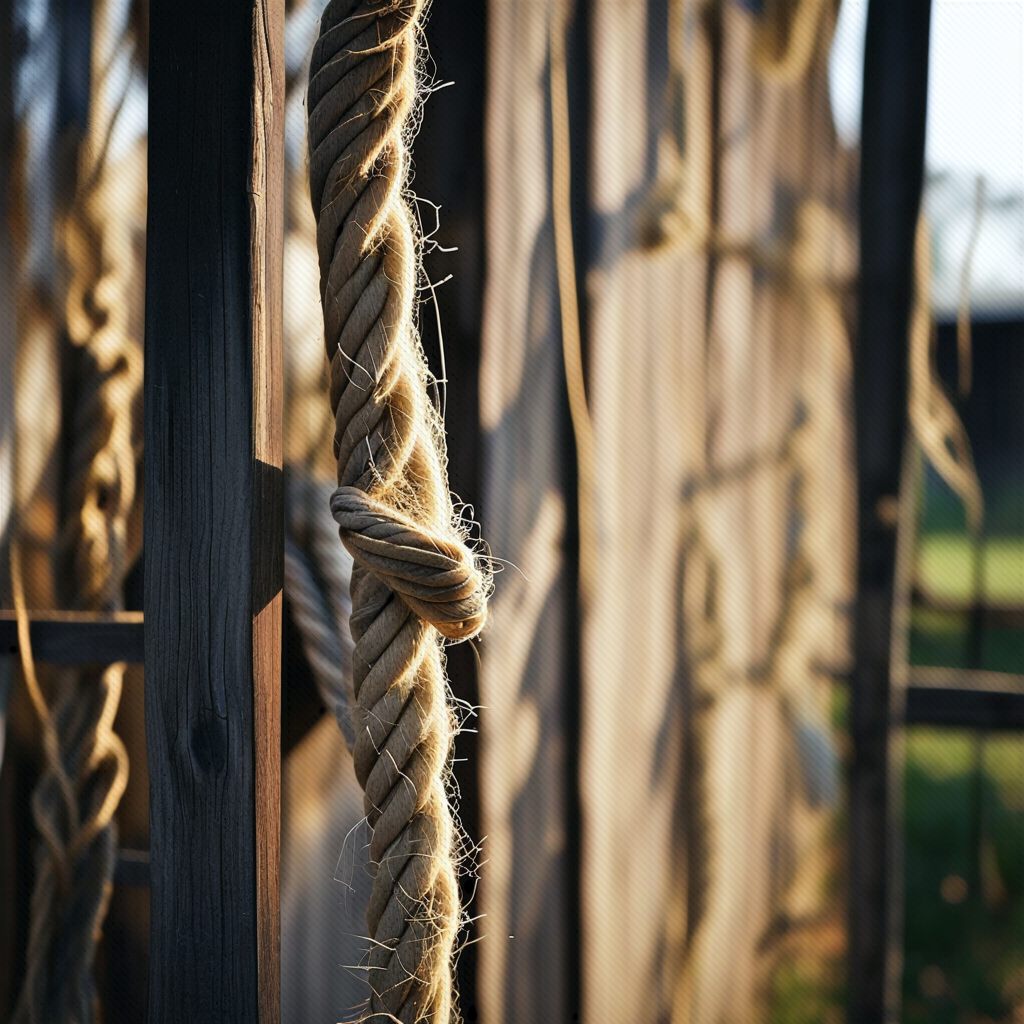}} &
\raisebox{-.5\height}{\includegraphics[width=\linewidth]{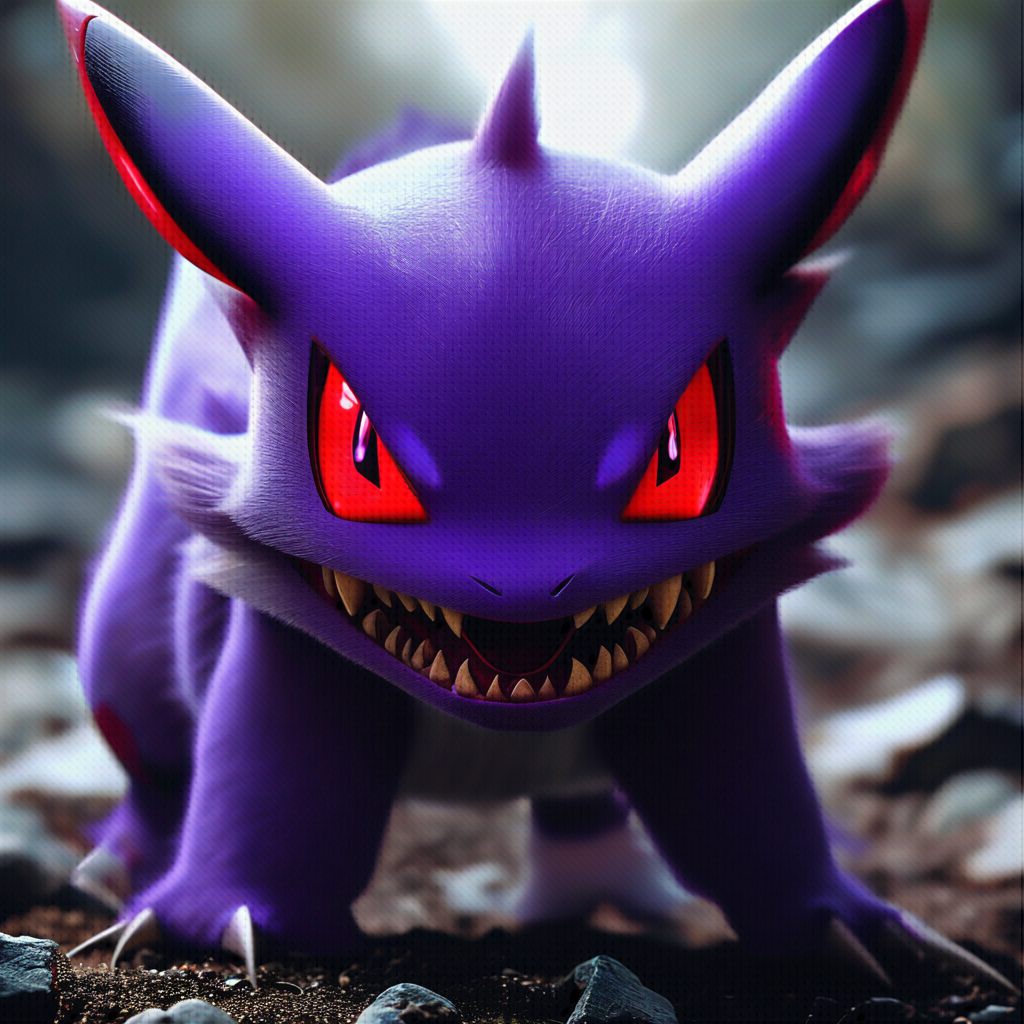}} \\
\bottomrule
\end{tabularx}
\caption{Benign visual fidelity examples across four datasets.}
\label{fig:visual_quality_examples}
\end{figure}

\subsection{Benign Visual Fidelity}
\label{sec:visual_quality}

Due to space limitations, the quantitative visual-quality metrics are provided in Appendix~\ref{Appendix:expresult}.
We evaluate \method{} across four models and four datasets. 
We first use the corresponding source model to generate 500 samples. The same latent representations are then reconstructed by both the clean decoder and the protected decoder, and the resulting outputs are used to compute the evaluation metrics.
Here, we present a visual comparison between protected and clean examples in this section.
As shown in Figure~\ref{fig:visual_quality_examples}, the composite visual constraints adopted by \method{} preserve visual quality effectively, introducing limited perceptible degradation across all four datasets.
This demonstrates the effectiveness of the composite visual constraints in preserving visual quality.

\begin{figure}[tbp]
    \centering
    \footnotesize
    \setlength{\tabcolsep}{1.4pt}
    \renewcommand{\arraystretch}{1.10}
    \begin{tabularx}{0.9\columnwidth}{@{}l*{3}{>{\centering\arraybackslash}X}@{}}
    \toprule
    \rowcolor{tablepanel}\multicolumn{4}{c}{\textbf{Cross-Dataset Generalization}} \\
    \midrule
    \rowcolor{tablehead}\textbf{Generation Set} & \textbf{Pokemon} & \textbf{Landscape} & \textbf{CelebA} \\
    \textbf{Example} & \raisebox{-.5\height}{\includegraphics[width=\linewidth]{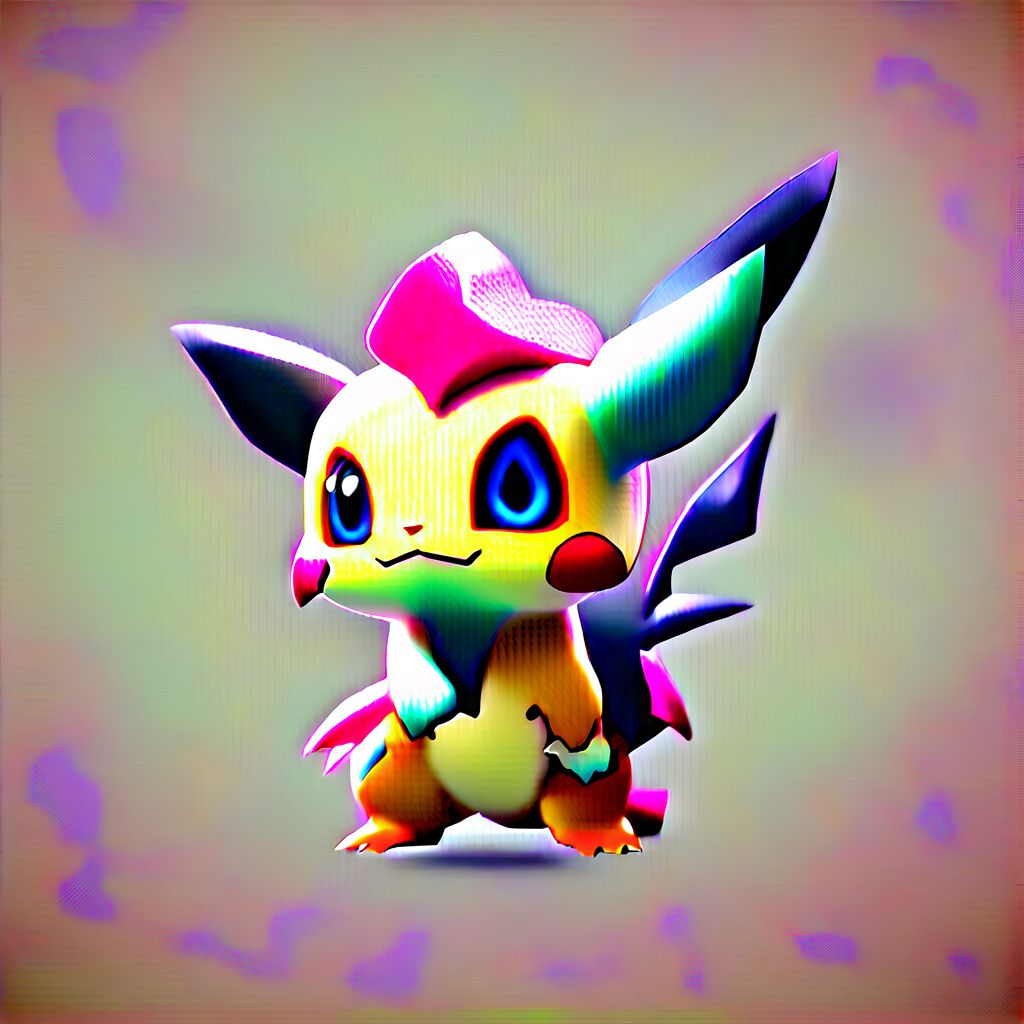}} &
    \raisebox{-.5\height}{\includegraphics[width=\linewidth]{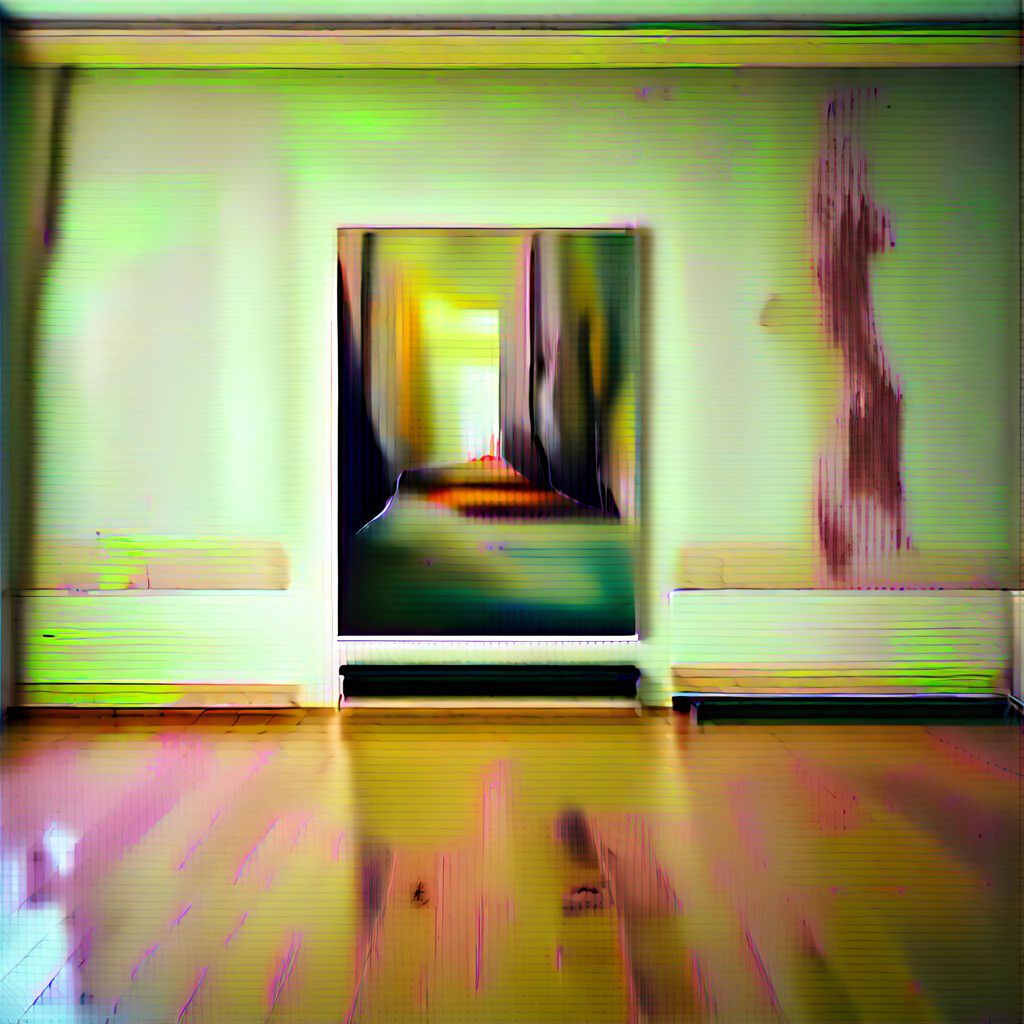}} &
    \raisebox{-.5\height}{\includegraphics[width=\linewidth]{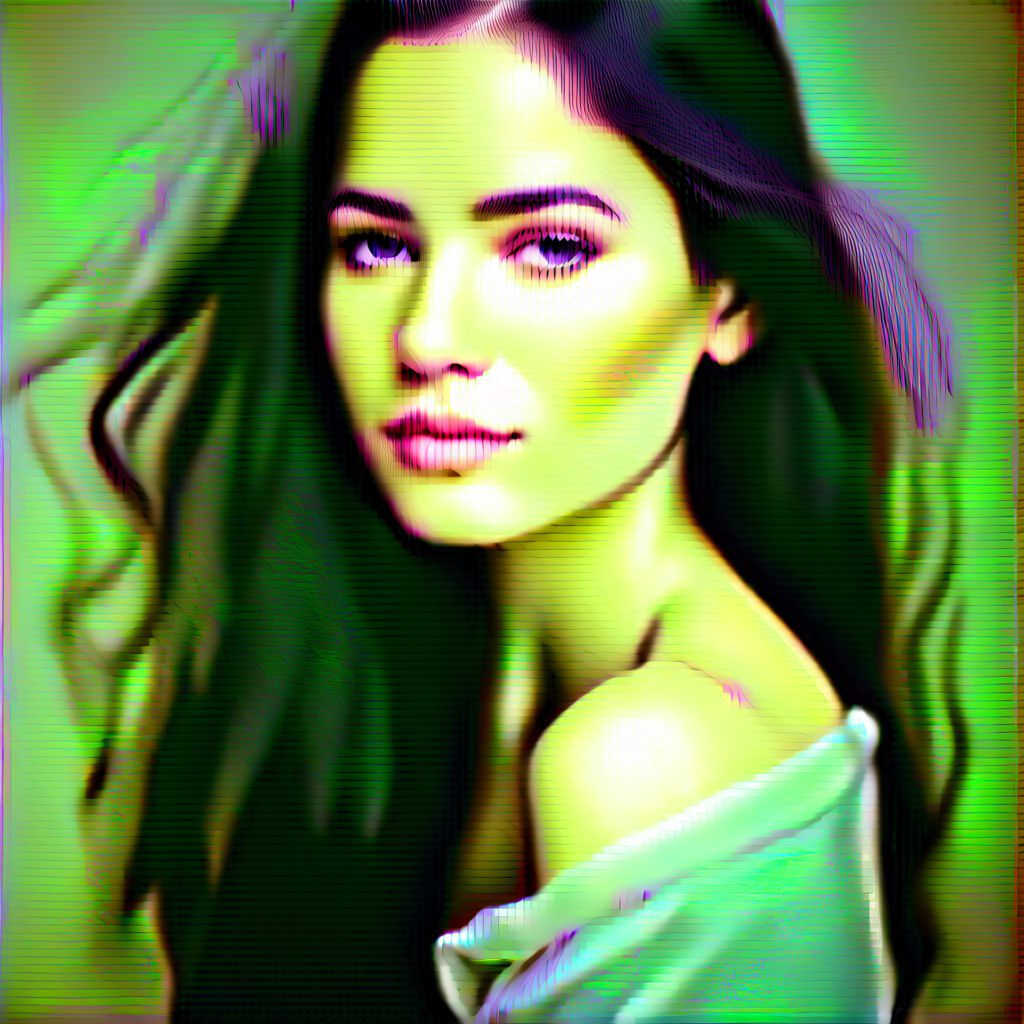}} \\
    \midrule
    \textbf{\mFID}$\uparrow$ & 91.54 & 132.61 & 103.68 \\
    \textbf{\mQA}$\downarrow$ & 2.02 & 1.78 & 1.75 \\
    \textbf{\mPK}$\downarrow$ & 21.57 & 20.43 & 19.17 \\
    \textbf{\mAES}$\downarrow$ & 5.14 & 4.98 & 5.29 \\
    \bottomrule
    \end{tabularx}
    \caption{Cross-dataset generalization when the decoder is trained on AFHQ and evaluated on Pokemon, Landscape, and CelebA generation sets.}
    \label{fig:ablation_dist_shift}
\end{figure}

\subsection{Cross-Dataset Generalization}

We evaluate cross-dataset generalization by fixing the protected decoder trained on AFHQ and collecting protected outputs from the Pokemon, Landscape, and CelebA generation sets. 
This setting reflects a realistic distillation scenario, where the attacker's query dataset is unlikely to be limited to the in-domain training set and may instead come from out-of-domain data sources. Therefore, a practical defense should retain effectiveness beyond the decoder-training distribution.
The results in Figure~\ref{fig:ablation_dist_shift} show that the protected decoder remains disruptive when outputs are collected from generation sets beyond the decoder-training distribution.
Specifically, QAlign remains low across Pokemon, Landscape, and CelebA, at 2.02, 1.78, and 1.75, respectively.
The images generated by the substitute also exhibit a substantial degradation in quality. 
We observe evident color shifts across all three datasets, together with pronounced artifacts around the foreground entities.
We speculate that this cross-dataset generalization arises because \method{} avoids external dependencies and instead encourages the shared decoder update to capture the general patterns of perturbations that are more generalizable across data distributions.
Such a commonality-oriented optimization objective encourages the model to identify a generalizable perturbation, rather than over-relying on external dependency.
Overall, \method{} realizes a generalizable perturbation that can remain effective against out-of-distribution downstream distillation data.\looseness=-1

\begin{figure}[tbp]
\centering
\scriptsize
\setlength{\tabcolsep}{2.2pt}
\renewcommand{\arraystretch}{1.10}
\newcommand{\pairhead}{\begin{tabular}{@{}>{\centering\arraybackslash}p{.49\linewidth}@{\hspace{1pt}}>{\centering\arraybackslash}p{.49\linewidth}@{}}\textbf{Clean} & \textbf{\method{}}\end{tabular}}
\newcommand{\pairimg}[2]{\begin{tabular}{@{}>{\centering\arraybackslash}p{.49\linewidth}@{\hspace{1pt}}>{\centering\arraybackslash}p{.49\linewidth}@{}}\includegraphics[width=\linewidth]{#1} & \includegraphics[width=\linewidth]{#2}\end{tabular}}
\newcommand{\pairnum}[2]{\begin{tabular}{@{}>{\centering\arraybackslash}p{.49\linewidth}@{\hspace{1pt}}>{\centering\arraybackslash}p{.49\linewidth}@{}}#1 & #2\end{tabular}}
\begin{tabularx}{\columnwidth}{@{}l*{3}{>{\centering\arraybackslash}X}@{}}
\toprule
\rowcolor{tablepanel}\multicolumn{4}{c}{\textbf{Different Training Methods}} \\
\midrule
\rowcolor{tablehead}\textbf{Metric} & \textbf{Full Parameters} & \textbf{Custom Diffusion} & \textbf{SVDiff} \\
\rowcolor{tablehead} & \pairhead & \pairhead & \pairhead \\
\textbf{Example} &
\pairimg{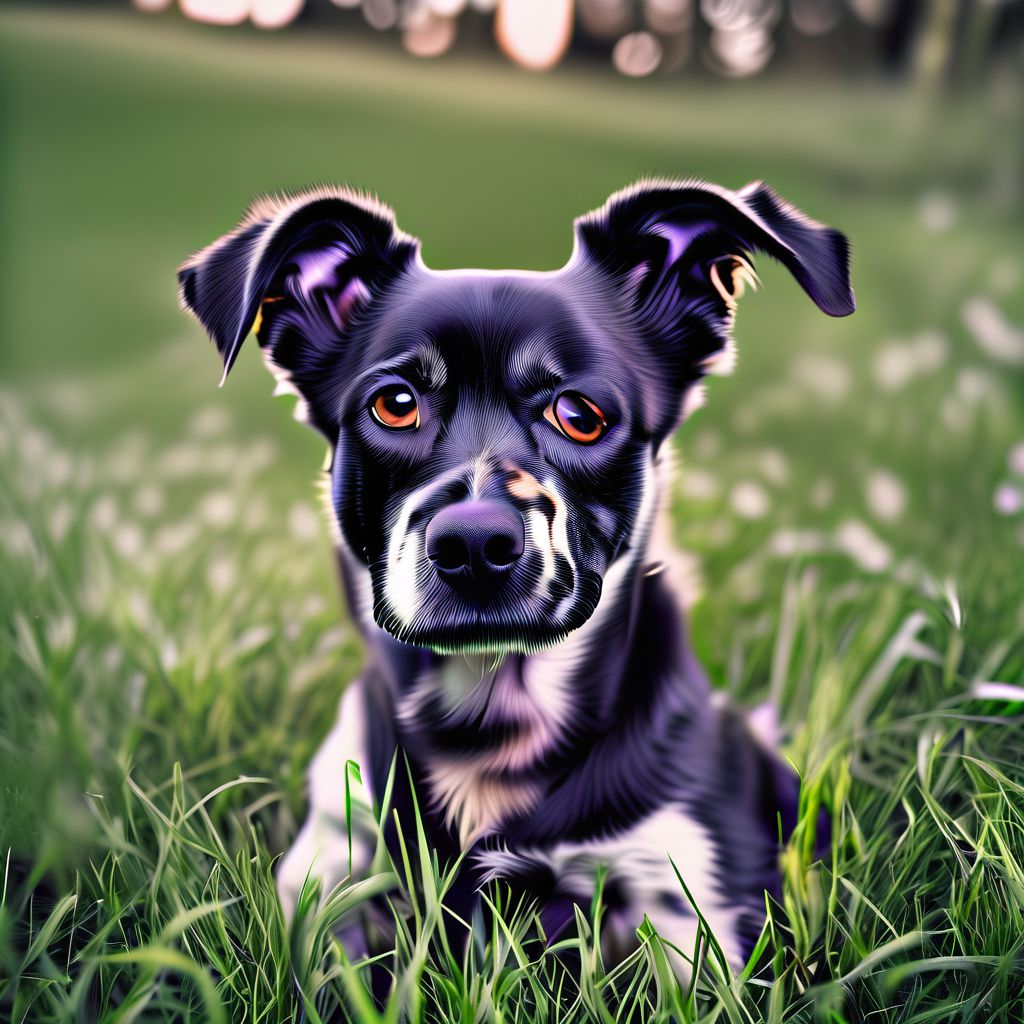}{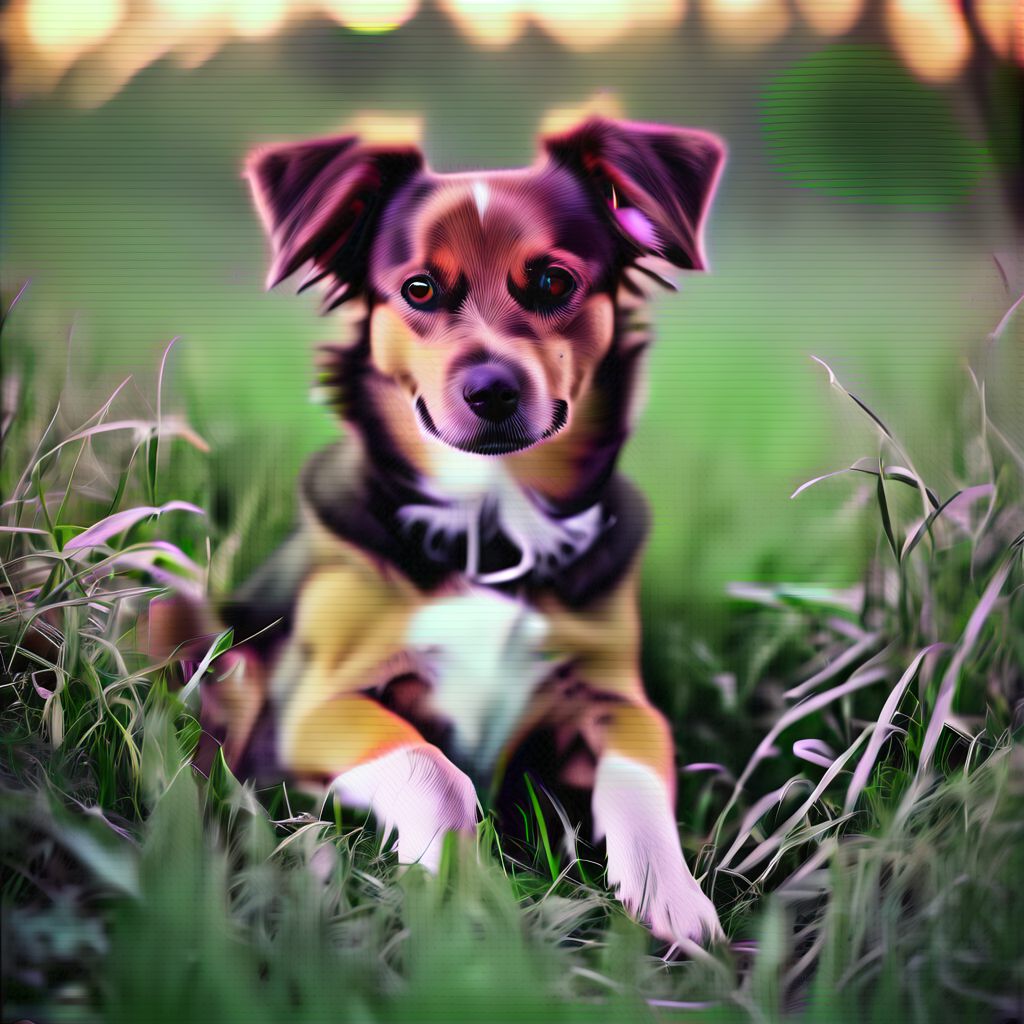} &
\pairimg{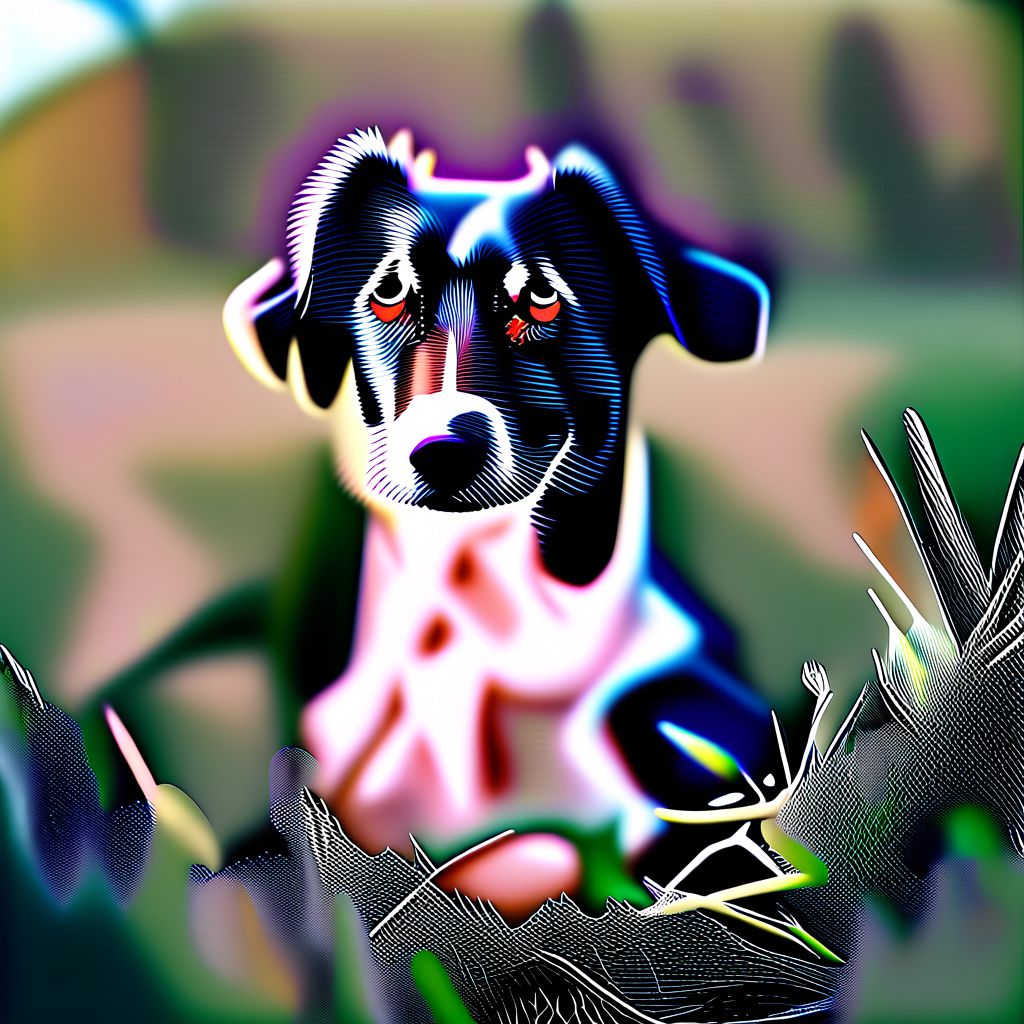}{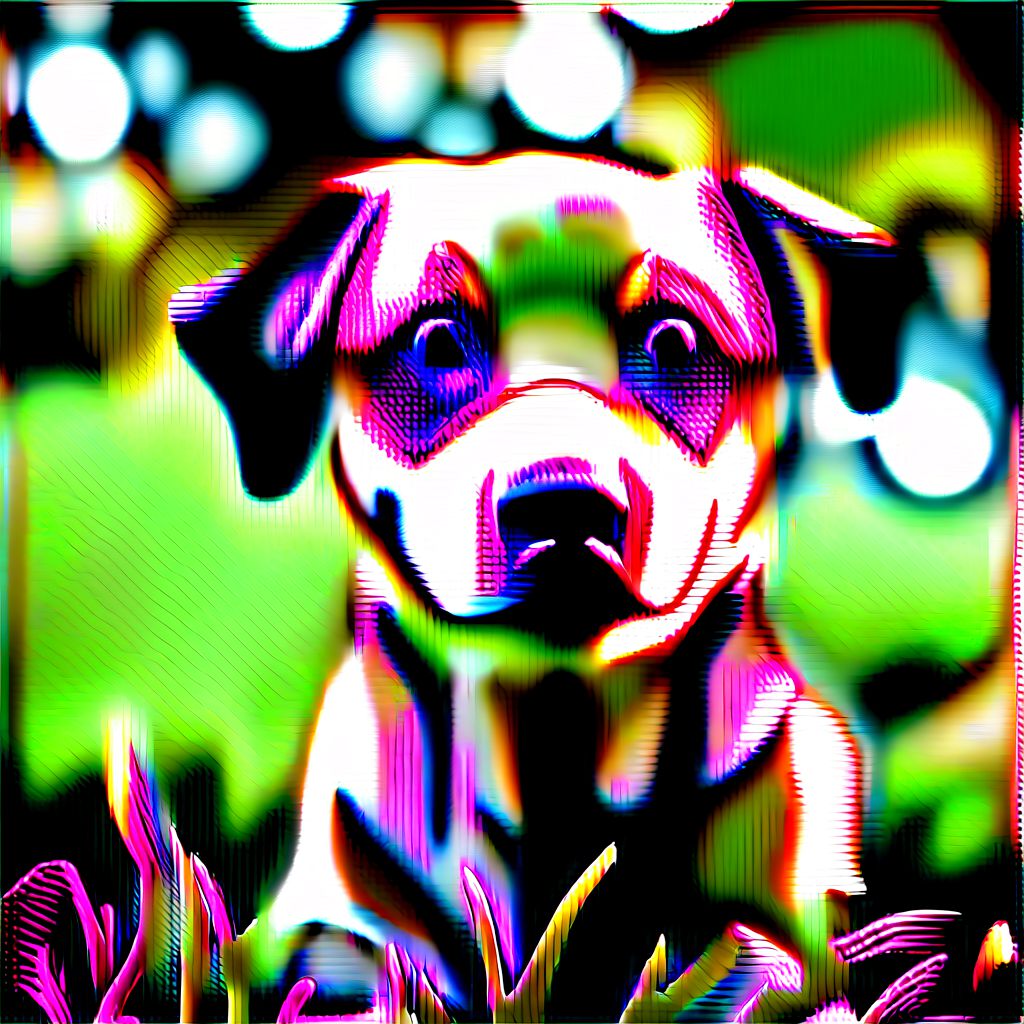} &
\pairimg{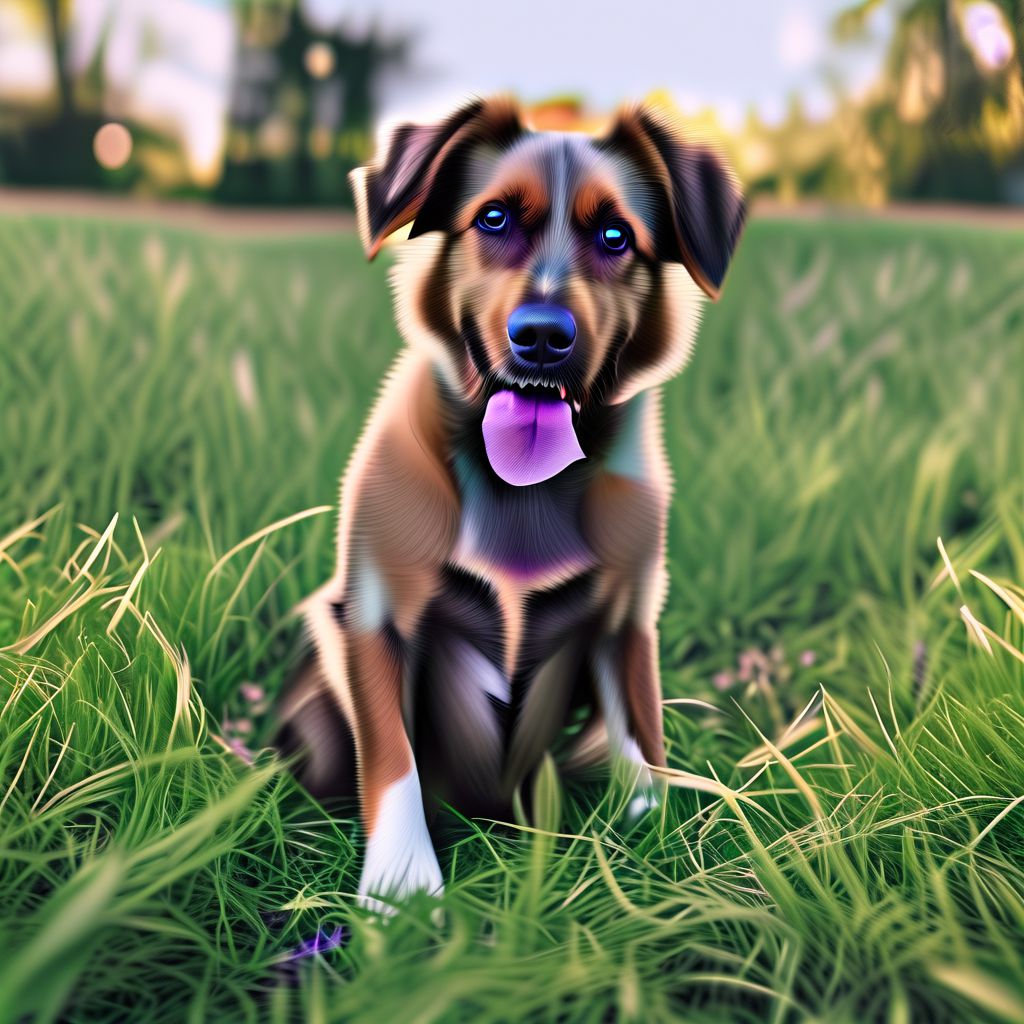}{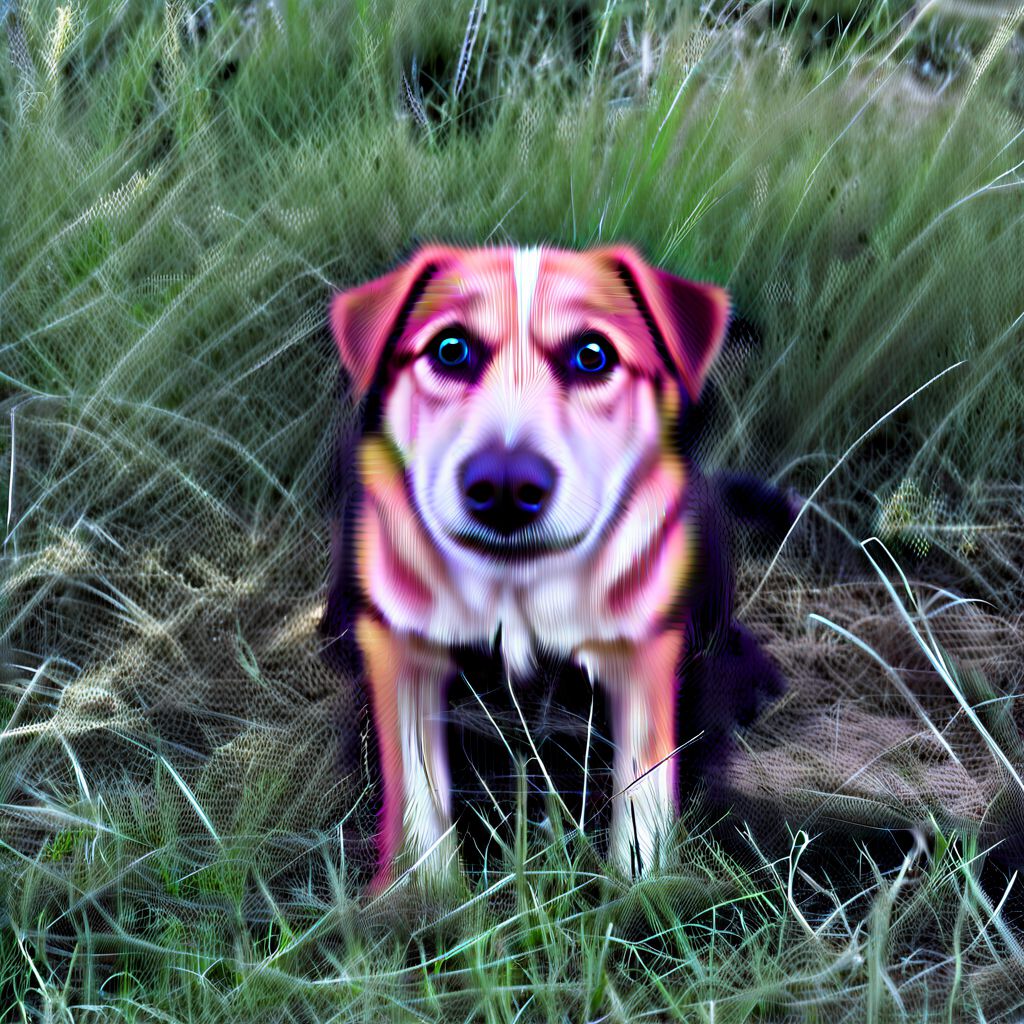} \\
\midrule
\textbf{\mFID}$\uparrow$ & \pairnum{56.00}{82.03} & \pairnum{178.86}{252.51} & \pairnum{57.10}{53.12} \\
\textbf{\mQA}$\downarrow$ & \pairnum{3.30}{2.58} & \pairnum{1.76}{1.14} & \pairnum{3.32}{2.59} \\
\textbf{\mPK}$\downarrow$ & \pairnum{21.05}{20.67} & \pairnum{19.44}{18.92} & \pairnum{20.94}{20.79} \\
\textbf{\mAES}$\downarrow$ & \pairnum{6.03}{5.80} & \pairnum{5.45}{4.90} & \pairnum{5.98}{5.66} \\
\bottomrule
\end{tabularx}
\caption{Effectiveness across three training methods.}
\label{fig:ablation_training_method}
\end{figure}

\subsection{Different Training Methods}

We evaluate three training methods used in model distillation: full parameter finetuning, Custom Diffusion~\cite{Customization}, and SVDiff~\cite{han2023svdiff}.
The results are shown in Figure~\ref{fig:ablation_training_method}.
Specifically, the selected Custom setting reduces QAlign from its clean reference value of 1.76 to 1.14, while also decreasing PickScore and Aesthetic from 19.44 and 5.45 to 18.92 and 4.90, respectively. Full finetuning remains vulnerable to distillation disruption: QAlign, PickScore, and Aesthetic decrease from 3.30, 21.05, and 6.03 to 2.58, 20.67, and 5.80, respectively, while FID increases from 56.00 to 82.03. SVDiff likewise reduces QAlign from 3.32 to 2.59.
The generated samples also show a clear decrease in image quality: the generated data exhibit noticeable artifacts and evident color discrepancies.
We attribute this to the fact that these training methods all rely on latent-space representations; therefore, perturbations introduced in the latent space can effectively generalize across different training methods. 
Overall, \method{} remains effective under all three common training methods, demonstrating its generalized disruptive capability. \looseness=-1

\subsection{Robustness}

We evaluate robustness against two black-box attacker variants: changing the collected data mixture and preprocessing protected images before substitute training.

\begin{figure}[tbp]
    \centering
    \footnotesize
    \setlength{\tabcolsep}{1.4pt}
    \renewcommand{\arraystretch}{1.10}
    \begin{tabularx}{\columnwidth}{@{}l*{5}{>{\centering\arraybackslash}X}@{}}
    \toprule
    \rowcolor{tablepanel}\multicolumn{6}{c}{\textbf{Robustness to Protected Data Ratio}} \\
    \midrule
    \rowcolor{tablehead}\textbf{Ratio} & \textbf{100\%} & \textbf{80\%} & \textbf{60\%} & \textbf{40\%} & \textbf{20\%} \\
    \textbf{Example} & \raisebox{-.5\height}{\includegraphics[width=\linewidth]{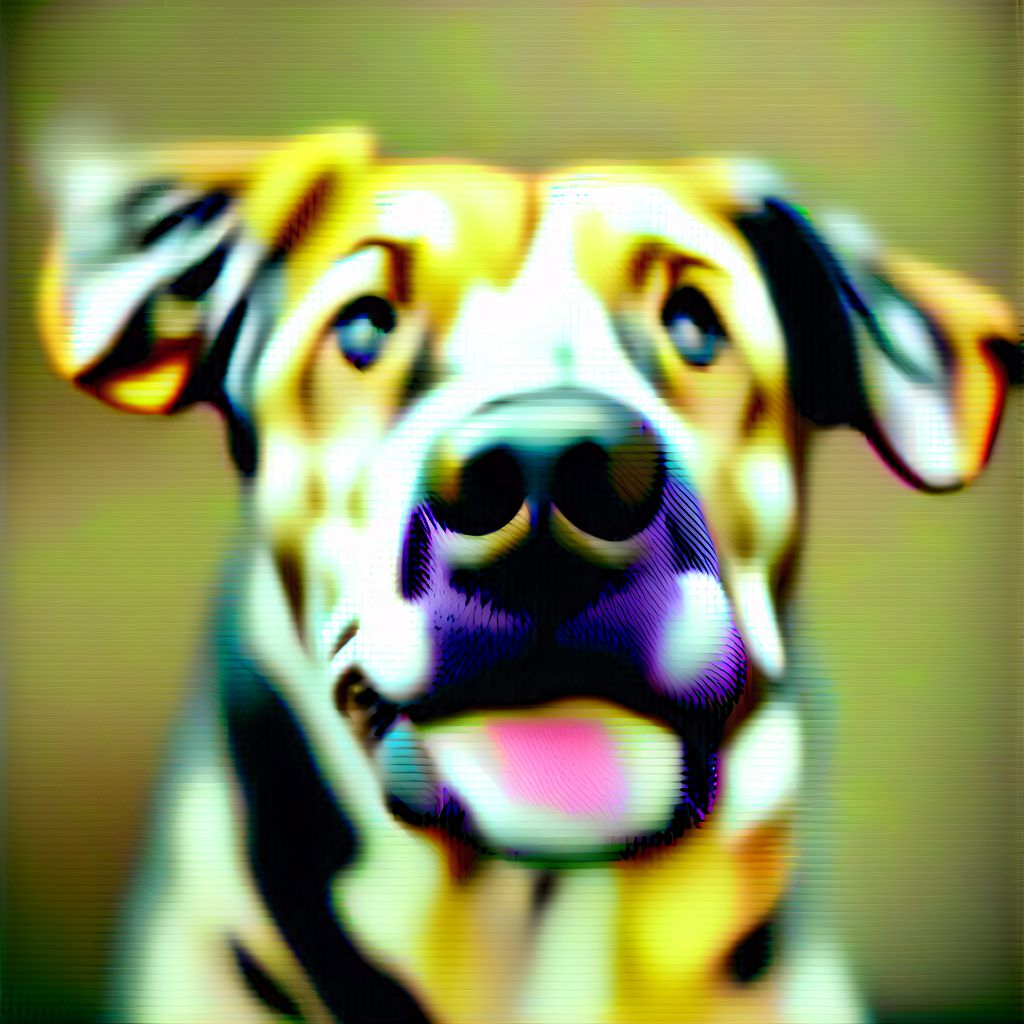}} &
    \raisebox{-.5\height}{\includegraphics[width=\linewidth]{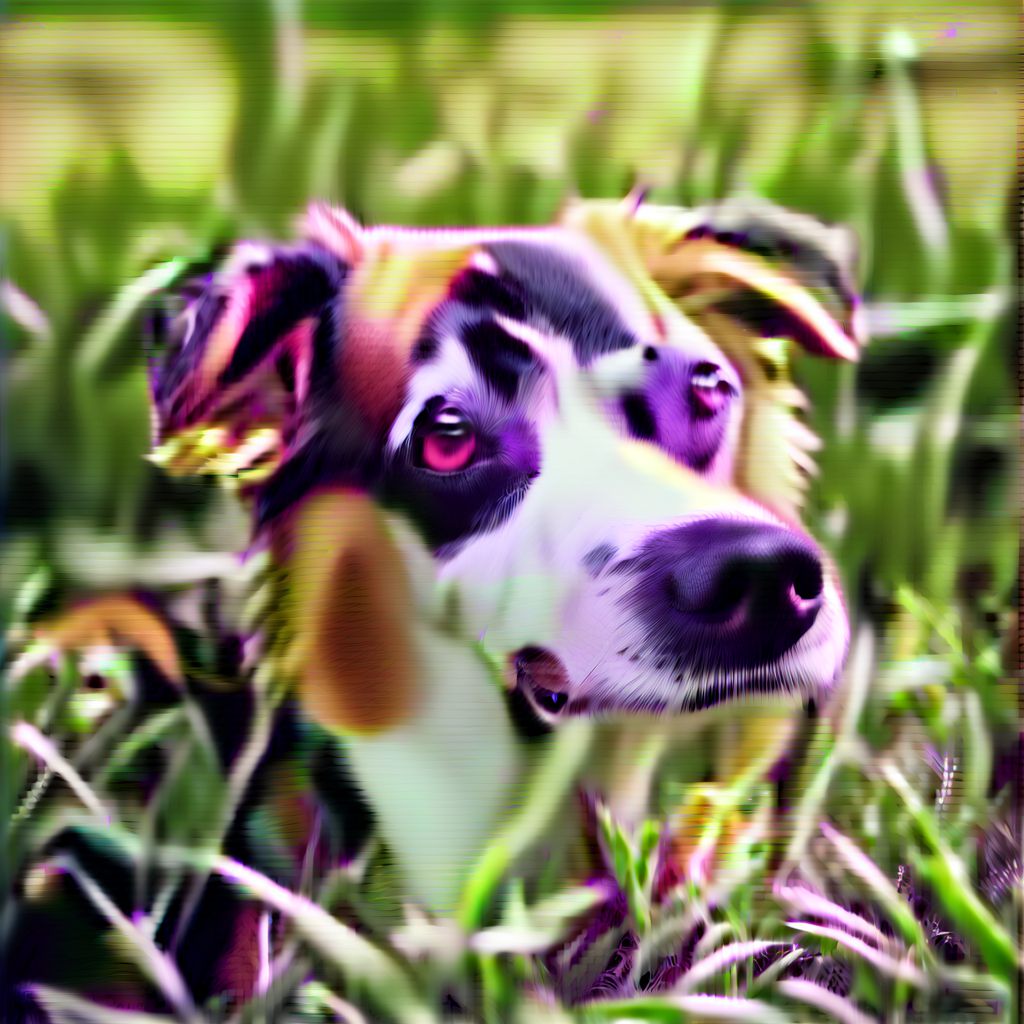}} &
    \raisebox{-.5\height}{\includegraphics[width=\linewidth]{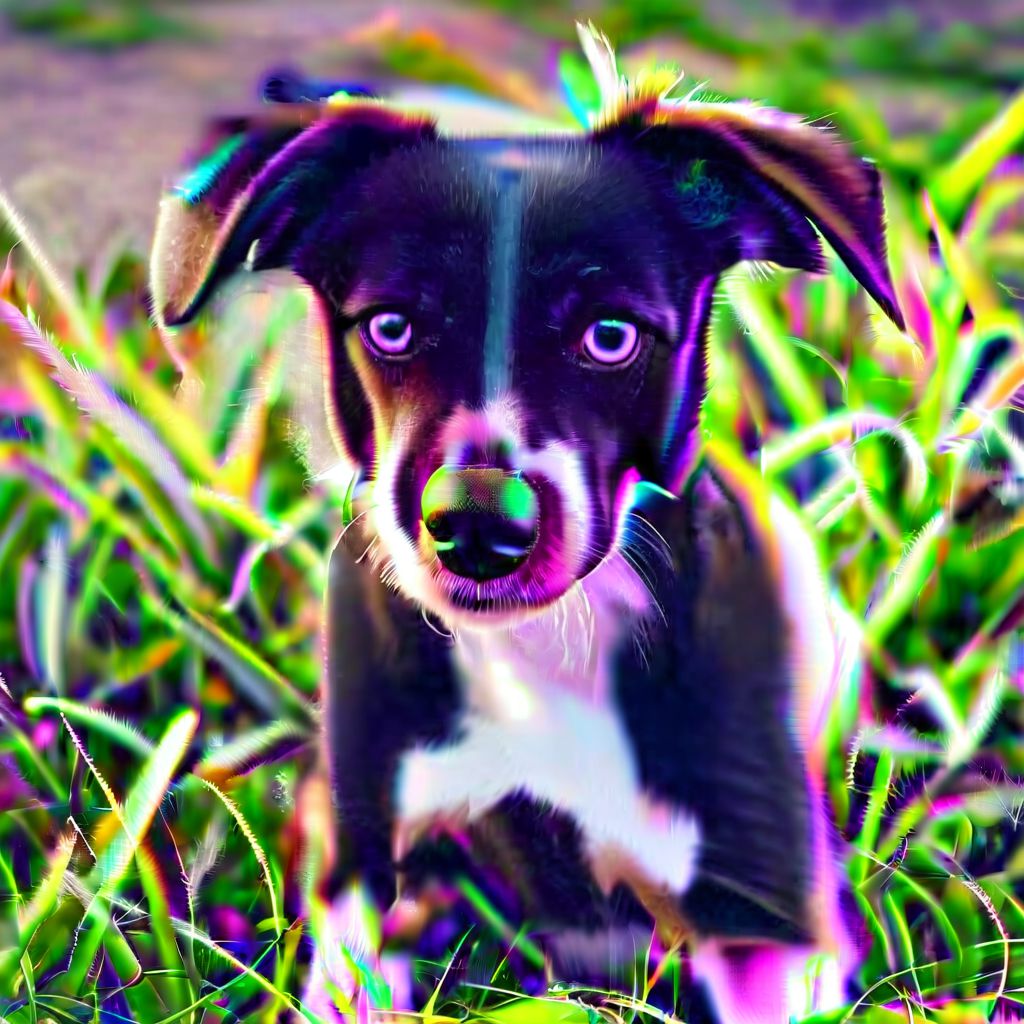}} &
    \raisebox{-.5\height}{\includegraphics[width=\linewidth]{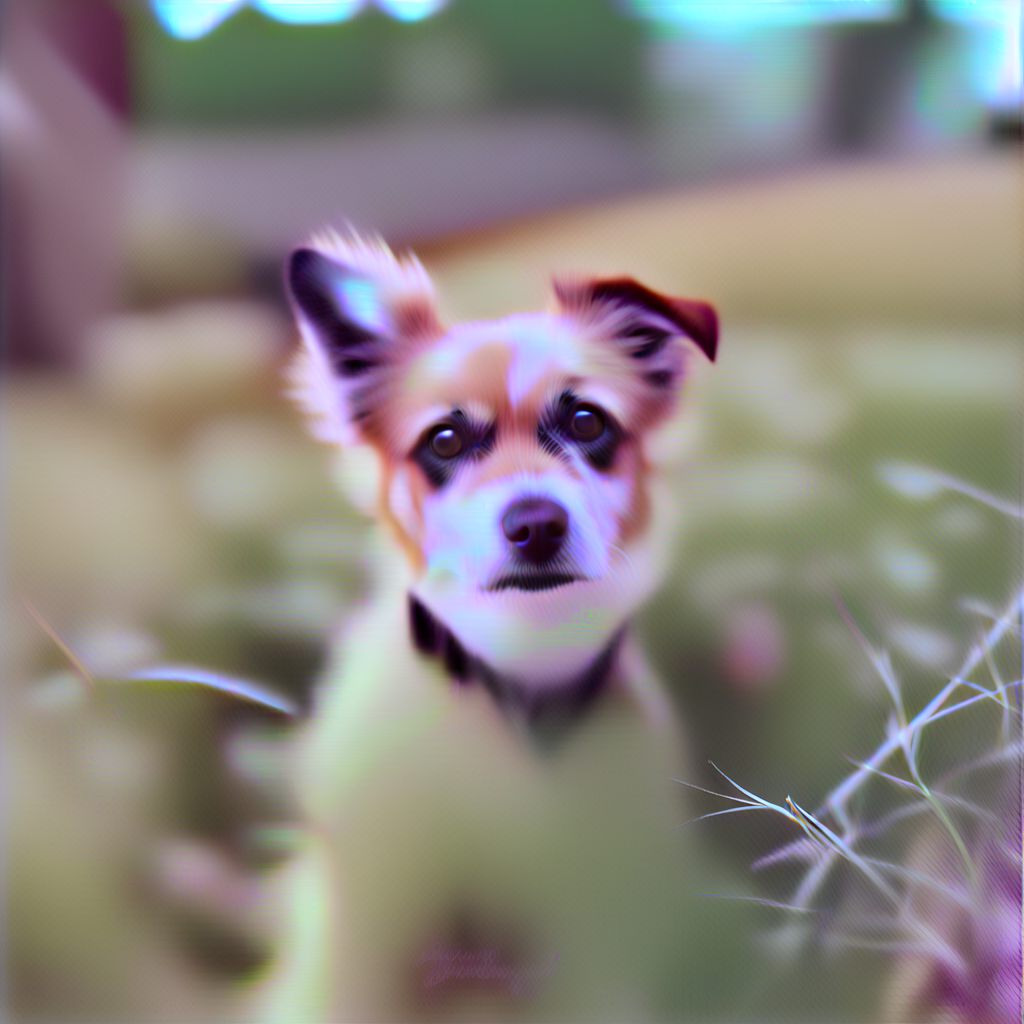}} &
    \raisebox{-.5\height}{\includegraphics[width=\linewidth]{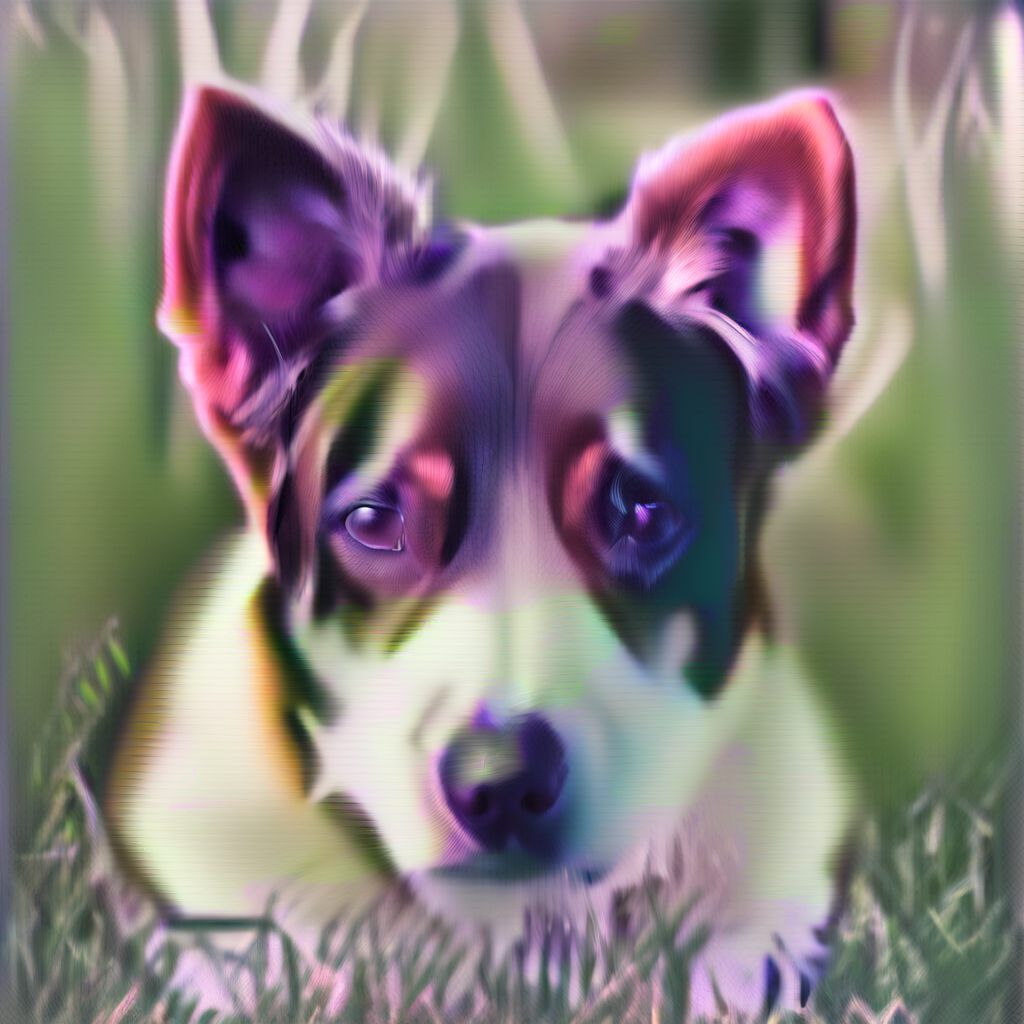}} \\
    \midrule
    \textbf{\mFID}$\uparrow$ & 132.62 & 93.03 & 105.24 & 94.75 & 87.31 \\
    \textbf{\mQA}$\downarrow$ & 1.59 & 2.23 & 2.16 & 2.47 & 2.54 \\
    \textbf{\mPK}$\downarrow$ & 19.75 & 20.41 & 20.24 & 20.47 & 20.57 \\
    \textbf{\mAES}$\downarrow$ & 5.25 & 5.48 & 5.57 & 5.81 & 5.83 \\
    \bottomrule
    \end{tabularx}
    \caption{Robustness to the protected-data ratio.}
    \label{fig:ablation_mixed_data}
\end{figure}

\mypara{Different Protected-Output Ratios.}
We consider a realistic mixed-data scenario in which the attacker's distillation dataset may not be obtained solely from the protected model, but may also contain clean training data from other sources.
We therefore evaluate the robustness of \method{} against mixed-data attackers by varying the protected-output ratio in the collected distillation dataset from 20\% to 100\%.
The results are shown in Figure~\ref{fig:ablation_mixed_data}.
Overall, as the proportion of protected data increases, the generation quality of the substitute model exhibits a decreasing trend.
Specifically, the 100\% protected-output setting gives the lowest QAlign, 1.59, whereas the weakest 20\% setting still keeps QAlign at 2.54, which is also a severe decrease compared with the Clean setting.
The results show that \method{} can effectively degrade generation quality even at a low protected-output ratio, demonstrating its robustness. \looseness=-1

\begin{figure}[tbp]
    \centering
    \footnotesize
    \setlength{\tabcolsep}{1.4pt}
    \renewcommand{\arraystretch}{1.10}
    \begin{tabularx}{\columnwidth}{@{}l*{4}{>{\centering\arraybackslash}X}@{}}
    \toprule
    \rowcolor{tablepanel}\multicolumn{5}{c}{\textbf{Robustness to Image Transformation}} \\
    \midrule
    \rowcolor{tablehead}\textbf{Transform} & \textbf{Clean} & \textbf{Center Crop} & \textbf{Gaussian} & \textbf{JPEG} \\
    \textbf{Example} & \raisebox{-.5\height}{\includegraphics[width=\linewidth]{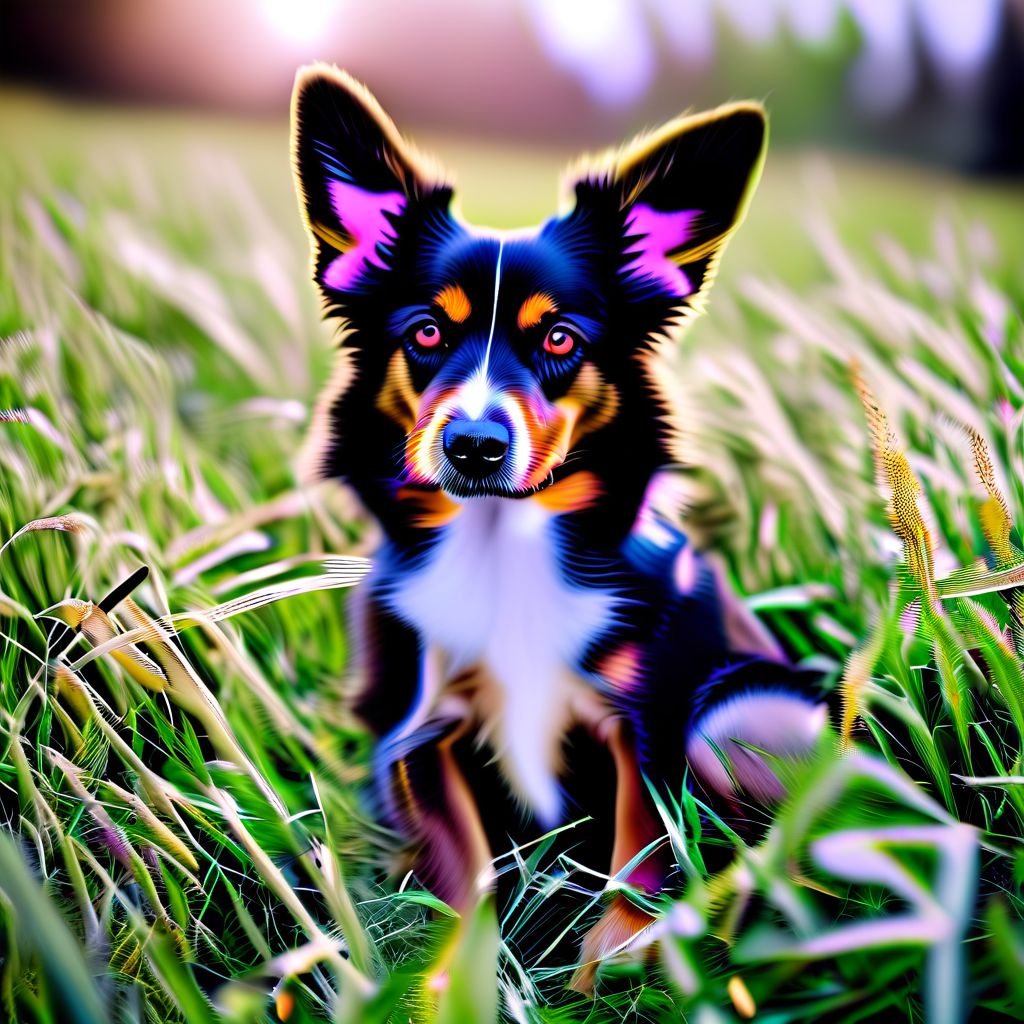}} &
    \raisebox{-.5\height}{\includegraphics[width=\linewidth]{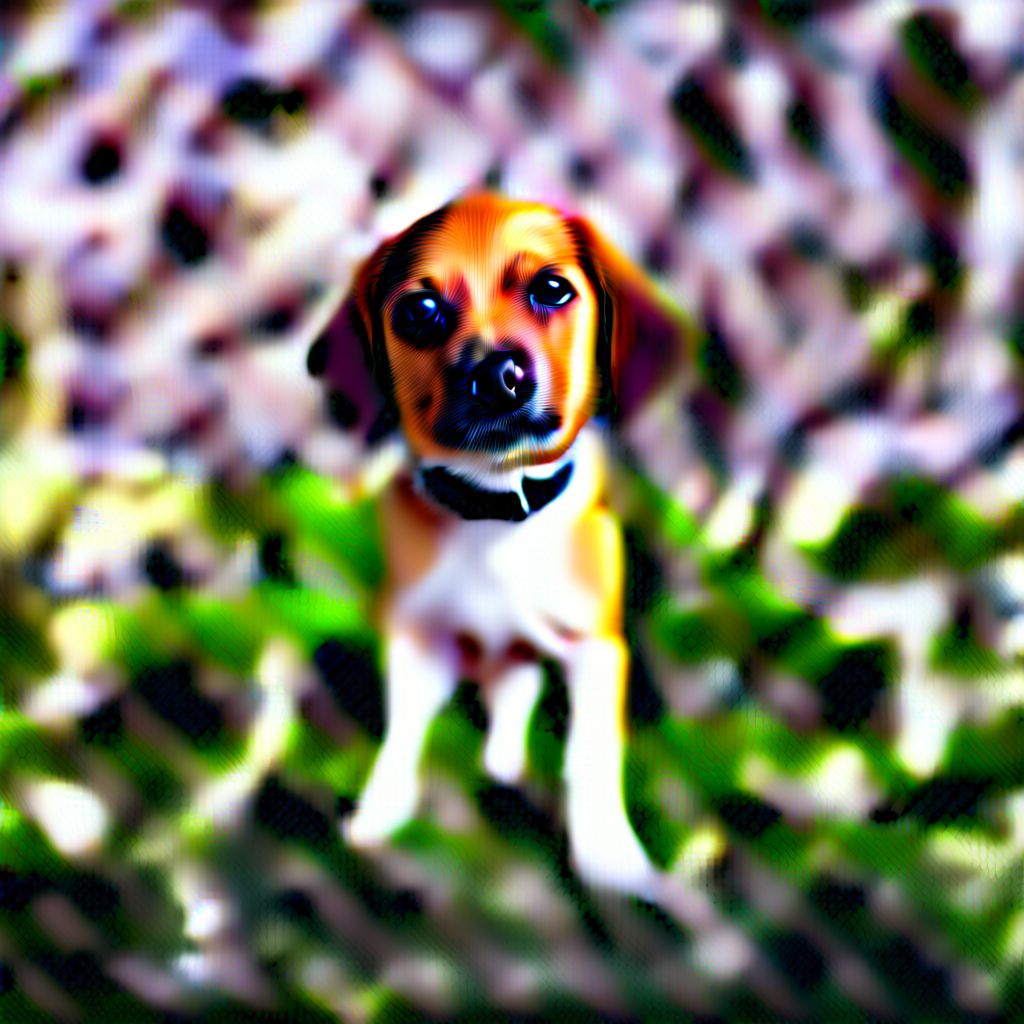}} &
    \raisebox{-.5\height}{\includegraphics[width=\linewidth]{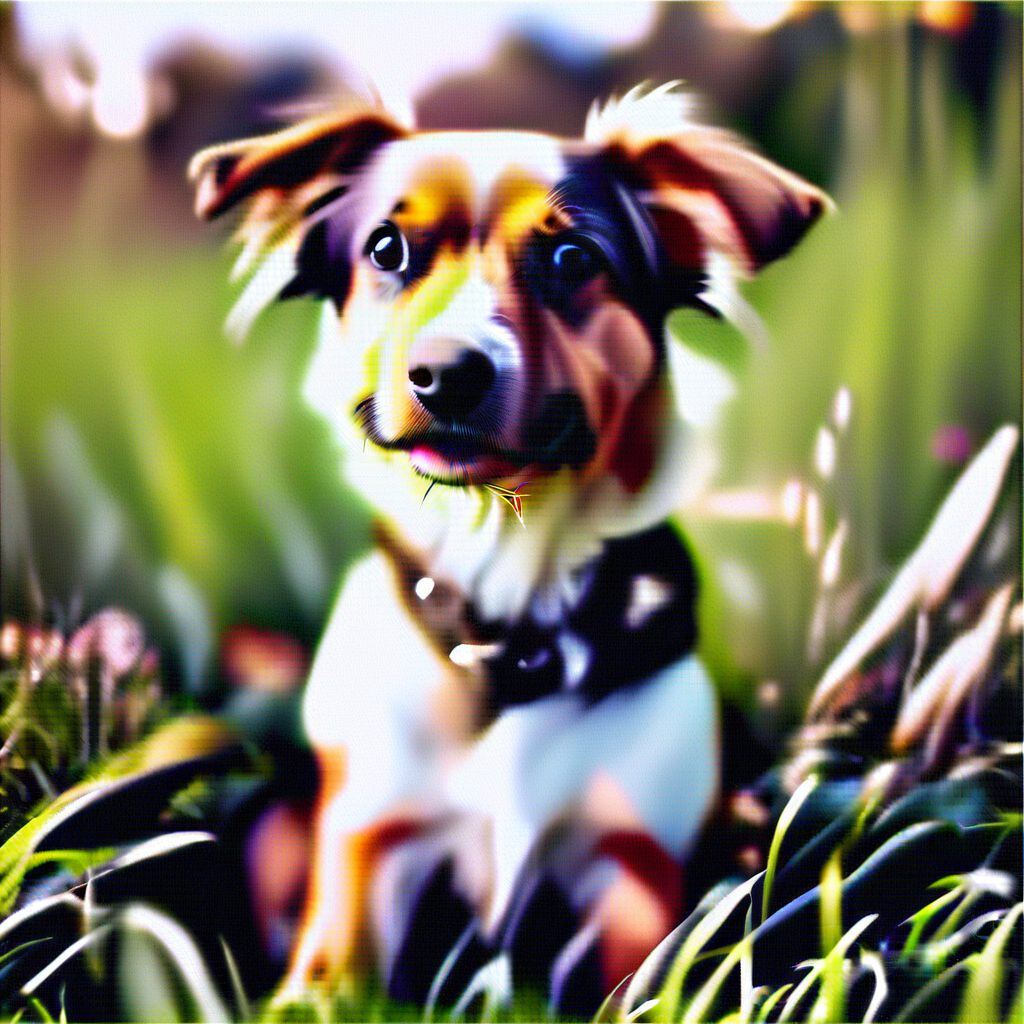}} &
    \raisebox{-.5\height}{\includegraphics[width=\linewidth]{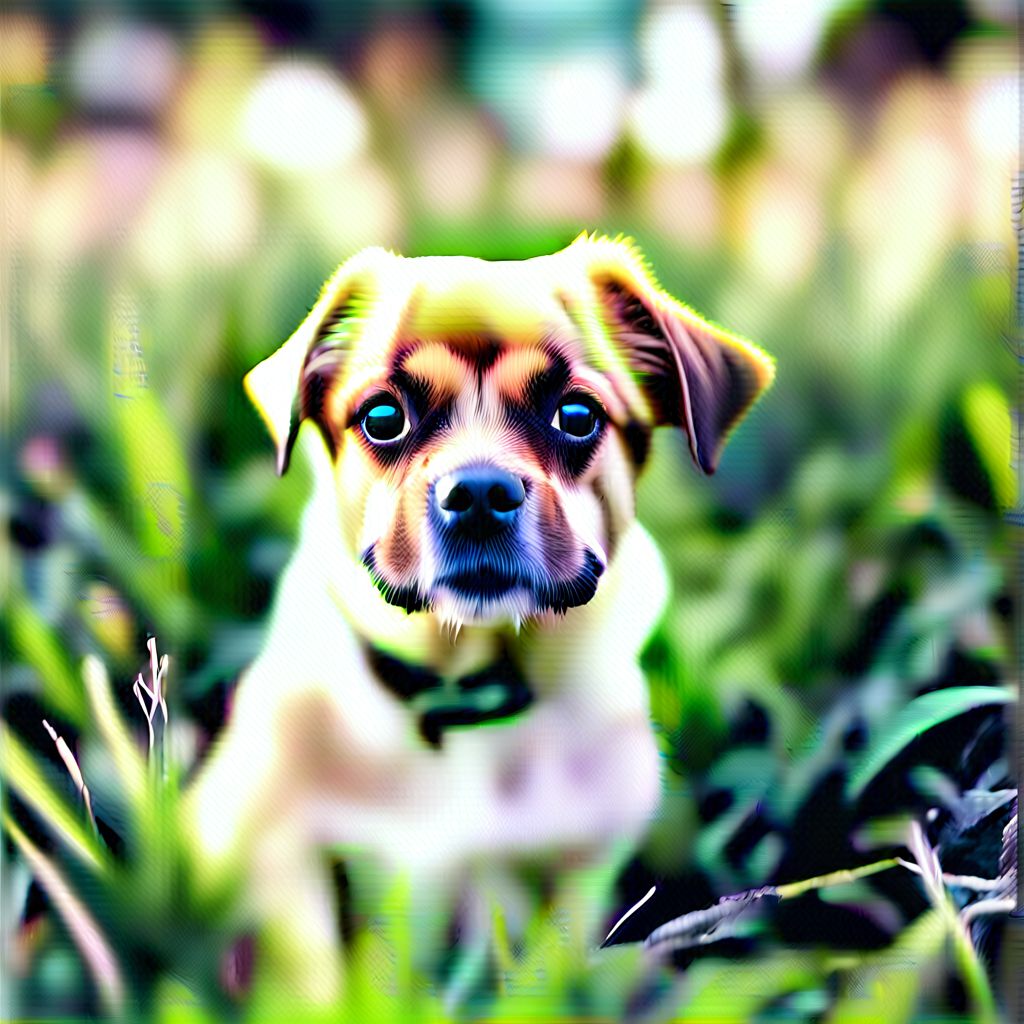}} \\
    \midrule
    \textbf{\mFID}$\uparrow$ & 60.13 & 115.10 & 108.14 & 84.88 \\
    \textbf{\mQA}$\downarrow$ & 3.10 & 1.87 & 2.74 & 2.59 \\
    \textbf{\mPK}$\downarrow$ & 20.80 & 19.74 & 20.78 & 20.68 \\
    \textbf{\mAES}$\downarrow$ & 5.97 & 5.34 & 5.92 & 5.76 \\
    \bottomrule
    \end{tabularx}
    \caption{Robustness against three common image transformations.}
    \label{fig:ablation_pretransform}
\end{figure}

\mypara{Different Preprocessing Schemes.}
We consider a realistic scenario in which, after obtaining the distillation data, the attacker may apply certain preprocessing transformations to the data.
We evaluate three attacker-side preprocessing schemes: Center crop (scaling factor = 0.9), Gaussian noise ($\sigma$ = 0.1), and JPEG compression (factor = 75)~\cite{liu2024metacloak,guo2018countering,xie2018mitigating};
Figure~\ref{fig:ablation_pretransform} shows the results.
Specifically, every preprocessing scheme keeps QAlign below the Clean value of 3.10: center crop reaches 1.87, Gaussian noise 2.74, and JPEG compression 2.59.
Thus, although common preprocessing can partially reduce the disruptive effect of \method{} on distillation training, \method{} still effectively degrades the generation quality of the distilled model. 
This demonstrates the robustness of \method{} against data transformations.\looseness=-1

\subsection{Ablation Study}

Finally, we investigate how surrogate models and objective variants influence the performance of \method{}. More ablation studies are presented in Appendix~\ref{Appendix:expresult}.

\begin{figure}[tbp]
    \centering
    \scriptsize
    \setlength{\tabcolsep}{1.2pt}
    \renewcommand{\arraystretch}{1.08}
    \newcommand{\spairhead}{\begin{tabular}{@{}>{\centering\arraybackslash}p{.49\linewidth}@{\hspace{1pt}}>{\centering\arraybackslash}p{.49\linewidth}@{}}\textbf{Clean} & \textbf{\method{}}\end{tabular}}
    \newcommand{\spairimg}[2]{\begin{tabular}{@{}>{\centering\arraybackslash}p{.49\linewidth}@{\hspace{1pt}}>{\centering\arraybackslash}p{.49\linewidth}@{}}\includegraphics[width=\linewidth]{#1} & \includegraphics[width=\linewidth]{#2}\end{tabular}}
    \newcommand{\spairnum}[2]{\begin{tabular}{@{}>{\centering\arraybackslash}p{.49\linewidth}@{\hspace{1pt}}>{\centering\arraybackslash}p{.49\linewidth}@{}}#1 & #2\end{tabular}}
    \begin{tabularx}{\columnwidth}{@{}>{\raggedright\arraybackslash}p{.15\columnwidth}*{2}{>{\centering\arraybackslash}X}@{}}
    \toprule
    \rowcolor{tablepanel}\multicolumn{3}{c}{\textbf{Ablation Study of Surrogate Model}} \\
    \midrule
    \rowcolor{tablehead}\makecell[l]{\textbf{}\\\textbf{Model}} & \textbf{PixArt} & \textbf{SDXL} \\
    \rowcolor{tablehead} & \spairhead & \spairhead \\
    \textbf{Example} &
    \spairimg{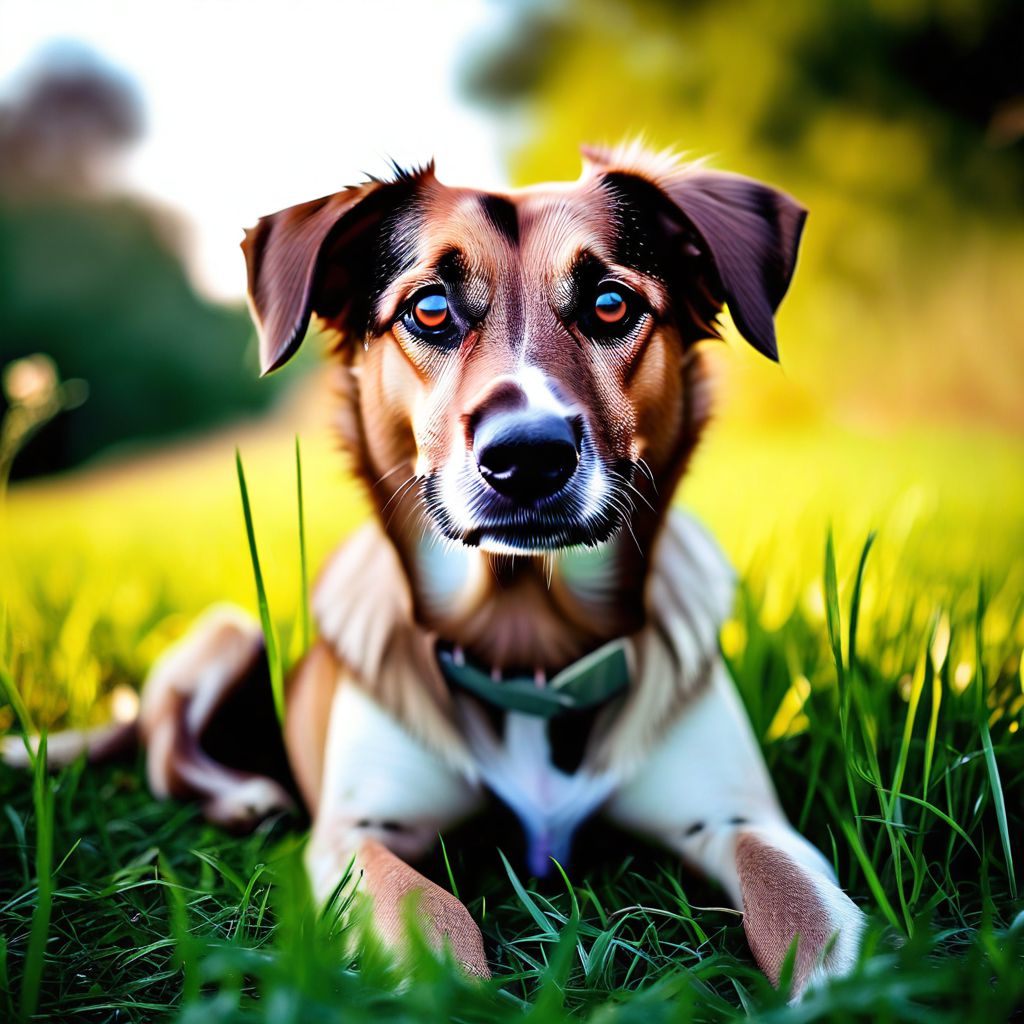}{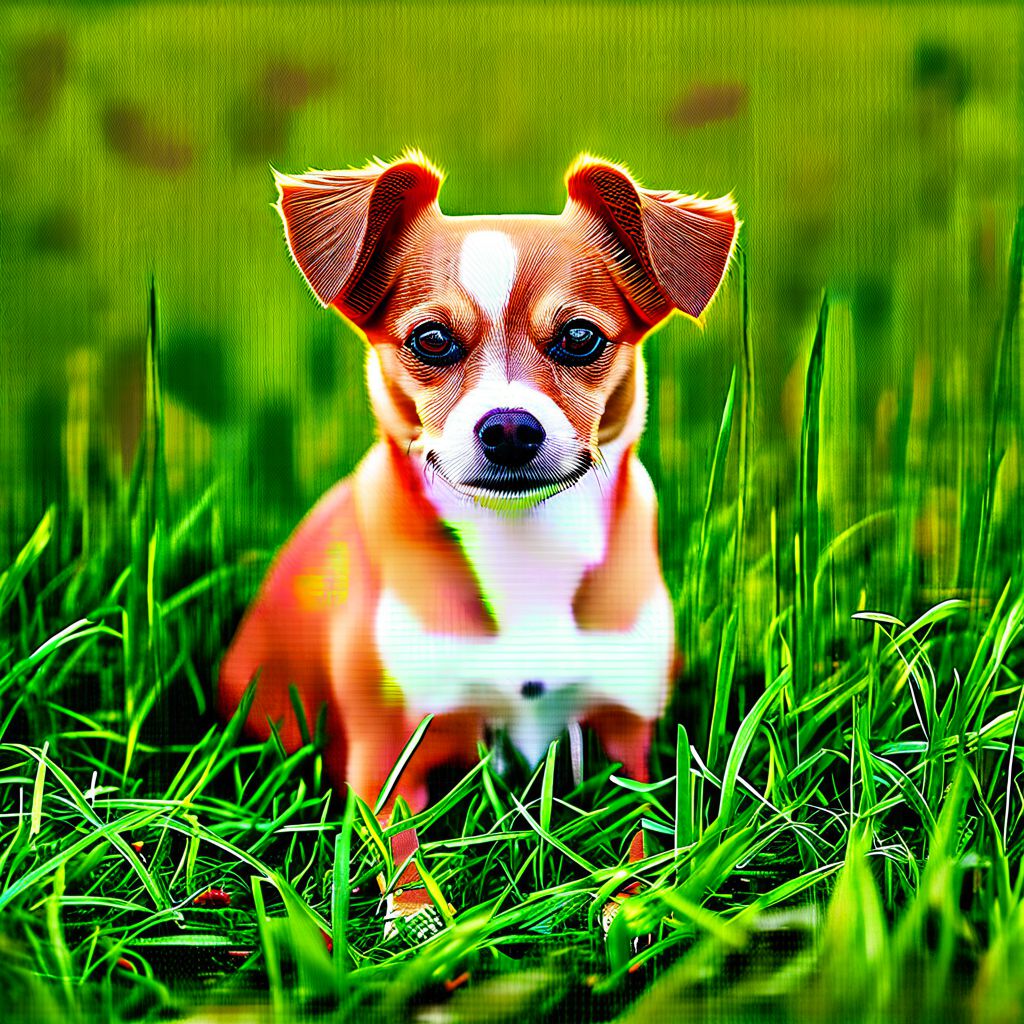} &
    \spairimg{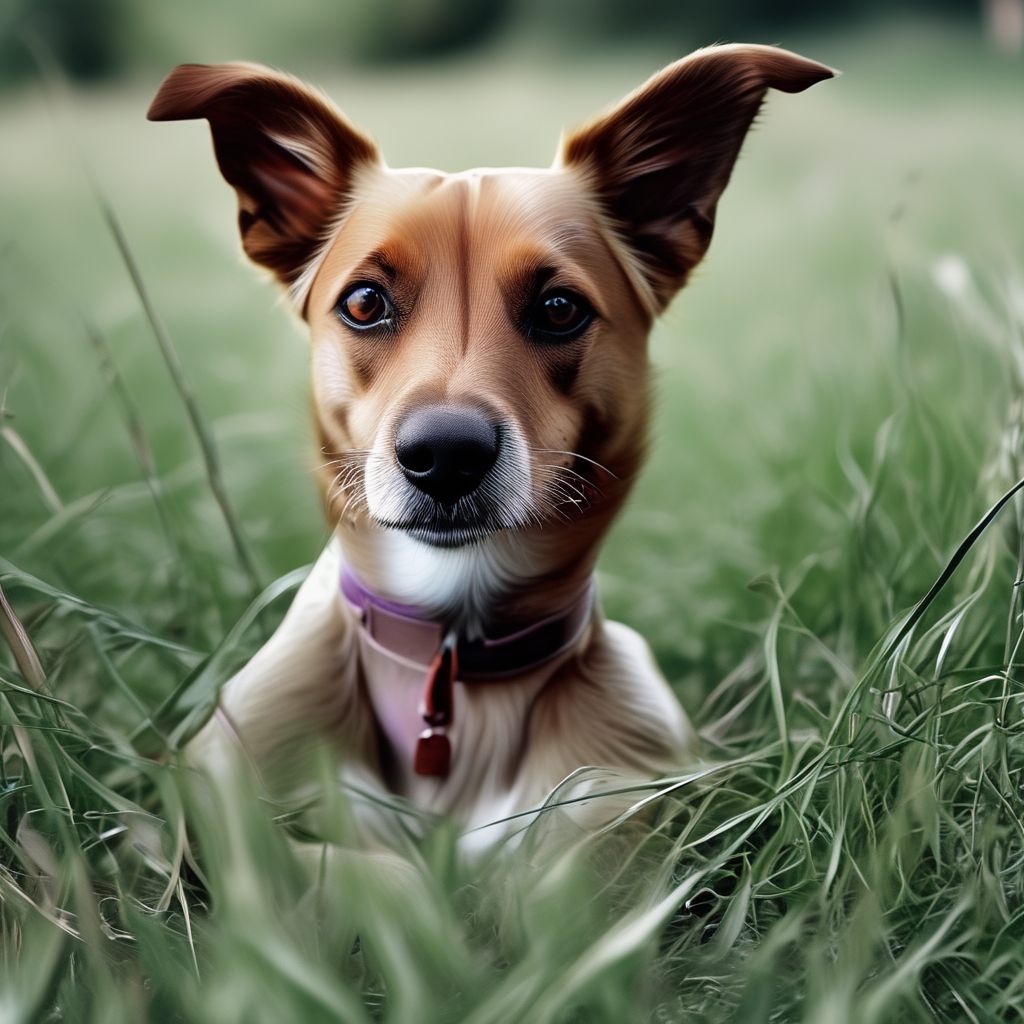}{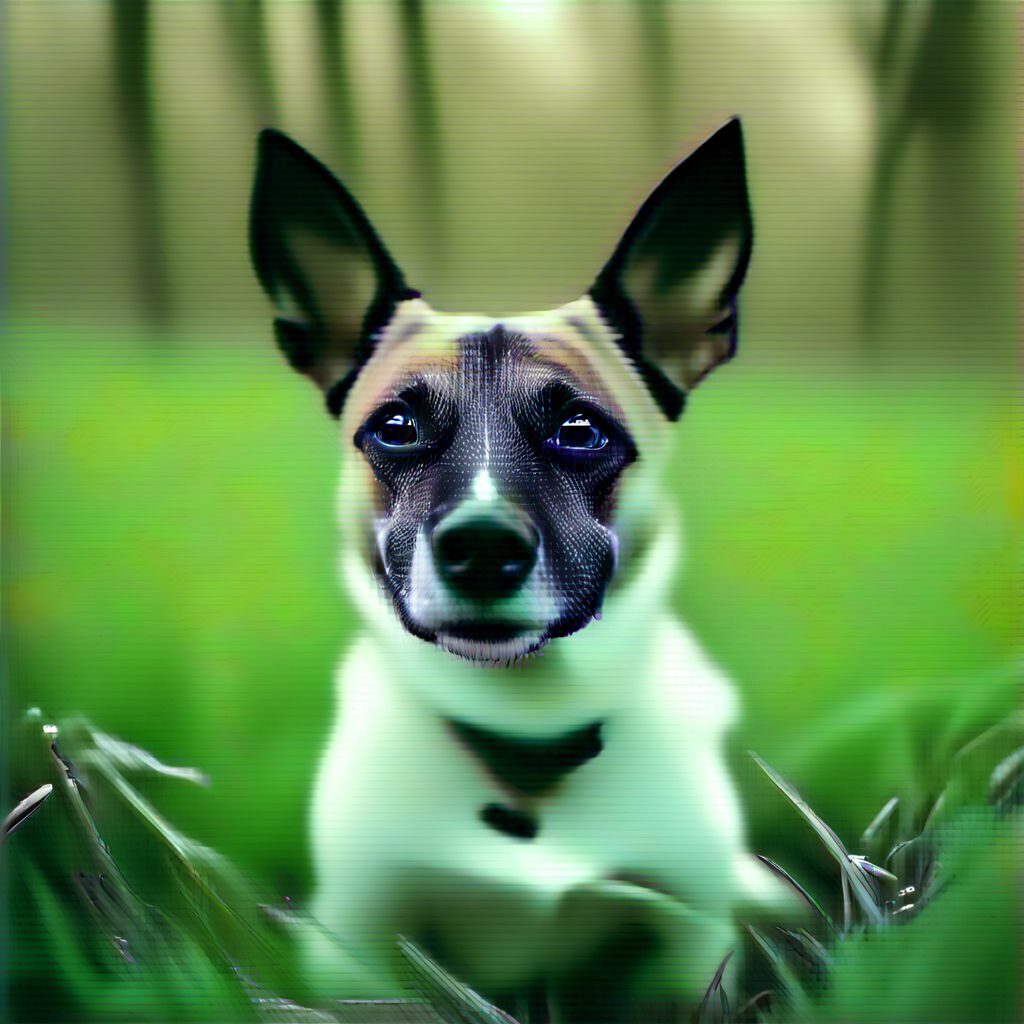} \\
    \midrule
    \textbf{\mFID}$\uparrow$ & \spairnum{51.15}{51.44} & \spairnum{43.99}{73.43} \\
    \textbf{\mQA}$\downarrow$ & \spairnum{3.27}{2.74} & \spairnum{4.31}{3.45} \\
    \textbf{\mPK}$\downarrow$ & \spairnum{21.24}{20.82} & \spairnum{21.44}{20.64} \\
    \textbf{\mAES}$\downarrow$ & \spairnum{6.15}{5.94} & \spairnum{6.13}{5.89} \\
    \bottomrule
    \end{tabularx}
    \caption{Surrogate model ablation on AFHQ with PixArt as source model.}
    \vspace{-6pt}

    \label{fig:ablation_surrogate_model}
\end{figure}

\mypara{Different Surrogate Models.}
We evaluate the effect of surrogate model choice on AFHQ by fixing PixArt as the source model and varying the attacker-side surrogate model between PixArt and SDXL.
Figure~\ref{fig:ablation_surrogate_model} shows the experimental results.
QAlign consistently decreases under both surrogate choices, from 3.27 to 2.74 with PixArt and from 4.31 to 3.45 with SDXL. We further observe that the generated images exhibit noticeable color shifts, resulting in degraded visual quality. 
These results indicate that \method{} remains effective across different surrogate models.

\begin{figure}[tbp]
\centering
\scriptsize
\setlength{\tabcolsep}{0.8pt}
\renewcommand{\arraystretch}{1.08}
\begin{tabularx}{\columnwidth}{@{}l*{6}{>{\centering\arraybackslash}X}@{}}
\toprule
\rowcolor{tablepanel}\multicolumn{7}{c}{\textbf{Ablation Study of Disruption Loss Design}} \\
\midrule
\rowcolor{tablehead}\textbf{Loss} & \textbf{Clean} & \makecell{\textbf{\mDE+Latent}\\\textbf{(Ours)}} & \textbf{Latent} & \textbf{\mDE} & \textbf{\mLP+Latent} & \textbf{L2+Latent} \\
\textbf{Example} & \raisebox{-.5\height}{\includegraphics[width=\linewidth]{figures_jpeg/ablation_baseline/5baselineafhq.jpg}} &
\raisebox{-.5\height}{\includegraphics[width=\linewidth]{figures_jpeg/ablation_visual_loss/ciede2000_latent.jpg}} &
\raisebox{-.5\height}{\includegraphics[width=\linewidth]{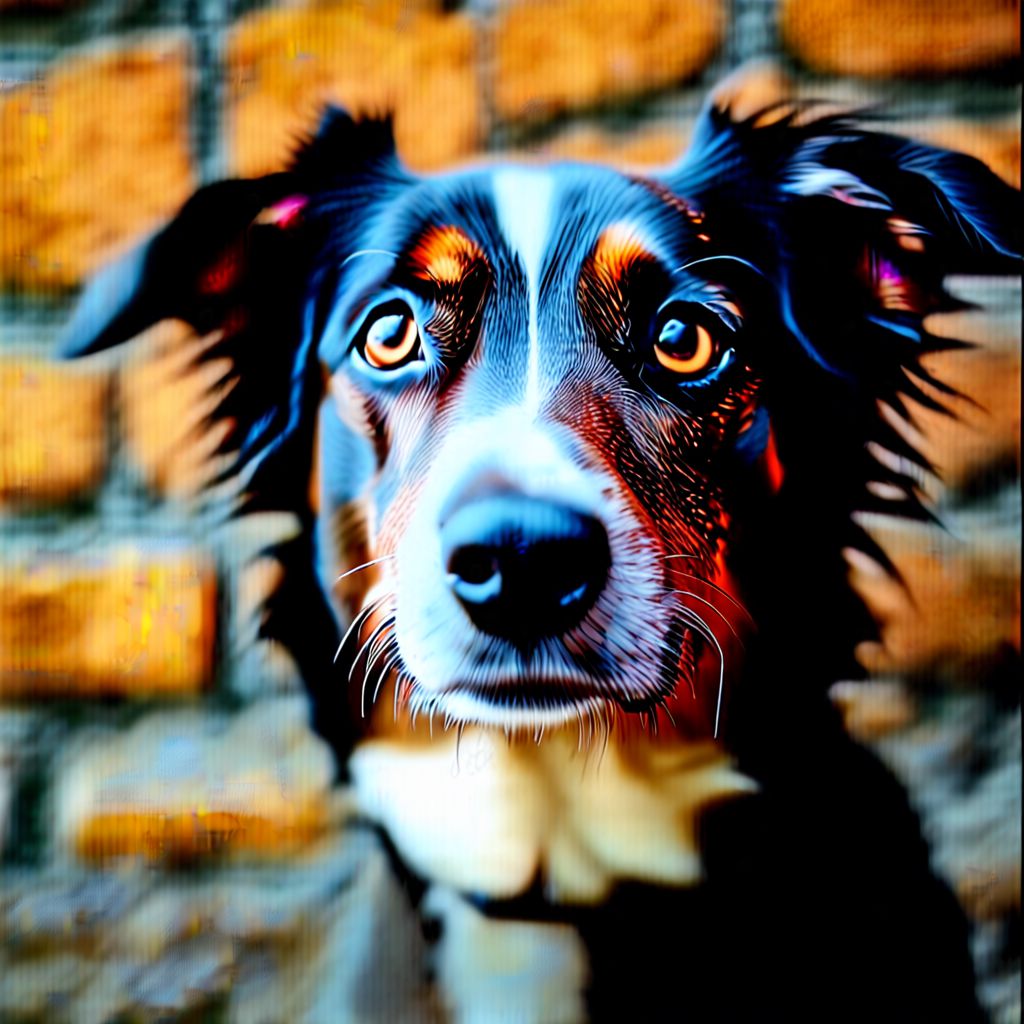}} &
\raisebox{-.5\height}{\includegraphics[width=\linewidth]{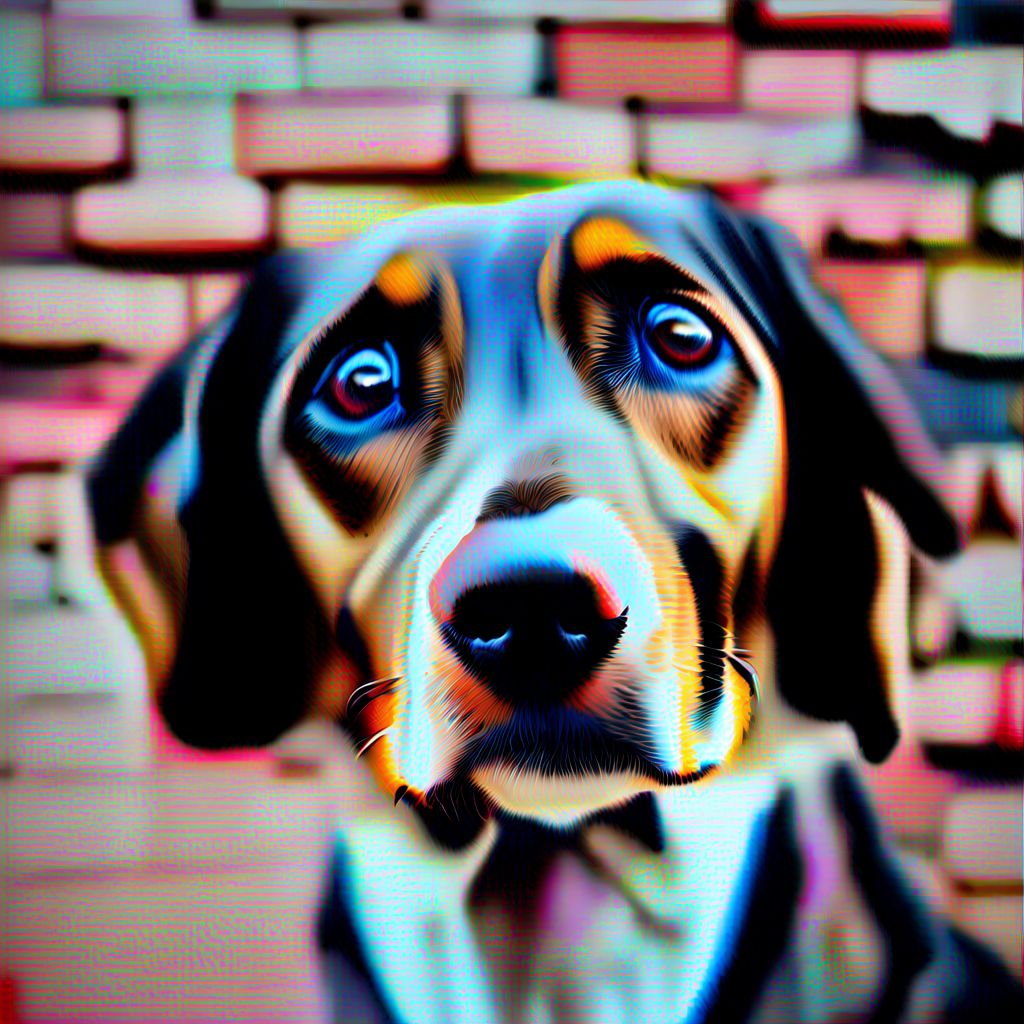}} &
\raisebox{-.5\height}{\includegraphics[width=\linewidth]{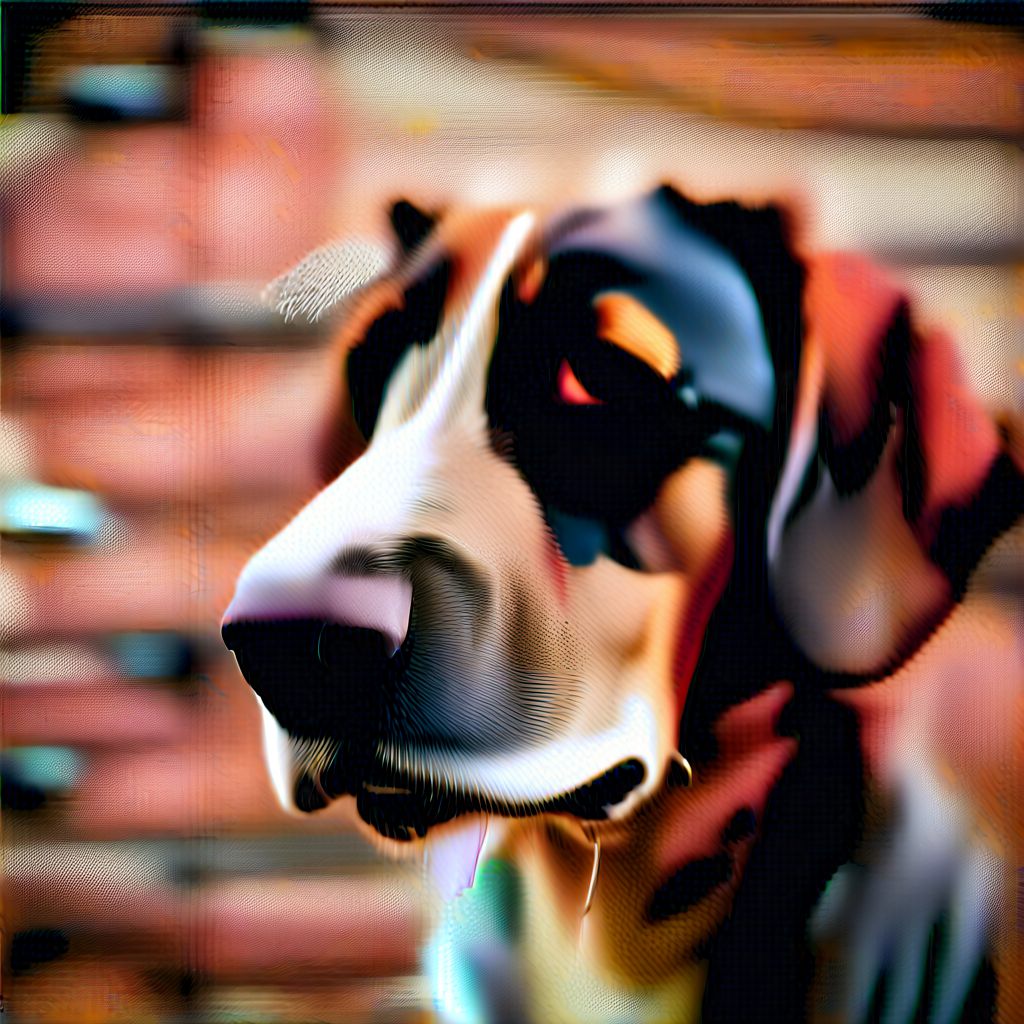}} &
\raisebox{-.5\height}{\includegraphics[width=\linewidth]{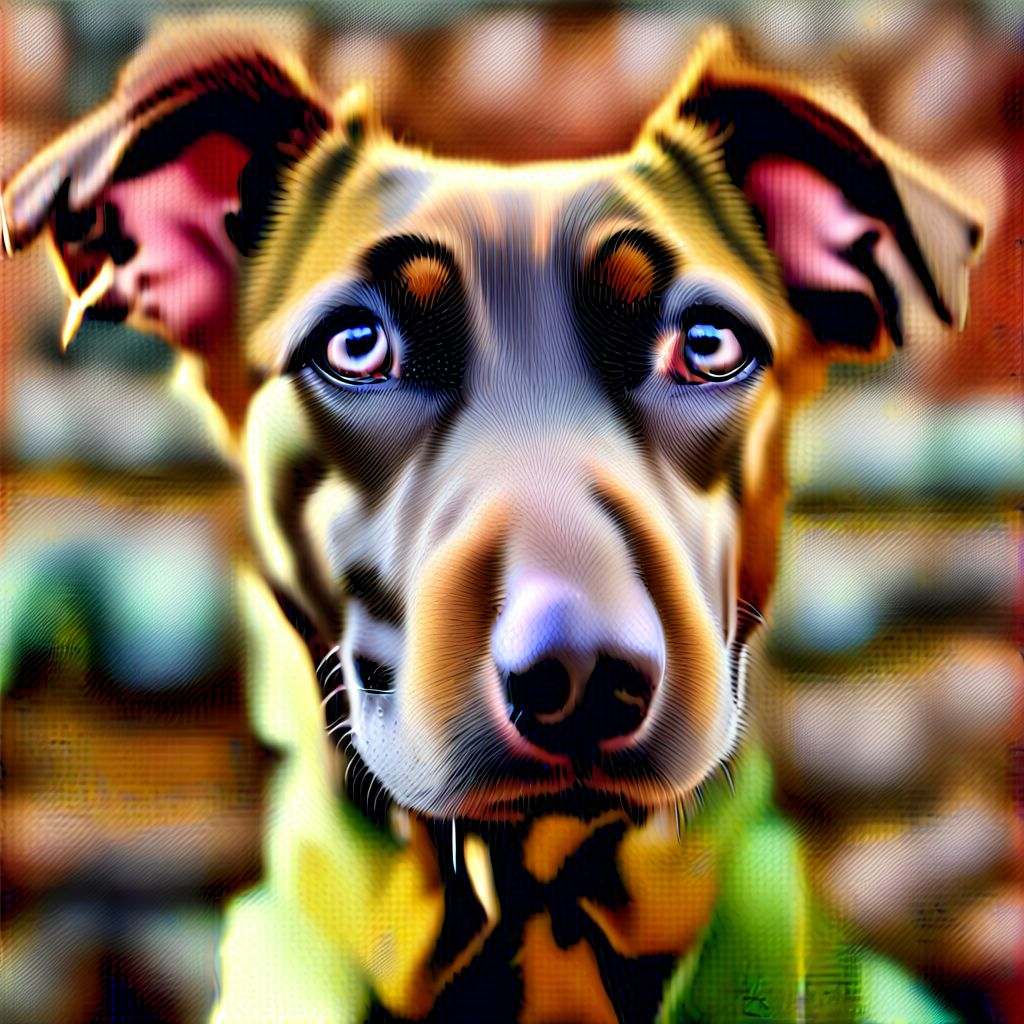}} \\
\midrule
\textbf{\mFID}$\uparrow$ & 60.13 & 132.62 & 101.69 & 91.47 & 102.12 & 90.56 \\
\textbf{\mQA}$\downarrow$ & 3.10 & 1.59 & 2.00 & 2.03 & 2.55 & 2.31 \\
\textbf{\mPK}$\downarrow$ & 20.80 & 19.75 & 20.26 & 20.43 & 20.63 & 20.46 \\
\textbf{\mAES}$\downarrow$ & 5.97 & 5.25 & 5.39 & 5.58 & 5.71 & 5.52 \\
\bottomrule
\end{tabularx}
\caption{Ablation study on disruption loss design. \mDE{} denotes CIEDE2000 color regularization, Latent denotes the self-referenced latent maximization, and LPIPS/L2 denote regularizations used to test latent-shortcut avoidance.}
\vspace{-6pt}
\label{fig:ablation_visual_loss}
\end{figure}

\mypara{Disruption Objective Design.}
We conduct an ablation study on the disruption loss design on AFHQ, using PixArt as the source model and SSD as the surrogate model.
This ablation evaluates whether reconstruction-guided color regularization avoids latent shortcuts and why other visual metrics are less suitable for this role than color regularization.
The experimental results are shown in Figure~\ref{fig:ablation_visual_loss}.
We observe that using only the latent representation as the optimization objective performs worse than \method{}.
In addition, the other reconstruction regularizers produce weaker quality degradation than CIEDE2000 color regularization. 
Using \mLtwo{} or \mLP{} as the visual regularization term even performs worse than the latent-only objective and is substantially inferior to \method{}. 
We also test a variant that removes the self-referenced latent loss and retains only the color regularization term; this variant also performs worse than both \method{} and the latent-only variant. 
Overall, these results demonstrate the effectiveness of \method{}'s self-referenced latent maximization loss, as well as the advantage of color regularization over other visual metrics as the regularization term. \looseness=-1

\section{Discussion}
The details of the adaptive attack and more discussion are shown in Appendix~\ref{Appendix:expsetup} and Appendix~\ref{Appendix:discussion}.

\mypara{Adaptive Attack.}
We discuss whether an adaptive attacker can mitigate \method{} under three adaptive attacker strategies: input detection, purification, and robust learning.\looseness=-1

\noindent\textit{Outlier Detection.}
We evaluate input-side detection by encoding samples with the CLIP image encoder and applying robust z-score outlier scoring based on median-centered absolute-deviation statistics~\cite{radford2021learning,rousseeuw1993alternatives,leys2013detecting}.
The results show that conventional CLIP feature-space outlier detection does not reliably separate protected samples from clean outputs.
Specifically, the robust z-score detector obtains an AUC of 0.5085, with a TPR of 0.5060 and an FPR of 0.4994, which is close to random discrimination.
This finding is consistent with the design goal of \method{}: protected samples become disruptive in the surrogate space used for distillation, while their generic CLIP feature distribution remains close to clean outputs because benign visual fidelity is preserved.
Thus, simple feature-space filtering is not sufficient to remove the protected training signal before substitute model training.

\begin{figure}[t]
\centering
\footnotesize
\setlength{\tabcolsep}{1.8pt}
\renewcommand{\arraystretch}{1.10}
\begin{tabularx}{0.5\textwidth}{@{}l*{4}{>{\centering\arraybackslash}X}@{}}
\toprule
\rowcolor{tablepanel}\multicolumn{5}{c}{\textbf{Adaptive Attacks}} \\
\midrule
\rowcolor{tablehead}\textbf{Setting} & \textbf{Clean} & \textbf{\makecell{\method{}\\(w/o Adapt.)}} & \textbf{Purification} & \textbf{\makecell{Robust\\Learning}} \\
\textbf{Example} & \raisebox{-.5\height}{\includegraphics[width=\linewidth]{figures_jpeg/ablation_baseline/5baselineafhq.jpg}} &
\raisebox{-.5\height}{\includegraphics[width=\linewidth]{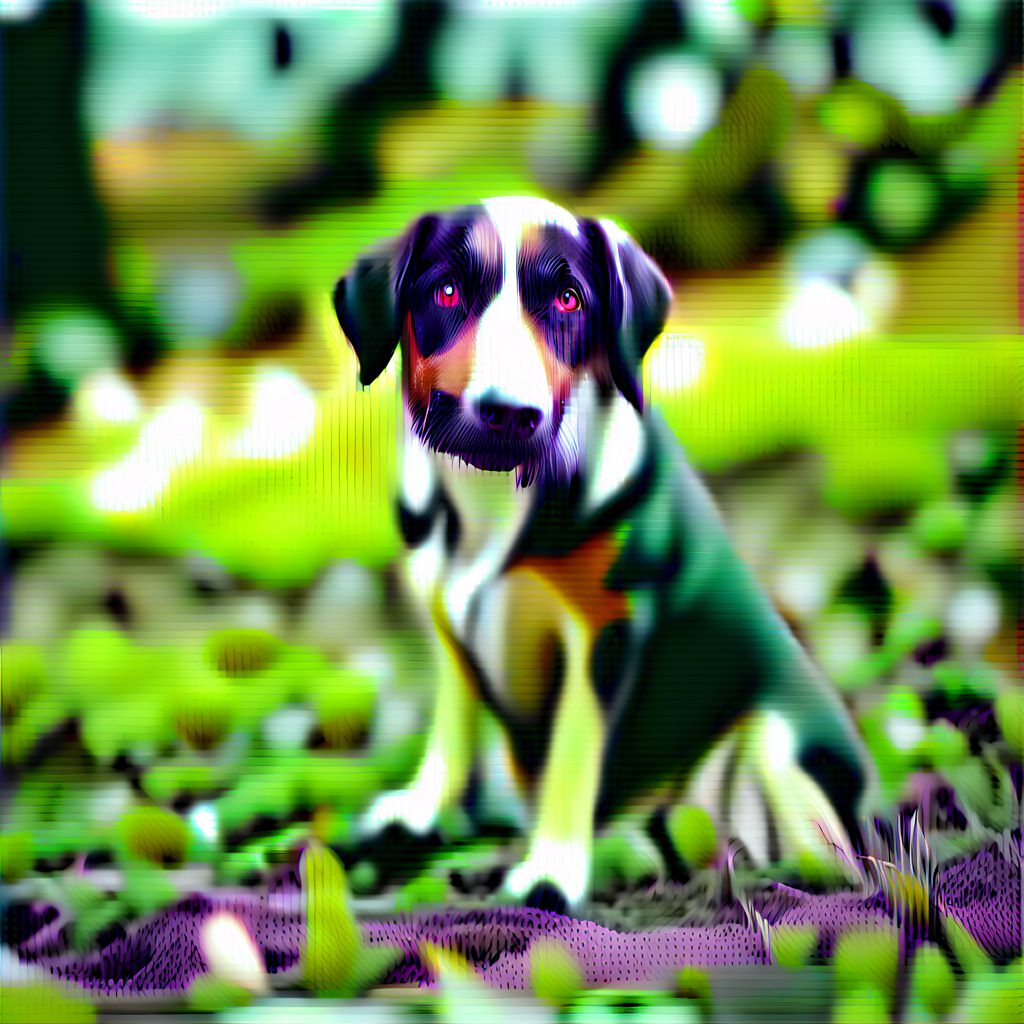}} &
\raisebox{-.5\height}{\includegraphics[width=\linewidth]{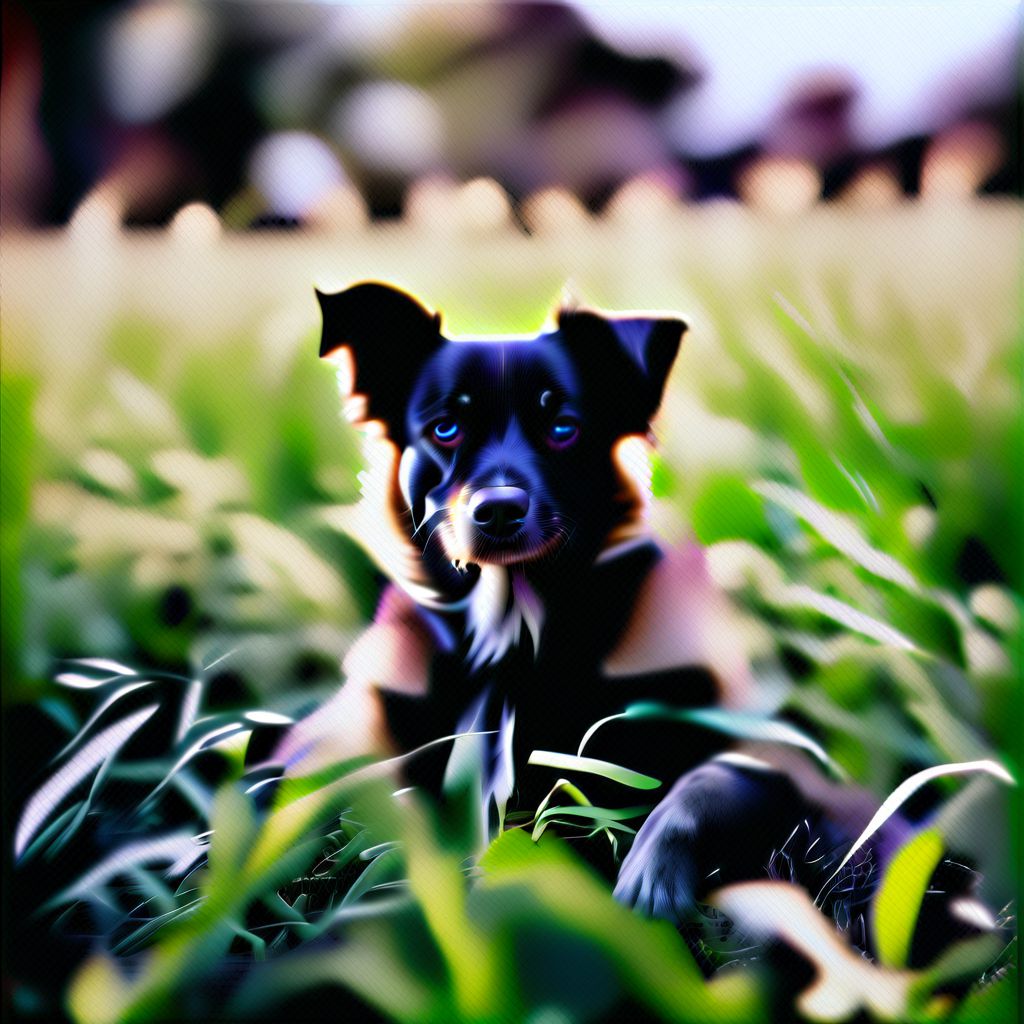}} &
\raisebox{-.5\height}{\includegraphics[width=\linewidth]{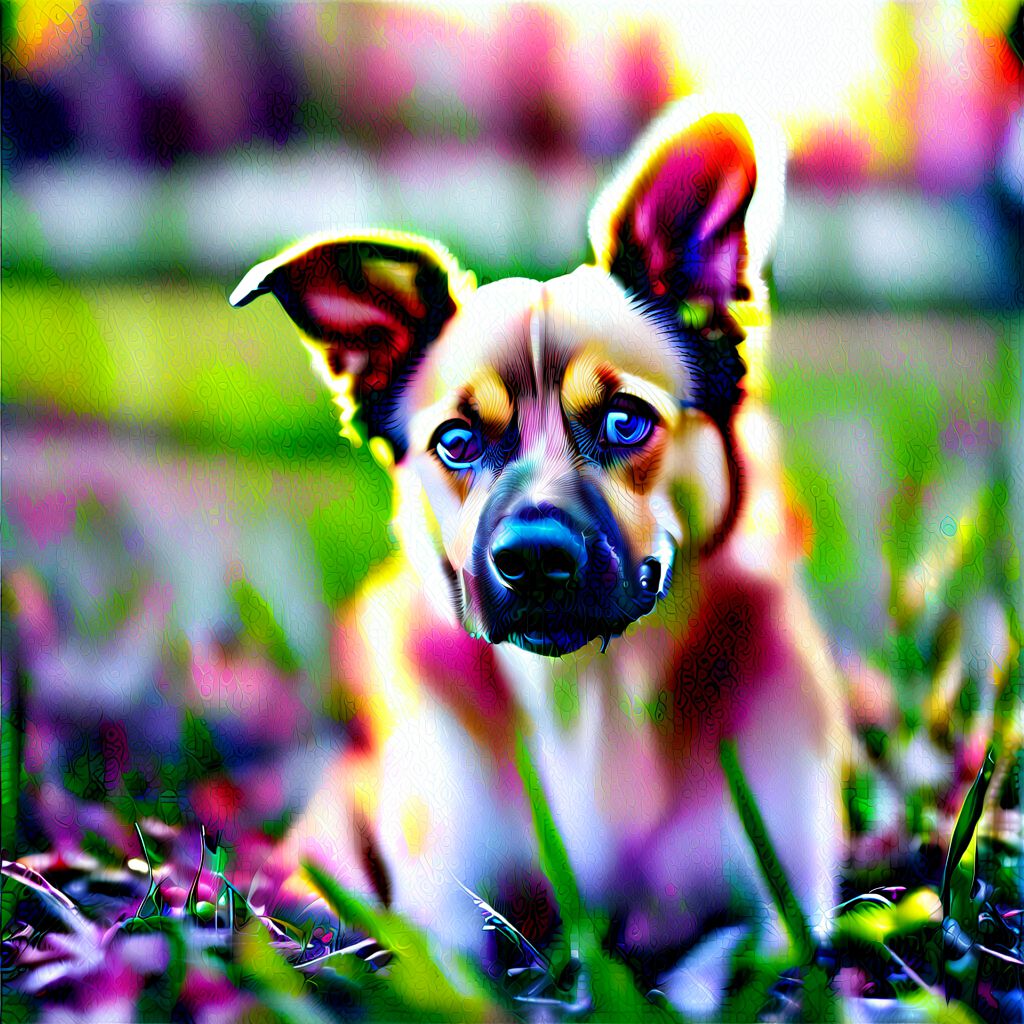}} \\
\midrule
\textbf{\mFID}$\uparrow$ & 60.13 & 132.62 & 95.50 & 66.47 \\
\textbf{\mQA}$\downarrow$ & 3.10 & 1.59 & 2.39 & 2.31 \\
\textbf{\mPK}$\downarrow$ & 20.81 & 19.75 & 20.31 & 20.53 \\
\textbf{\mAES}$\downarrow$ & 5.97 & 5.25 & 5.82 & 5.71 \\
\bottomrule
\end{tabularx}
\caption{Adaptive attack results under VAE-based data purification and robust learning. The Clean and \method{} columns are shared by both attacks; \method{} (w/o Adapt.) denotes training without attacker-side adaptation.}
\label{fig:adaptive_vae_purification}
\label{fig:adaptive_robust_learning}
\end{figure}

\noindent\textit{VAE Purification.}
We consider a scenario in which the attacker applies VAE-based reconstruction to remove subtle perturbations from the protected images, thereby attempting to circumvent the defense provided by \method{}. 
We evaluate this VAE-based purification attack on AFHQ, using PixArt as the source model, SSD as the substitute model, and SD-v2.1 VAE~\cite{sd21} as the purification module.
The results in Figure~\ref{fig:adaptive_vae_purification} show that VAE purification fails to recover clean-level substitute quality. Specifically, after purification, QAlign remains below the Clean setting, decreasing from 3.10 to 2.39. Compared with the unpurified \method{} setting, purification increases QAlign from 1.59 to 2.39 and partially reduces the FID disruption from 132.62 to 95.50. 
These results suggest that VAE-based purification cannot completely remove the defensive perturbations, and therefore does not recover clean-level quality \method{}. \looseness=-1

\noindent\textit{Robust Learning.}
We further evaluate a robust-learning attacker inspired by prior work by Radiya-Dixit et al.~\cite{radiyadixit2022data}, which shows that an adaptive trainer can use adversarial robust training to reduce the effect of collected poisoned images.
In our setting, the attacker trains the SSD substitute on the collected protected outputs with adversarially augmented robust learning, using the same AFHQ and PixArt source model setting.
Figure~\ref{fig:adaptive_robust_learning} reports the results.
Robust learning partially restores the quality of the substitute model compared with training without robust learning under the \method{} setting, increasing QAlign from 1.59 to 2.31.
However, it does not fully restore clean-level substitute quality: QAlign remains below the clean value, reaching 2.31 compared with 3.10 under the Clean setting, and the qualitative example still exhibits visible degradation relative to the clean substitute.
These results indicate that robust learning is a strong adaptive strategy, but it still leaves measurable quality degradation.\looseness=-1


\mypara{Limitations and Future Work.}
\noindent\textit{T2I Model-family Scope.}
\method{} is designed for LDM-based T2I services, in which the service-side VAE decoder introduces perturbations to induce disruption in the substitute model's latent space. As a result, it is not directly applicable to the small class of pixel-space diffusion models that do not rely on a VAE latent space.
Defending against distillation against such models may require poisoning or unlearnable examples~\cite{huang2021unlearnable}. 
If the deployed service itself performs pixel-space denoising, a similar service-internal protection principle may be explored by injecting protection into the denoising component rather than the VAE decoder, potentially requiring full denoising runs for training. 
We leave this setting as future work.

\noindent\textit{Broader Latent Space Architectures.}
\method{} may also extend to other generative services whose public outputs are decoded from private latent representations, such as latent audio/music generation, image-to-video, and text-to-video systems. 
In principle, a modality-specific version of \method{} could reduce the distillation value of released outputs while preserving perceptual quality. 
Nevertheless, extending our approach to these domains may require task-specific designs, such as mechanisms for handling spatiotemporal video latents and optimization reachability.
We leave distillation defenses for latent audio, music, and video generation as future work. \looseness=-1

\section{Conclusion}

Unauthorized distillation exposes a central weakness of public T2I services: the images that make these systems useful can also serve as supervision for transferring proprietary generation capability into unlicensed substitute models.
This paper shows that this risk cannot be effectively addressed by conventional sample-wise optimization under online service constraints.
A practical alternative is to internalize the defense within the model itself, enabling protection to be produced as an inherent part of generation while preserving the latency, throughput, and visual quality required for online deployment.
\method{} follows this direction by moving anti-distillation protection from sample-wise optimization into the service-side decoder.
Its key lesson is that service-level protection requires objectives achievable through a shared model update, rather than externally prescribed perturbations optimized independently for each image.
By using self-referenced latent maximization with reconstruction-guided color regularization, \method{} achieves disruption effectiveness against distillation while preserving close fidelity to clean service images.
\method{} establishes a new paradigm for real-time protection against unauthorized distillation in deployed T2I systems. \looseness=-1

\bibliographystyle{IEEEtran}
\bibliography{references}

\appendices

\section{Additional Experimental Setup}\label{Appendix:expsetup}
\subsection{Implementation Details}
In the main experiment, we set the protection loss weights to $\lambda_v=1000$, $\lambda_z=10/11$, and $\lambda_c=1/11$. For the visual constraints, we use $\tau_{cie}=2.0$, $\tau_{\mathrm{lpips}}=0.05$, $\tau_{\mathrm{tv}}=0.01$, and $\tau_{patch}=3.0$. Unless otherwise specified, the relative weights among the visual metrics are all set to $1$, we train the decoder with full parameters with a batch size of 1 and an AdamW learning rate of $10^{-5}$.
During decoder training, we query the corresponding protected model with these prompts and use the resulting generated images as training data, so that the decoder-training data follows the same distribution as images produced during generation. 
In the distillation training, standard LoRA finetuning is performed to reduce computational cost; only LoRA adapters on the denoiser are trained. 
The default learning rate is $10^{-4}$, the LoRA rank is 4, the batch size is 1,
Optimization uses AdamW with a cosine learning-rate schedule and warmup. 
For the CIEDE2000 patch-level visual constraint, we set the patch size to $64 \times 64$, and the perceptual color-difference threshold to $\tau_{patch}=3.0$. 
For each image pair, we divide the reconstructed image and the reference reconstruction into non-overlapping square patches, compute the average CIEDE2000 color difference within each patch, and penalize only patches whose color difference exceeds the threshold. 
Here, $N$ denotes the number of patches and $\Delta E_{00}^{(i)}$ is the mean CIEDE2000 distance of the $i$-th patch. The global visual loss coefficient is set to $1.0$. \looseness=-1


\subsection{Details for Adaptive Attack}

All adaptive-attack experiments use AFHQ with PixArt as the source model and SSD as the substitute model, following the same prompts, resolution, LoRA setting, optimizer, schedule, and evaluation metrics as the main experiments.
We consider a defense-aware attacker who knows that service outputs may contain a training-disruptive signal, but has no access to the clean decoder, protected decoder, or paired clean/protected outputs.
For attacker-side outlier detection, we embed 500 clean and 500 protected samples with a frozen CLIP ViT-B/32 encoder, score them using robust z-scores with contamination ratio $\gamma=0.5$, flag the 500 highest-scoring samples as protected, and report TPR, FPR, and ROC-AUC.
For VAE purification, each protected image is encoded and decoded by the SD-v2.1 VAE before substitute training, and the reconstructed image replaces the original protected output.
For robust learning, the attacker uses PGD-augmented denoising training with random initialization, $\ell_\infty$ budget $\epsilon=16/255$, step size $\alpha=4/255$, and $K=5$ inner steps, projecting perturbations to satisfy $\|\delta\|_\infty\leq\epsilon$.

\begin{table}[t]
\centering
\caption{Metric-based benign visual fidelity. \mLP, \mSS, \mDE, and \mPN{} denote LPIPS, SSIM, CIEDE2000, and PSNR, respectively.}
\label{tab:visual_quality_metrics}
\setlength{\tabcolsep}{3.0pt}
\renewcommand{\arraystretch}{1.10}
\resizebox{0.7\columnwidth}{!}{%
\begin{tabular}{@{}llcccc@{}}
\toprule
\rowcolor{tablepanel}\multicolumn{6}{c}{\textbf{Results of Benign Visual Fidelity}} \\
\midrule
\rowcolor{tablehead}\textbf{Model} & \textbf{Metric} & \textbf{AFHQ} & \textbf{CelebA} & {\scriptsize\textbf{Landscape}} & \textbf{Pokemon} \\
\midrule
\multirow{4}{*}{\textbf{SSD}}
 & \mLP$\downarrow$ & 0.047 & 0.047 & 0.044 & 0.044 \\
 & \mSS$\uparrow$ & 0.857 & 0.848 & 0.867 & 0.899 \\
 & \mDE$\downarrow$ & 1.931 & 1.917 & 1.839 & 1.903 \\
 & \mPN$\uparrow$ & 33.74 & 34.54 & 34.25 & 33.86 \\
\addlinespace[0.2em]
\multirow{4}{*}{\textbf{PixArt}}
 & \mLP$\downarrow$ & 0.044 & 0.043 & 0.041 & 0.042 \\
 & \mSS$\uparrow$ & 0.879 & 0.900 & 0.893 & 0.908 \\
 & \mDE$\downarrow$ & 1.770 & 1.551 & 1.646 & 1.714 \\
 & \mPN$\uparrow$ & 33.84 & 36.34 & 34.66 & 33.43 \\
\addlinespace[0.2em]
\multirow{4}{*}{\textbf{SDXL}}
 & \mLP$\downarrow$ & 0.045 & 0.046 & 0.040 & 0.045 \\
 & \mSS$\uparrow$ & 0.900 & 0.884 & 0.902 & 0.907 \\
 & \mDE$\downarrow$ & 1.918 & 1.887 & 1.838 & 1.884 \\
 & \mPN$\uparrow$ & 36.35 & 36.37 & 35.96 & 36.29 \\
\addlinespace[0.2em]
\multirow{4}{*}{\textbf{Shuttle3}}
 & \mLP$\downarrow$ & 0.044 & 0.044 & 0.041 & 0.041 \\
 & \mSS$\uparrow$ & 0.910 & 0.909 & 0.915 & 0.929 \\
 & \mDE$\downarrow$ & 1.915 & 1.880 & 1.794 & 1.862 \\
 & \mPN$\uparrow$ & 34.74 & 34.97 & 33.81 & 34.29 \\
\bottomrule
\end{tabular}}
\end{table}
\subsection{Baselines}
We compare \method{} with five representative protection methods. 
Because the original methods are designed as sample-wise perturbation optimizers, we implement each baseline by training a separate protected VAE decoder with the corresponding objective.
For fairness, all decoder-adapted baselines use the same protected-decoder architecture, data, optimizer, training, and evaluation pipeline as \method{}.\looseness=-1

\noindent\textit{AdvDM~\cite{liang2023advdm}.}
AdvDM is adapted by setting the baseline objective mode to diffusion, turning it into a decoder-level diffusion-loss maximization objective.
Following AdvDM, for each protected output, the image is encoded by the surrogate VAE, noised at a randomly sampled diffusion timestep, and passed through a surrogate diffusion model.
The objective value is the MSE between the surrogate model's predicted noise and the sampled true noise.
We sample diffusion timesteps from the full range $[0,999]$; this adversarial term is combined with the same visual fidelity penalties used by the other decoder-adapted baselines.\looseness=-1

\noindent\textit{Anti-DreamBooth~\cite{vanle2023antidreambooth}.}
Anti-DreamBooth is adapted by preserving its bilevel optimization between a surrogate diffusion learner and the protected decoder.
During the surrogate step, current decoder outputs are used to train the surrogate diffusion model with its denoising loss; during the decoder step, the decoder is updated to maximize the surrogate denoising loss evaluated on the current protected outputs.
The surrogate UNet is updated with a learning rate $10^{-5}$.\looseness=-1

\noindent\textit{Glaze~\cite{shan2023glaze}.}
Glaze is adapted with precomputed style-transferred targets rather than running T2I style transfer during decoder training.
In our experiments, these targets are generated with SD-v2.1~\cite{sd21} in the Van Gogh style using the prompt ``An oil painting in the style of Van Gogh''.
We use style strength 0.5 and style guidance scale 5.0.
For each pair, the current decoder output and the styled target are both encoded by the surrogate VAE, and the objective minimizes the MSE between their surrogate latents.\looseness=-1

\noindent\textit{Nightshade~\cite{shan2024nightshade}.}
Nightshade is adapted as a class-wise targeted objective.
The target images are sampled from the ``person'' category of the MS-COCO dataset~\cite{cocodataset}. Following the main experimental protocol, we use BLIP-Large~\cite{blip} to generate captions/prompts for the target images.
For each training batch, both the protected output and the selected target image are encoded by the surrogate VAE, and the protected decoder is optimized to minimize the MSE between their latent representations. \looseness=-1

\noindent\textit{PhotoGuard~\cite{salman2023photoguard}.}
We adapt PhotoGuard using its black-target objective variant. Specifically, we construct a full-black image, encode it once with the surrogate VAE, and cache the resulting black latent. During decoder training, the protected decoder is optimized to minimize the MSE between the latent representation of the current protected output and the cached black latent. \looseness=-1


\section{Additional Experimental Results}\label{Appendix:expresult}

\subsection{Benign Visual Fidelity}
We evaluate benign visual fidelity by comparing each protected output with its clean service output, as reported in Table~\ref{tab:visual_quality_metrics} and Figure~\ref{fig:visual_quality_examples}.
The results show that the protection preserves benign output quality across the evaluated service models and image domains.
For example, using PixArt as the source model and CelebA as the dataset, \method{} achieves LPIPS 0.043, SSIM 0.900, CIEDE2000 1.551, and PSNR 36.34, indicating close agreement with the clean service output.
The protected outputs preserve the main subject, layout, color composition, and local structure of the clean references.
Thus, \method{} reduces the training value of public outputs for attackers while keeping protected outputs close to clean service outputs under the reported fidelity metrics.\looseness=-1


\subsection{Additional Metrics}
Table~\ref{tab:appendix_pixart_kid_qaaes} reports the two additional metrics for the PixArt main-experiment setting.
KID measures distributional mismatch from the reference set, while QAlign-Aesthetic (QAAES) measures the aesthetic subscore of QAlign.
Thus, higher KID and lower QAAES indicate stronger degradation of the attacker-trained substitute.
Across all four datasets, \method{} achieves the largest KID and the lowest QAAES, which is consistent with the main-table trends in FID, QAlign, PickScore, and Aesthetic score.

\begin{table}[t]
\centering
\caption{Additional results on KID and QAAES.}
\label{tab:appendix_pixart_kid_qaaes}
\scriptsize
\setlength{\tabcolsep}{1.3pt}
\renewcommand{\arraystretch}{1.08}
\resizebox{\columnwidth}{!}{%
\begin{tabular}{@{}llcccccccc@{}}
\toprule
\rowcolor{tablepanel}\multicolumn{10}{c}{\textbf{Supplementary PixArt Main Results}} \\
\midrule
\rowcolor{tablehead}\textbf{Dataset} & \textbf{Metric} & \textbf{Ref.} & \textbf{Clean} & \textbf{Ours} & \textbf{Adv} & \textbf{Anti} & \textbf{Glaze} & \textbf{Night} & \textbf{PG} \\
\midrule
\multirow{2}{*}{\textbf{AFHQ}}
 & KID$\uparrow$ & -- & 0.0099 & \best{0.0671} & 0.0257 & 0.0296 & 0.0422 & 0.0302 & \second{0.0522} \\
 & QAAES$\downarrow$ & 4.08 & 3.45 & \best{2.16} & 3.37 & 3.49 & 3.10 & 3.37 & \second{2.92} \\
\addlinespace[0.2em]
\multirow{2}{*}{\textbf{CelebA}}
 & KID$\uparrow$ & -- & 0.0222 & \best{0.0480} & 0.0163 & 0.0236 & \second{0.0300} & 0.0271 & 0.0269 \\
 & QAAES$\downarrow$ & 4.28 & 4.11 & \best{2.54} & 4.21 & 4.19 & 3.43 & \second{3.42} & 3.84 \\
\addlinespace[0.2em]
\multirow{2}{*}{\textbf{Landscape}}
 & KID$\uparrow$ & -- & 0.0064 & \best{0.0289} & 0.0079 & 0.0082 & 0.0101 & \second{0.0178} & 0.0095 \\
 & QAAES$\downarrow$ & 3.96 & 4.11 & \best{2.57} & 4.07 & 4.15 & 3.96 & 3.62 & \second{3.53} \\
\addlinespace[0.2em]
\multirow{2}{*}{\textbf{Pokemon}}
 & KID$\uparrow$ & -- & 0.0104 & \best{0.0296} & 0.0118 & 0.0121 & 0.0122 & \second{0.0200} & 0.0194 \\
 & QAAES$\downarrow$ & 3.95 & 3.52 & \best{2.03} & 3.33 & 3.40 & 3.24 & 2.72 & \second{2.63} \\
\bottomrule
\end{tabular}}
\end{table}

\subsection{Ensemble Surrogates Enhance Transferability}
A natural question is whether protection learned against one surrogate model can transfer to substitute models from different model families. 
Consistent with prior observations on adversarial transferability, we find that single-surrogate transfer is limited when the substitute model belongs to a distant VAE family~\cite{adversarialtransferability1,adversarialtransferability2,adversarialtransferability3}. 
This limitation is expected: different VAEs may encode, smooth, and reconstruct the same protected image in different ways, so a protected output optimized for one VAE bottleneck may no longer induce the intended latent or reconstruction mismatch under another.
However, we find that using an ensemble of surrogate VAEs during decoder protection training is effective for enhancing the transferability.
Specifically, we train our protected decoder using \method{} across multiple frozen surrogate VAEs, and the other settings are aligned with those of the main experiment.
This design encourages the same protected output to remain disruptive under several plausible VAE bottlenecks, reducing overfitting to one surrogate family without assuming knowledge of the exact attacker model.

\begin{figure}[tbp]
\centering
\scriptsize
\setlength{\tabcolsep}{1.2pt}
\renewcommand{\arraystretch}{1.08}
\newcommand{\tpairhead}{\begin{tabular}{@{}>{\centering\arraybackslash}p{.49\linewidth}@{\hspace{1pt}}>{\centering\arraybackslash}p{.49\linewidth}@{}}\textbf{Clean} & \textbf{\method{}}\end{tabular}}
\newcommand{\tpairimg}[2]{\begin{tabular}{@{}>{\centering\arraybackslash}p{.49\linewidth}@{\hspace{1pt}}>{\centering\arraybackslash}p{.49\linewidth}@{}}\includegraphics[width=\linewidth]{#1} & \includegraphics[width=\linewidth]{#2}\end{tabular}}
\begin{tabularx}{\columnwidth}{@{}*{2}{>{\centering\arraybackslash}X}@{}}
\toprule
\rowcolor{tablepanel}\multicolumn{2}{c}{\textbf{Transferability with Ensemble Surrogates}} \\
\midrule
\rowcolor{tablehead}\textbf{FLUX2klein} & \textbf{SSD} \\
\rowcolor{tablehead}\tpairhead & \tpairhead \\
\tpairimg{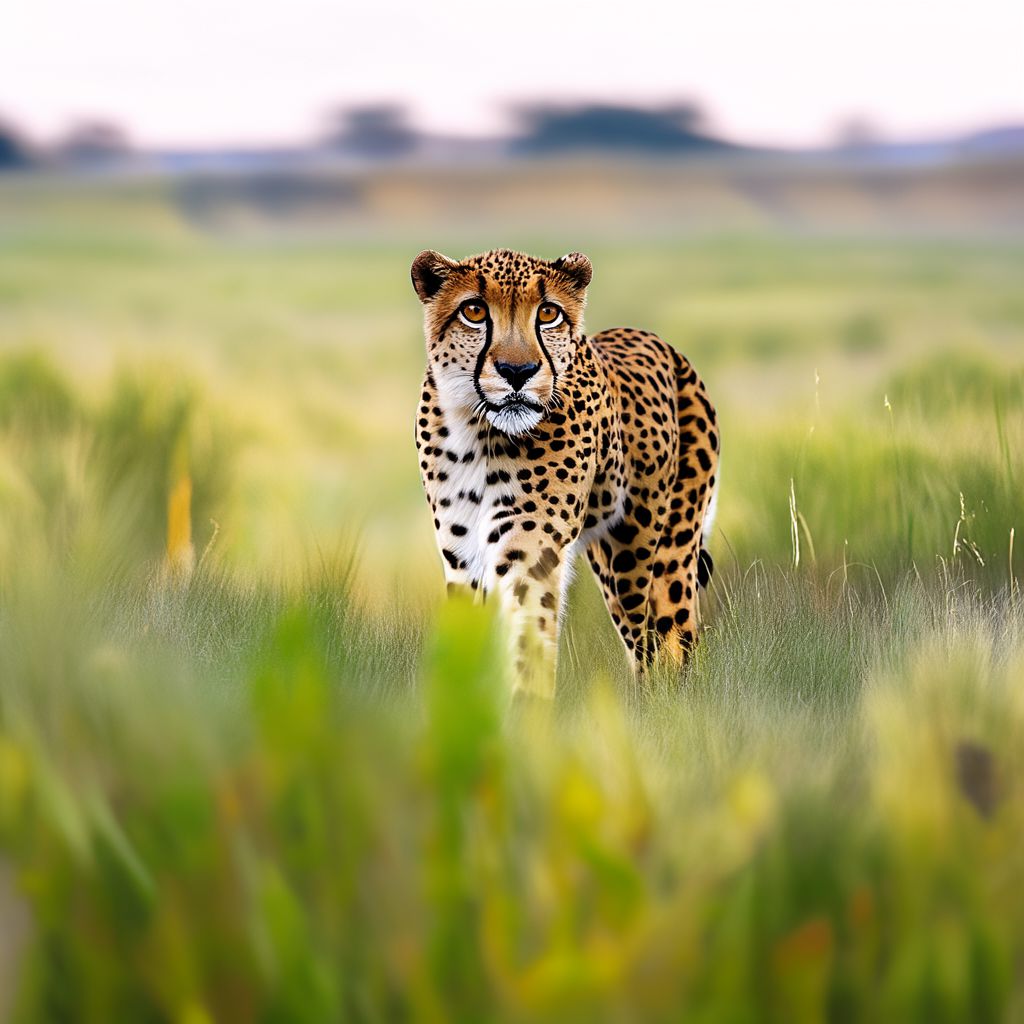}{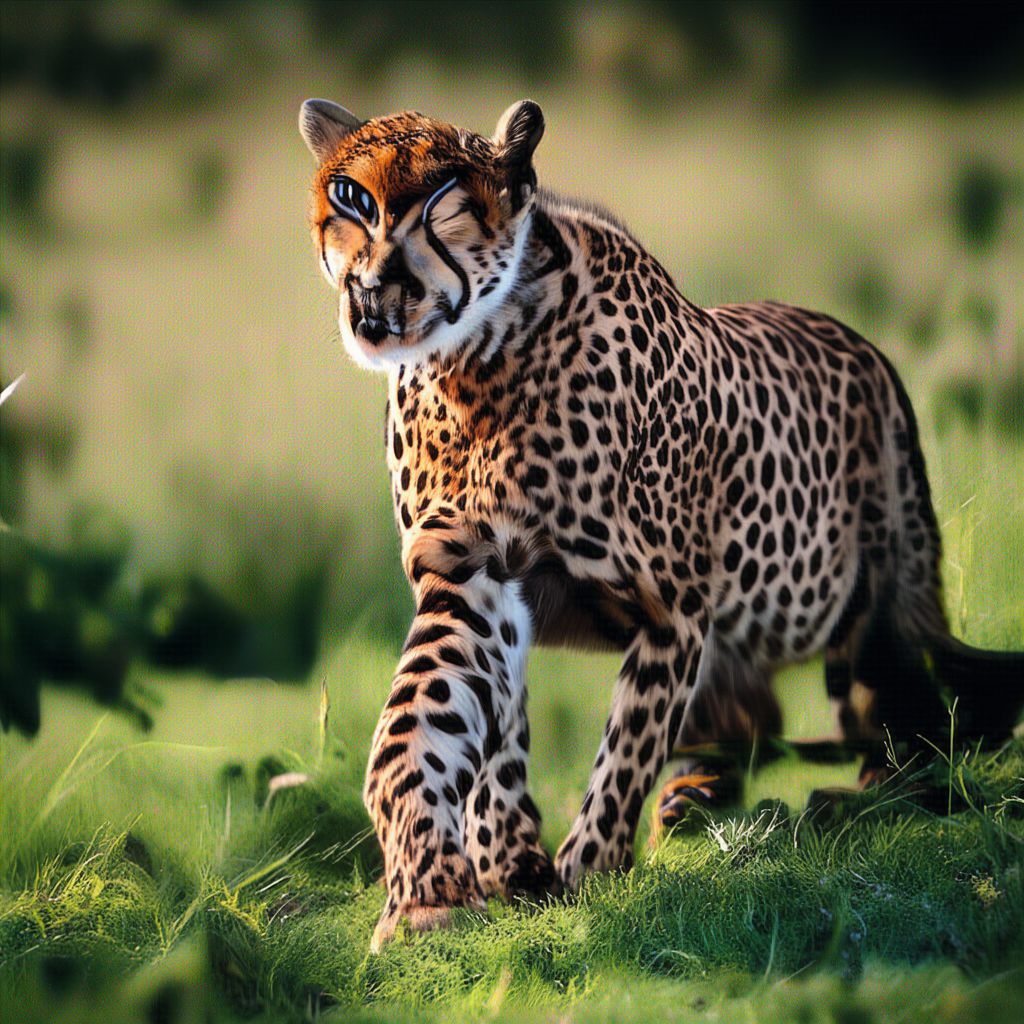} &
\tpairimg{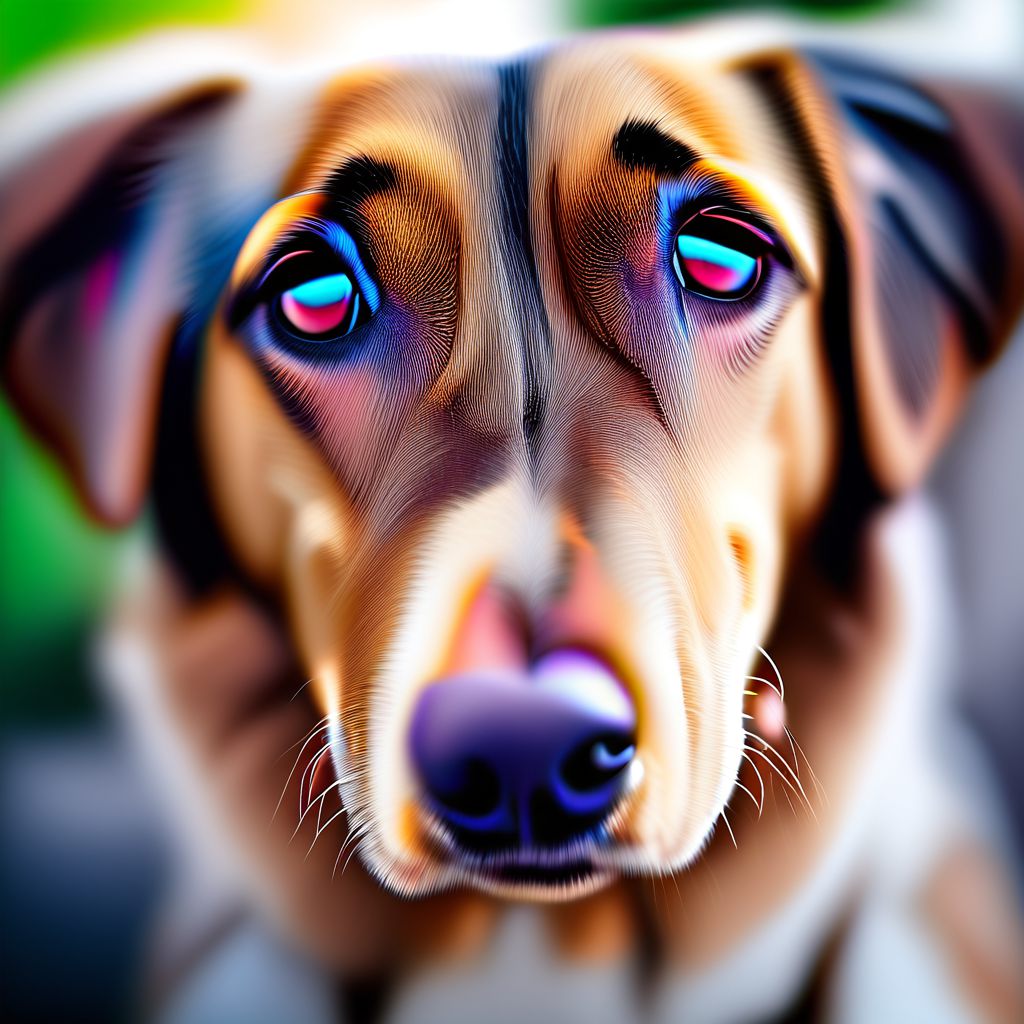}{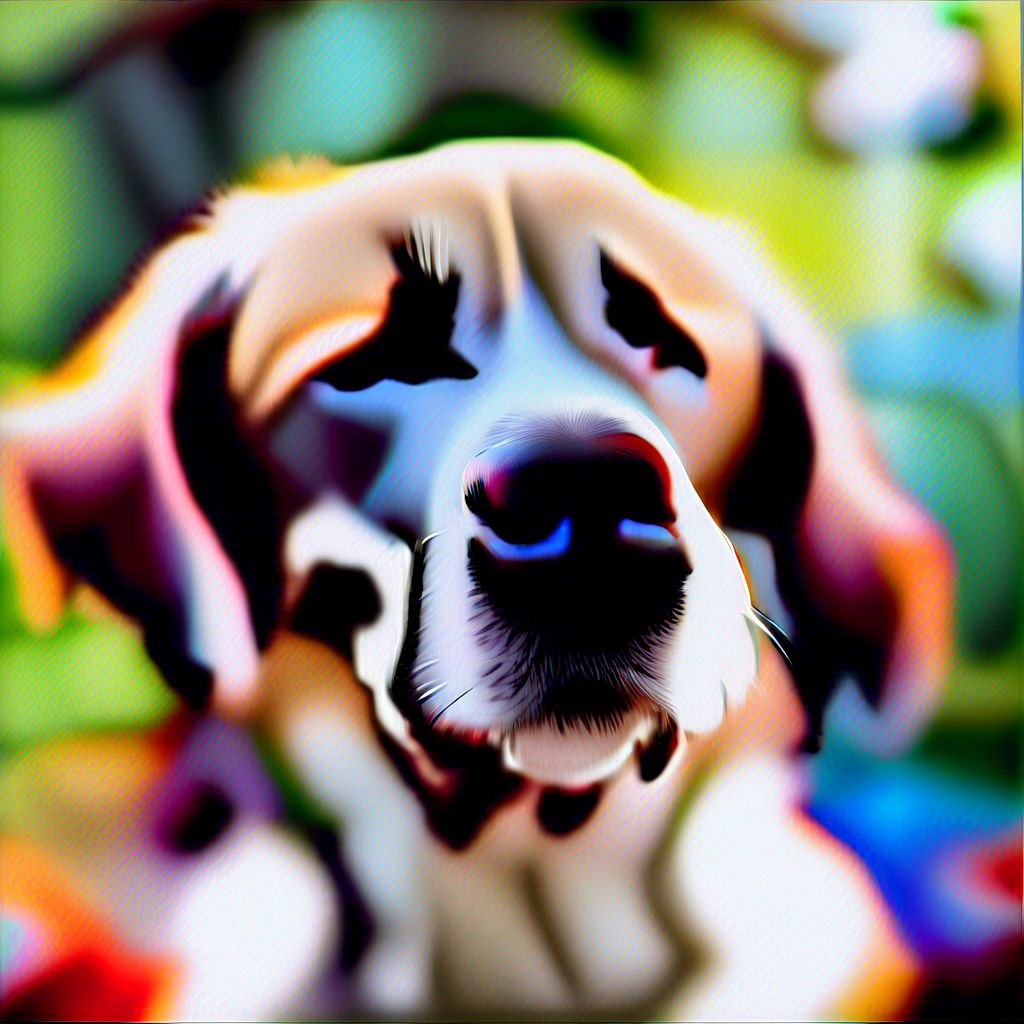} \\
\bottomrule
\end{tabularx}
\caption{Experimental transferability case study on AFHQ with PixArt as the source model. 
FLUX2klein and SSD are downstream substitute models trained on the same protected distillation dataset. }
\label{fig:transferability_ensemble}
\end{figure}

We evaluate the ensemble setting on AFHQ~\cite{choi2020starganv2} with PixArt~\cite{chen2024pixartsigma} as the source model using FLUX2klein~\cite{flux2} and SSD~\cite{segmind2023ssd1b} downstream as the substitute models; the other settings follow the main experiments.
As shown in Figure~\ref{fig:transferability_ensemble}, substitutes trained on ensemble-protected outputs exhibit clear visual degradation compared with their clean substitutes.
The FLUX2klein result shows structural distortion and color artifacts in the background, while the SSD result shows severe structural distortion and color shift.
These results suggest that ensemble surrogate training can substantially broaden the transferability of \method{} beyond a single surrogate family.
Importantly, this strategy is practical in the current T2I ecosystem: while the number of models is large, they often rely on a relatively small set of VAE backbones or closely related VAE families. 
As a result, training against an ensemble of representative surrogate VAEs allows \method{} to cover a much broader range of plausible substitute models, thereby expanding the effective defense surface without assuming knowledge of the attacker's exact model.\looseness=-1


\begin{table}[!t]
    \centering
    \caption{Visual-metric ablation. \mLP, \mSS, \mDE, and \mPN{} denote LPIPS, SSIM, CIEDE2000, and PSNR, respectively.}
    \label{tab:ablation_visual_metrics}
    \scriptsize
    \setlength{\tabcolsep}{2.2pt}
    \renewcommand{\arraystretch}{1.10}
    \begin{tabularx}{0.8\columnwidth}{@{}>{\raggedright\arraybackslash}Xcccc@{}}
    \toprule
    \rowcolor{tablepanel}\multicolumn{5}{c}{\textbf{Ablation Study of Composite Visual Constraint}} \\
    \midrule
    \rowcolor{tablehead}\textbf{Visual loss} & \textbf{\mLP}$\downarrow$ & \textbf{\mSS}$\uparrow$ & \textbf{\mDE}$\downarrow$ & \textbf{\mPN}$\uparrow$ \\
    \midrule
    $\mathcal{L}_{cie}$ & 0.376 & 0.647 & 1.940 & 28.73 \\
    $\mathcal{L}_{cie}+\mathcal{L}_{lpips}$ & 0.046 & 0.768 & 1.903 & 33.14 \\
    $\mathcal{L}_{cie}+\mathcal{L}_{lpips}+\mathcal{L}_{tv}$ & 0.046 & 0.857 & 1.850 & 34.76 \\
    $\mathcal{L}_{cie}+\mathcal{L}_{lpips}+\mathcal{L}_{tv}+\mathcal{L}_{patch}$ & 0.043 & 0.900 & 1.551 & 36.34 \\
    \bottomrule
    \end{tabularx}
\end{table}



\begin{figure}[tbp]
    \centering
    \includegraphics[width=\columnwidth]{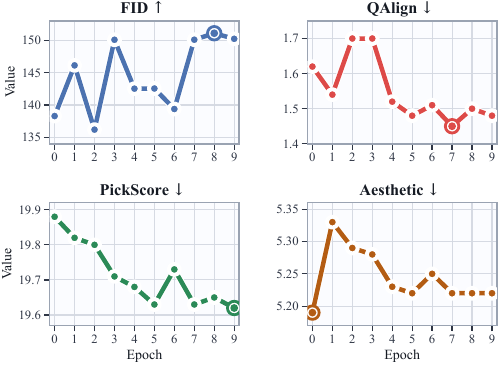}
    \caption{Ablation study of distillation training epochs.}
    \label{fig:ablation_epoch_trend}
\end{figure}

\begin{figure}[t]
\centering
\scriptsize
\setlength{\tabcolsep}{1.0pt}
\renewcommand{\arraystretch}{1.06}
\newcommand{\epochimg}[1]{\raisebox{-.5\height}{\includegraphics[width=\linewidth]{figures_jpeg/ablation_epoch/#1}}}
\begin{tabularx}{\columnwidth}{@{}>{\raggedright\arraybackslash}p{.14\columnwidth}*{5}{>{\centering\arraybackslash}X}@{}}
\toprule
\rowcolor{tablepanel}\multicolumn{6}{c}{\textbf{Ablation Study of Decoder-Training Epochs}} \\
\midrule
\rowcolor{tablehead}\textbf{Epoch} & \textbf{0} & \textbf{1} & \textbf{2} & \textbf{3} & \textbf{4} \\
\textbf{Example} & \epochimg{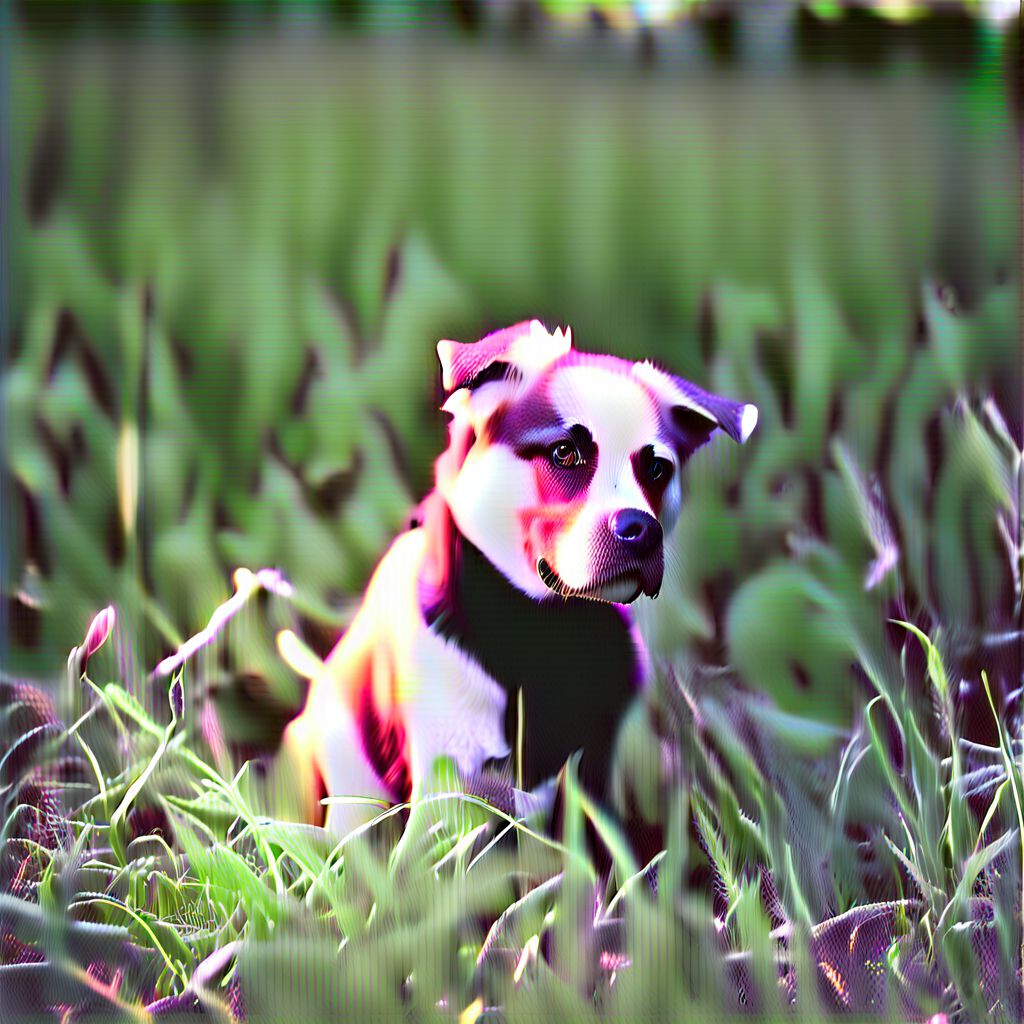} & \epochimg{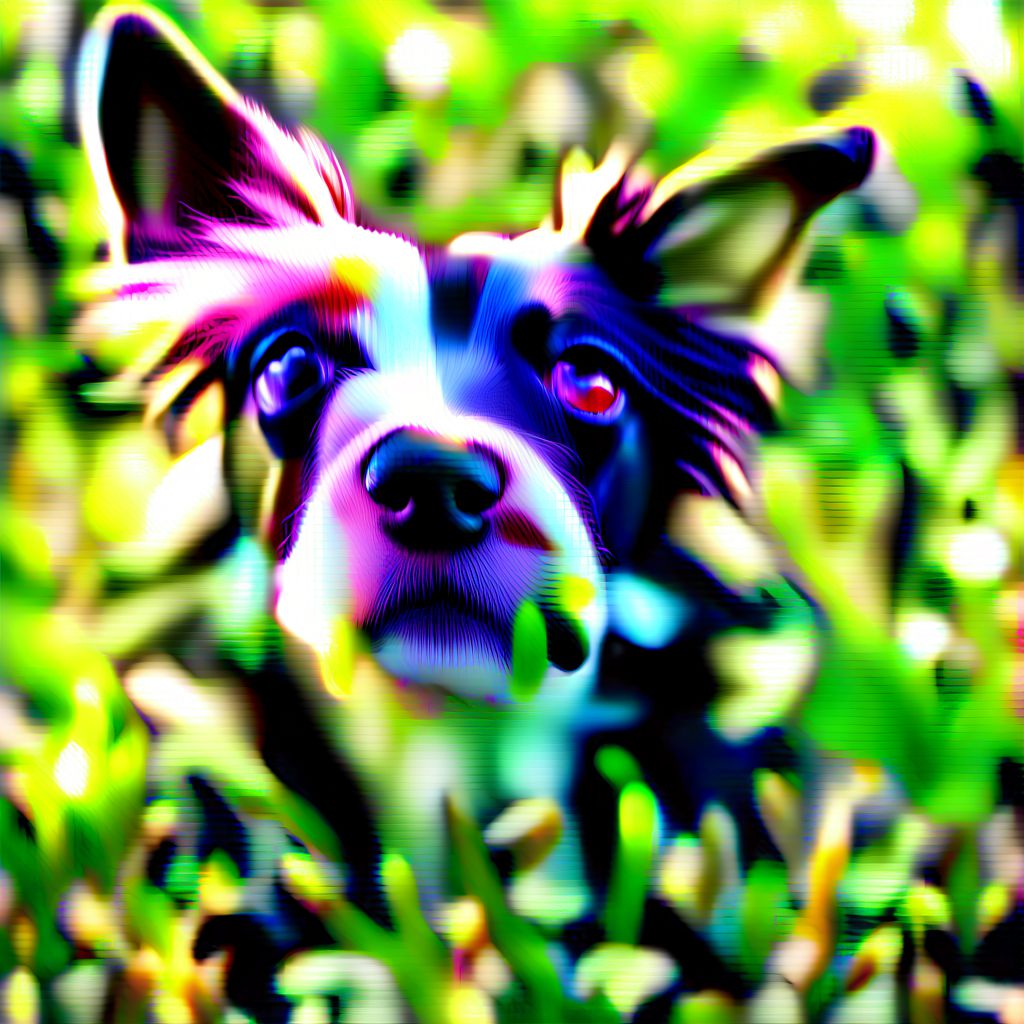} & \epochimg{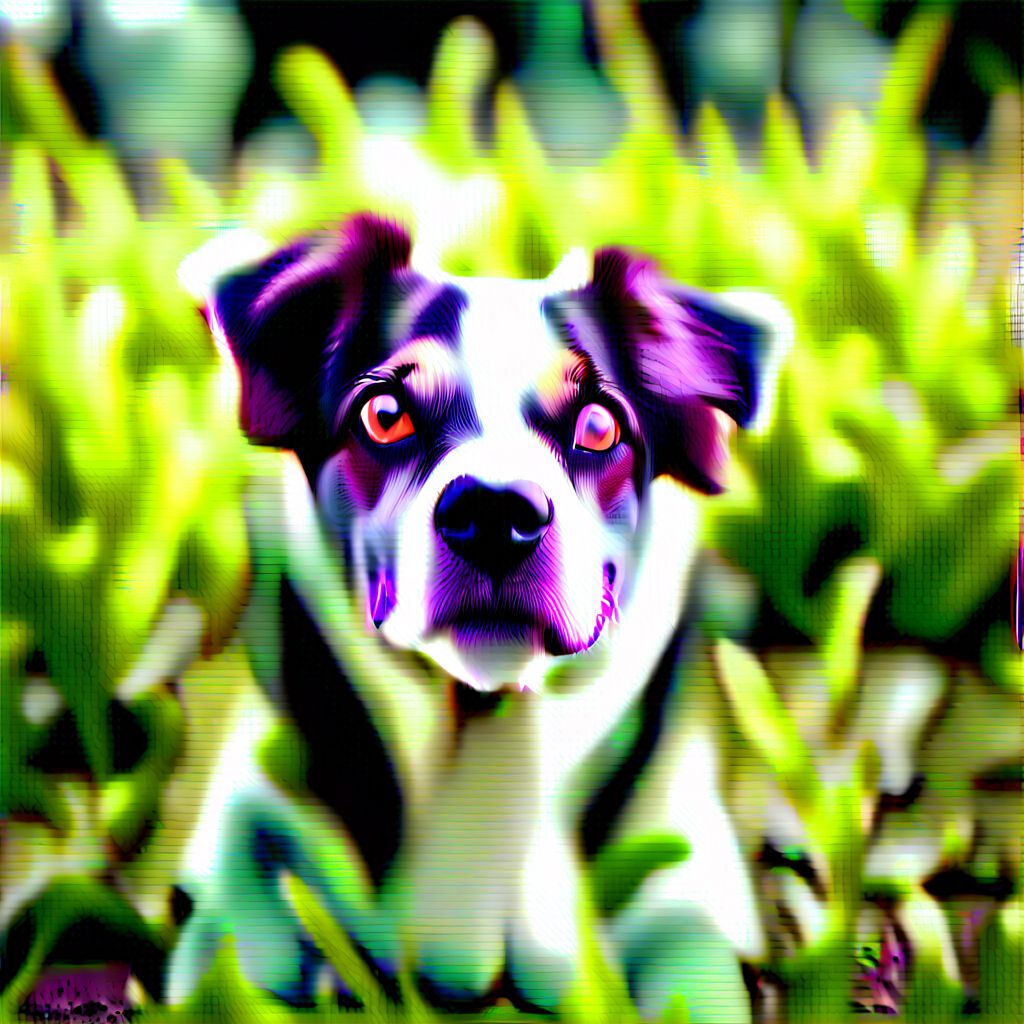} & \epochimg{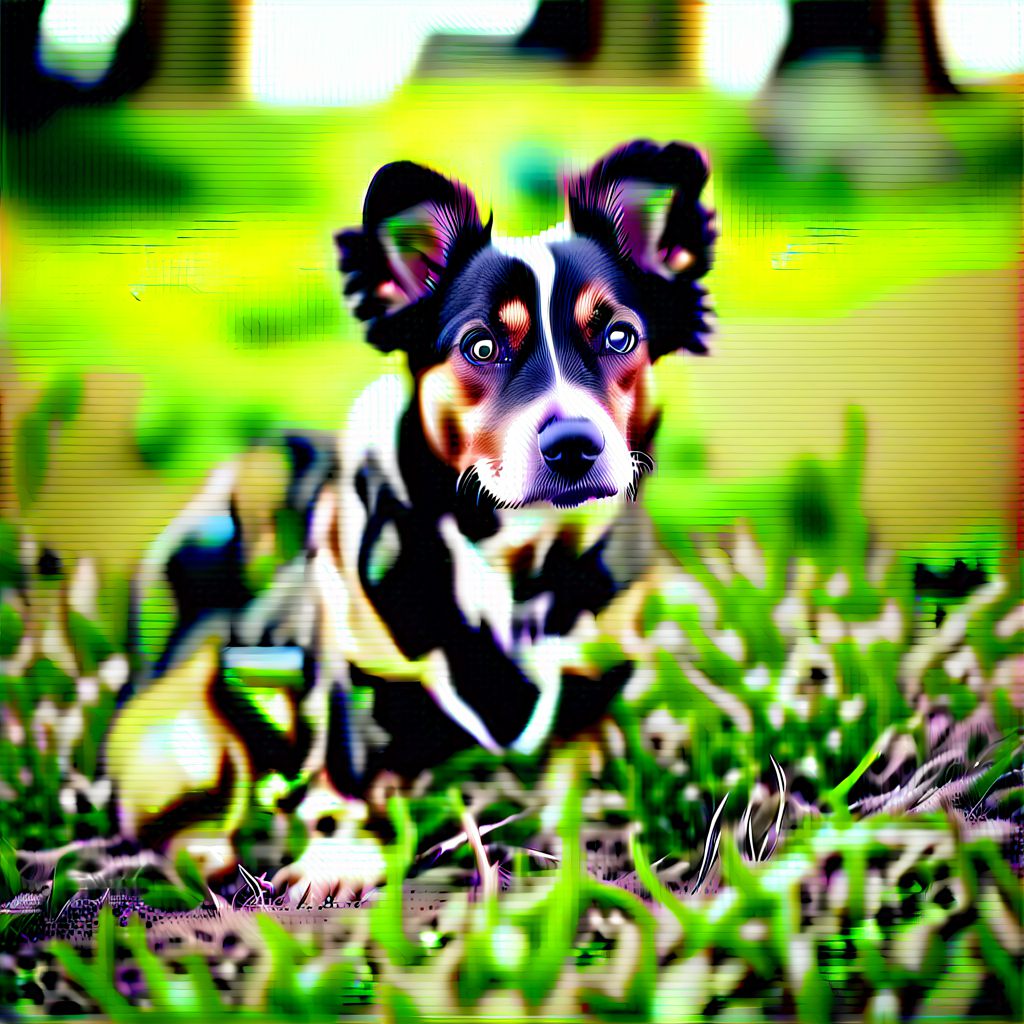} & \epochimg{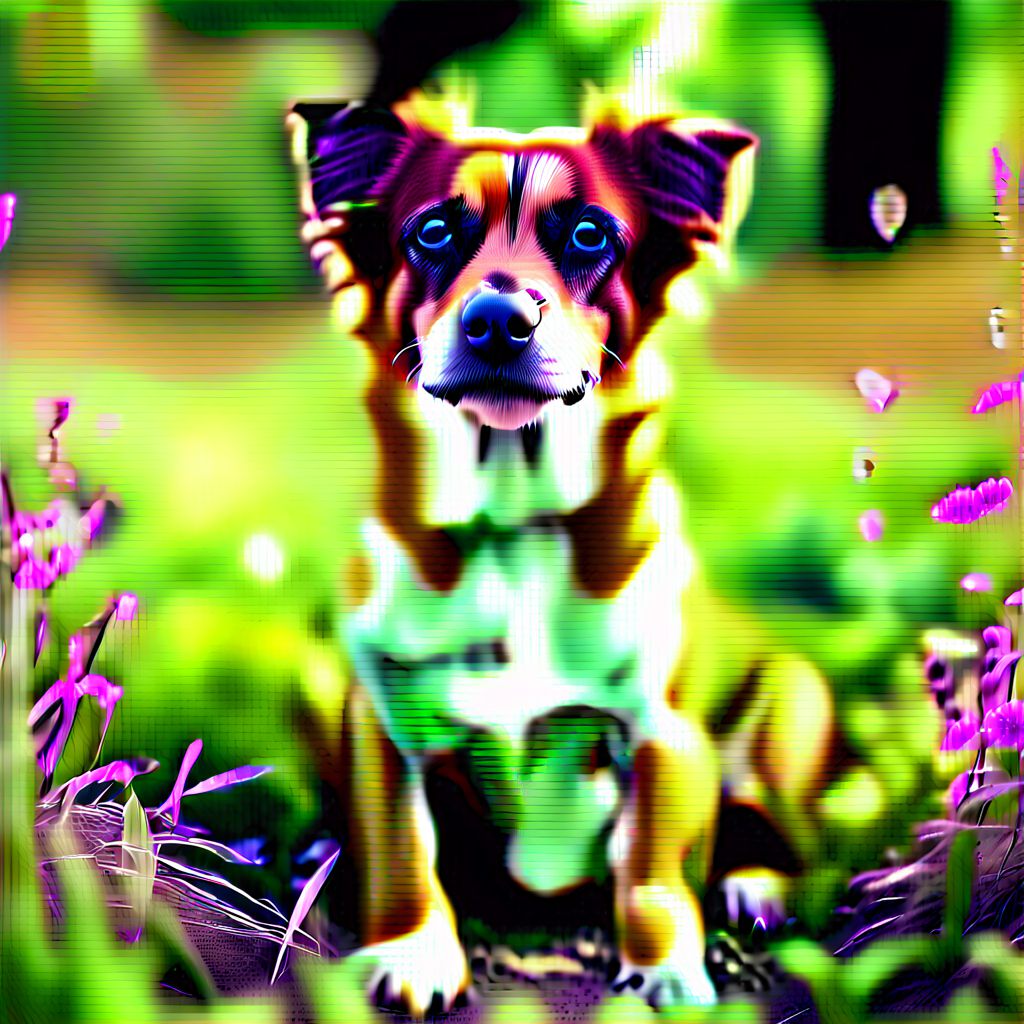} \\
\midrule
\textbf{\mFID}$\uparrow$ & 138.29 & 146.13 & 136.19 & 150.09 & 142.53 \\
\textbf{\mQA}$\downarrow$ & 1.62 & 1.54 & 1.70 & 1.70 & 1.52 \\
\textbf{\mPK}$\downarrow$ & 19.88 & 19.82 & 19.80 & 19.71 & 19.68 \\
\textbf{\mAES}$\downarrow$ & 5.19 & 5.33 & 5.29 & 5.28 & 5.23 \\
\midrule
\rowcolor{tablehead}\textbf{Epoch} & \textbf{5} & \textbf{6} & \textbf{7} & \textbf{8} & \textbf{9} \\
\textbf{Example} & \epochimg{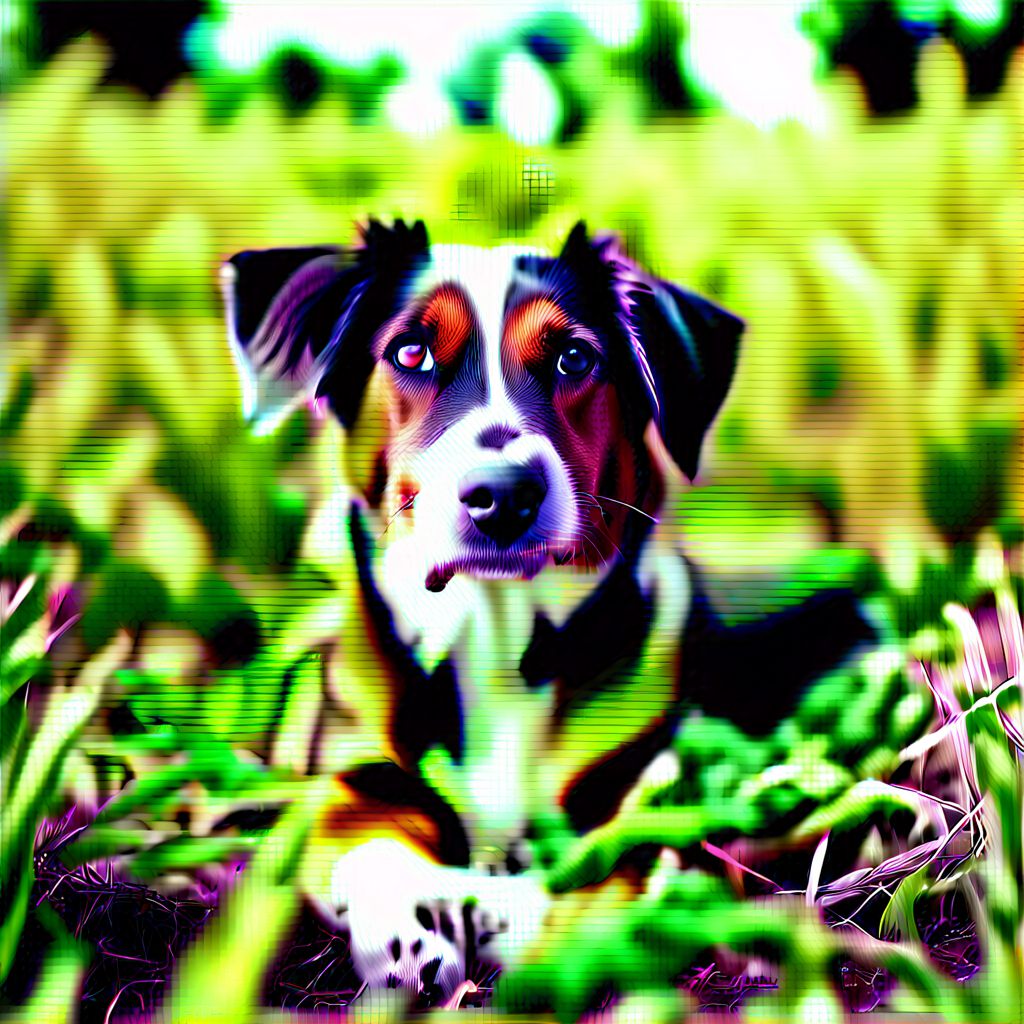} & \epochimg{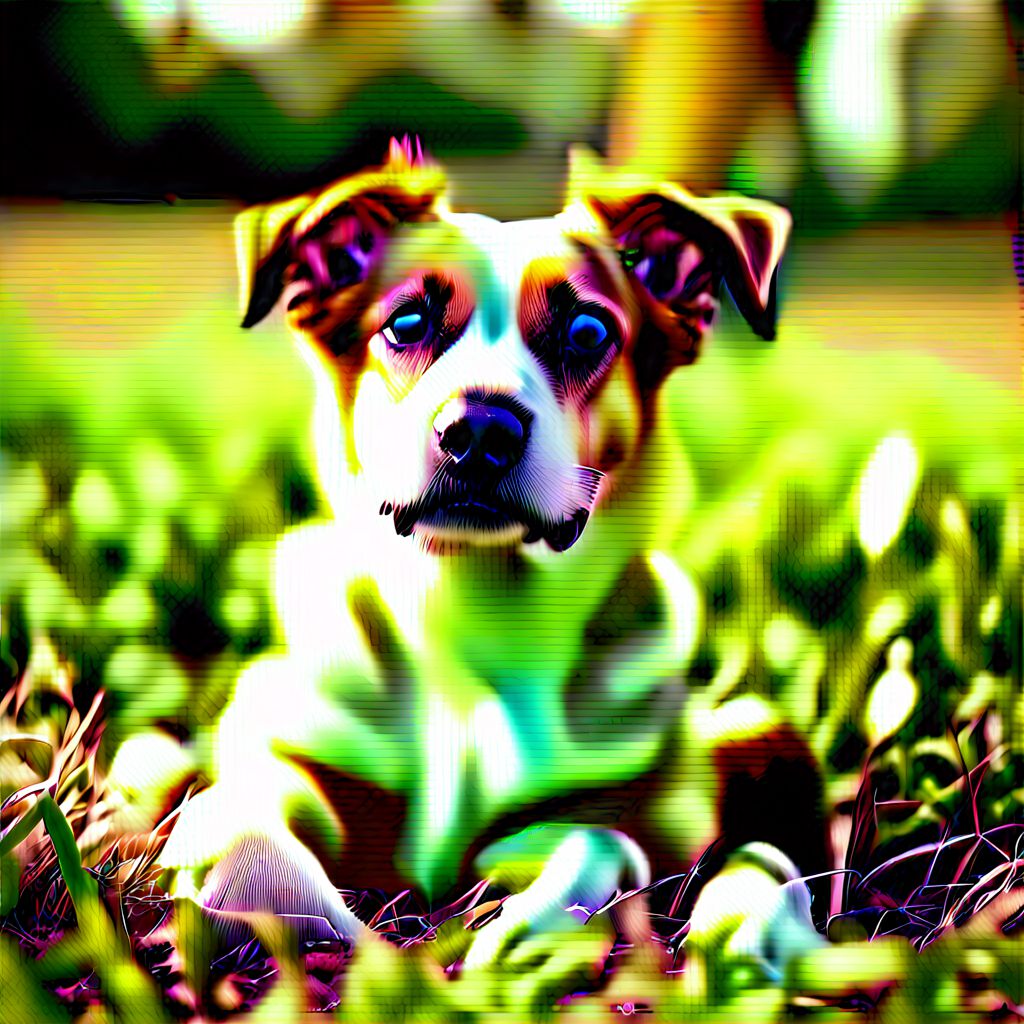} & \epochimg{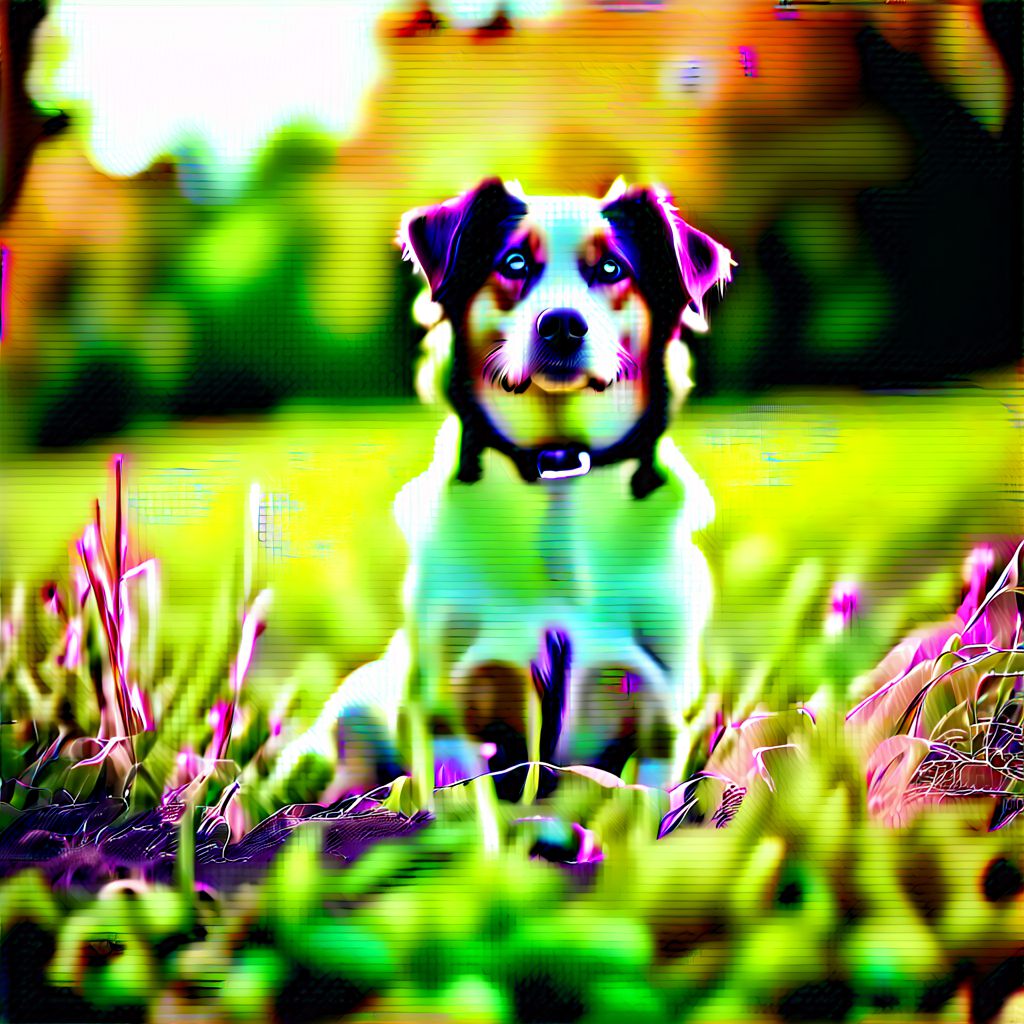} & \epochimg{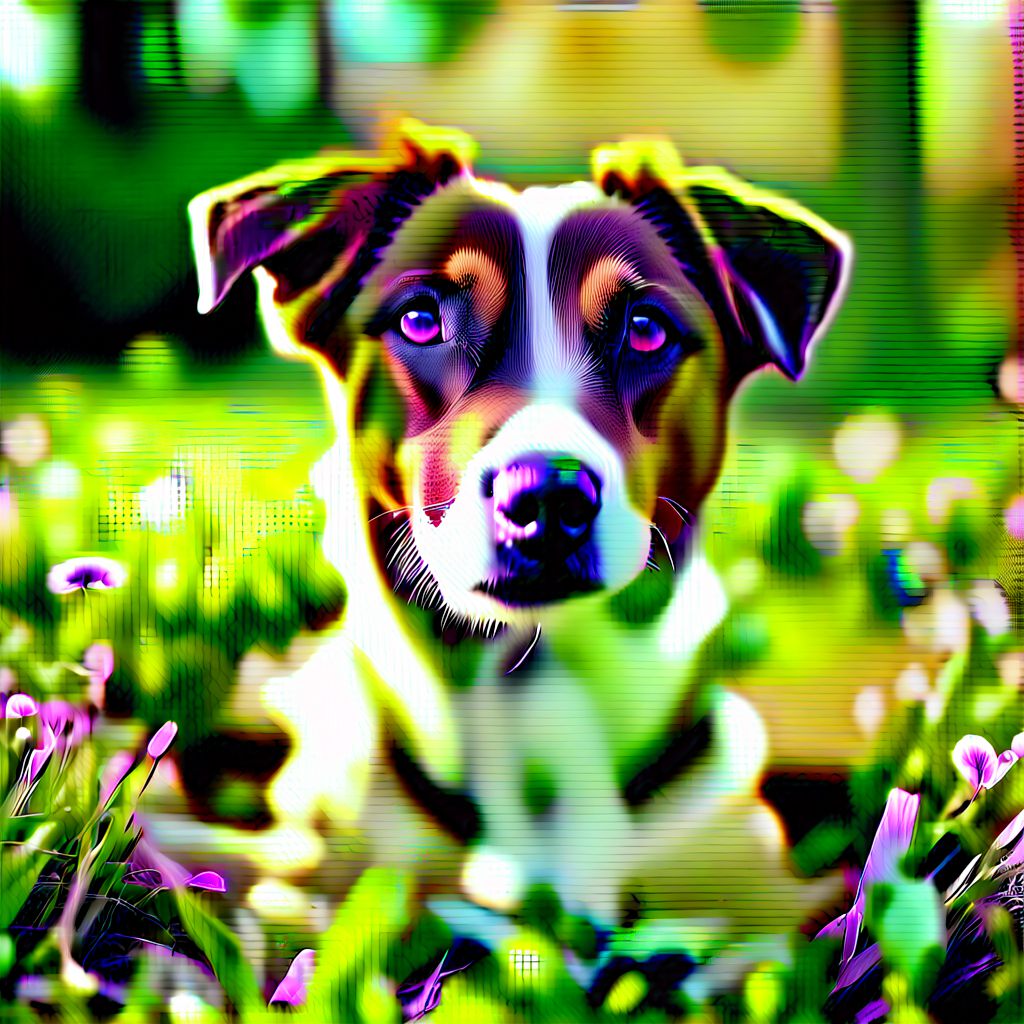} & \epochimg{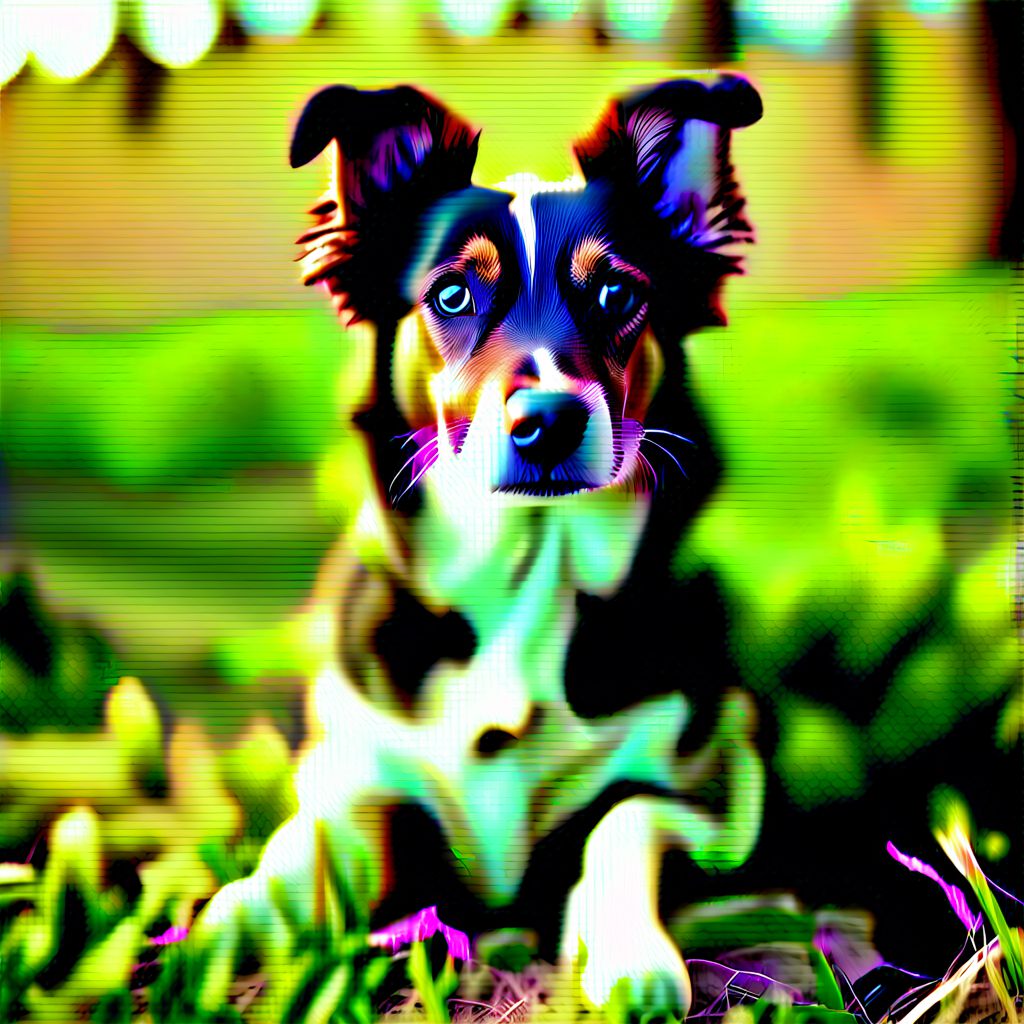} \\
\midrule
\textbf{\mFID}$\uparrow$ & 142.55 & 139.38 & 150.10 & 151.10 & 150.24 \\
\textbf{\mQA}$\downarrow$ & 1.48 & 1.51 & 1.45 & 1.50 & 1.48 \\
\textbf{\mPK}$\downarrow$ & 19.63 & 19.73 & 19.63 & 19.65 & 19.62 \\
\textbf{\mAES}$\downarrow$ & 5.22 & 5.25 & 5.22 & 5.22 & 5.22 \\
\bottomrule
\end{tabularx}
\caption{Ablation study of decoder-training epochs.}
\label{fig:ablation_epoch}
\end{figure}

\subsection{Ablation Study}
\mypara{Different Decoder Training Epochs.}
We evaluate training duration on AFHQ with SSD downstream evaluation.
Figure~\ref{fig:ablation_epoch} shows that training duration controls the strength and type of downstream degradation.
Specifically, QAlign reaches its lowest value at epoch 7, with a score of 1.45.
The best epoch depends on the metric: epoch 7 gives the lowest QAlign, epoch 8 gives the highest FID at 151.10, epoch 9 gives the lowest PickScore at 19.62, and epoch 0 gives the lowest Aesthetic score at 5.19.
From a visual perspective, we further inspect the generated samples at different epochs. 
We observe that the samples at epoch 0 exhibit relatively better visual quality, but their quality is still clearly inferior to that under the Clean setting. 
After epoch 0, the generated images remain consistently poor in visual quality, with only minor differences across subsequent epochs. Starting from epoch 1, the generated samples show pronounced color shifts, along with noticeable noise and grid-like artifacts in the background. 
This suggests that the model may have largely converged after epoch 1.
Figure~\ref{fig:ablation_epoch_trend} illustrates the epoch-wise performance trends. 
While the metrics are not strictly monotonic, the overall results show that increasing the number of training epochs generally degrades the generation quality of the substitute model, with the degradation tending to stabilize after the seventh epoch.
This finding demonstrates that the disruptive capability of \method{} increases with training depth, suggesting that its defensive effectiveness becomes more pronounced in large-scale industrial model distillation.

\mypara{Composite Visual Constraint Design.}
We conduct an ablation study on the composite visual constraint design to evaluate its effectiveness on the CelebA dataset using PixArt as the source model and SSD as the surrogate model.
We find that visual quality can be maintained within a desired range only when all four loss terms are used jointly. 
In addition, when using only the CIEDE2000 loss $\mathcal{L}_{cie}$, it achieves a low color difference, which is 1.94, but LPIPS still remains very poor at 0.376. 
This indicates that relying on a single visual constraint is insufficient and supports the need for composite visual constraints.\looseness=-1

\section{Additional Discussion}\label{Appendix:discussion}
\mypara{Defense Surface for Open-Source Distillation.}
\method{} can also support defenses for open-source T2I models when misuse still relies on image-level supervision.
We consider an open-source T2I model provider that releases an open-source model while seeking to discourage unauthorized distillation or downstream finetuning using generated images.
Because \method{} is integrated into the generation path, the protective signal is intrinsic to each released image before attacker-side collection begins.
This gives it a broader defense surface than post-processing defenses, which can be skipped, replaced, or inconsistently applied after generation.
However, we do not claim that our defense can withstand soft distillation attacks. 
If an attacker bypasses pixel-level training and directly accesses teacher-side logits, denoising predictions, latent trajectories, or intermediate features, the defense may become ineffective. 
Nevertheless, as an intrinsic defense, \method{} substantially expands the protected surface across output-collection pipelines, while not claiming to protect interfaces that expose internal teacher signals.

\end{document}